\documentclass[prd,aps,
tightenlines,
superscriptaddress,showpacs,nofootinbib
,eqsecnum,amsfonts,amsmath
,twocolumn
]{revtex4-1}
\usepackage{bm,graphics,graphicx,epsfig,color
,subfigure,rotating,latexsym,textcomp
}
\usepackage{amsthm}
\usepackage{amssymb}
\usepackage{url}
\usepackage{fancyhdr}
\usepackage{alltt}
\usepackage{acronym}

\def\be{\begin{equation}}
\def\ee{\end{equation}}

\def\bea{\begin{eqnarray}}
\def\eea{\end{eqnarray}}

\def\cross{\times}

\begin{document}

\title{A targeted coherent search for gravitational waves from compact binary
coalescences}

\author{I.\ W.\ Harry}
\email{ian.harry@astro.cf.ac.uk}
\affiliation{School of Physics and Astronomy, Cardiff University, Queens 
Buildings, The Parade, Cardiff, CF24 3AA, UK}
\author{S.\ Fairhurst}
\email{stephen.fairhurst@astro.cf.ac.uk}
\affiliation{School of Physics and Astronomy, Cardiff University, Queens 
Buildings, The Parade, Cardiff, CF24 3AA, UK}

\begin{abstract}

We present the details of a method for conducting a targeted, coherent
search for compact binary coalescences.  The search is tailored to be
used as a followup to electromagnetic transients such as Gamma Ray
Bursts.  We derive the coherent search statistic for Gaussian detector
noise and discuss
the benefits of a coherent, multi-detector search over coincidence
methods.  To mitigate the effects of non-stationary data, we introduce a
number of signal consistency tests, including the null SNR, amplitude
consistency and several $\chi^{2}$ tests.  We demonstrate the search
performance on Gaussian noise and on data from LIGO's fourth science run
and verify that the signal consistency tests are capable of removing the
majority of noise transients and the search gives an efficiency
comparable to that achieved in Gaussian noise.

\end{abstract}
\maketitle

\acrodef{LIGO}[LIGO]{Laser Interferometer Gravitational-wave Observatory}
\acrodef{CBC}[CBC]{Compact binary coalescences}
\acrodef{S6}[S6]{LIGO's sixth science run}
\acrodef{VSR23}[VSR2 and VSR3]{Virgo's second and third science runs}
\acrodef{EM}[EM]{electromagnetic}
\acrodef{GW}[GW]{gravitational wave}
\acrodef{NS}[NS]{neutron star}
\acrodef{BNS}[BNS]{binary neutron stars}
\acrodef{NSBH}[NSBH]{neutron star--black hole binaries}
\acrodef{GRB}[GRB]{gamma-ray burst}
\acrodef{S5}[S5]{LIGO's fifth science run}
\acrodef{S4}[S4]{LIGO's fourth science run}
\acrodef{VSR1}[VSR1]{Virgo's first science run}
\acrodef{PSD}[PSD]{power spectral density}
\acrodef{VSR3}[VSR3]{Virgo's third science run}
\acrodef{BBH}[BBH]{binary black holes}
\acrodef{SNR}[SNR]{signal-to-noise ratio}
\acrodef{SPA}[SPA]{stationary-phase approximation}
\acrodef{LHO}[LHO]{LIGO Hanford Observatory}
\acrodef{LLO}[LLO]{LIGO Livingston Observatory}
\acrodef{LSC}[LSC]{LIGO Scientific Collaboration}
\acrodef{PN}[PN]{Post-Newtonian}
\acrodef{DQ}[DQ]{data quality}
\acrodef{IFO}[IFO]{interferometer}

\section{Introduction}
\label{sec:introduction}

There has been excellent progress towards gravitational wave astronomy
over recent years.  The first generation of large scale gravitational
wave interferometers reached unprecedented sensitivities and have
undertaken extended science runs.  The U.S. \ac{LIGO}
\cite{Abbott:2009li}, the French--Italian Virgo \cite{Acernese:2006bj}
and the German--British GEO600 \cite{Willke:2007zz} detectors now form a
collaborative network of interferometers.  The data from these detectors
has been analyzed for gravitational waves from compact binary
coalescence \cite{Abadie:2010yba}, stochastic background
\cite{Abbott:2009ws}, unmodelled burst \cite{Abadie:2010mt} and pulsar
\cite{Collaboration:2009rfa} sources.  \ac{S6} and \ac{VSR23} ended in
October 2010 and yielded the most sensitive data yet taken; the analysis
of this data is ongoing.  In the meantime, the detectors are being
upgraded to their advanced configurations \cite{Harry:2010zz,avlligowebsite,
advvirgowebsite}, with the expectation of a ten fold improvement in
sensitivity.  With these sensitivities, it is expected that
gravitational waves will be observed regularly \cite{Temp:2010cfa}.
Furthermore, with a proposed advanced detector in Japan
\cite{lcgtwebsite}, a possible detector in Australia \cite{aigowebsite},
and 3rd generation detectors on the horizon \cite{Hild:2008ng}, future
prospects are promising.

As the gravitational wave community matures it is essential that a
relationship is built between \ac{GW} and \ac{EM} astronomers.  The
\ac{GW} emission from a source is likely to provide complementary
information to emission in various \ac{EM} bands, and a joint
observation is significantly more likely to answer outstanding
astrophysical questions.  Already this relationship is beginning to
mature.  A number of \ac{EM} transients have already been followed up in
\ac{GW} data \cite{Abbott:2007rh, Abadie:2010uf, Collaboration:2009kk}.
Additionally, infrastructure is also being put in place to allow for
\ac{EM} follow-up of \ac{GW} observations \cite{Kanner:2008zh}.

\ac{CBC} are one of the most promising sources of gravitational waves,
and also an ideal candidate for joint \ac{GW}-\ac{EM} astronomy.  During
the late stages of inspiral and merger, a compact binary emits a
distinctive, ``chirping'' gravitational wave signal.  Furthermore,
\ac{CBC}s containing at least one \ac{NS} are expected to emit
electromagnetically.  Specifically, \ac{BNS} and \ac{NSBH} mergers are
the preferred progenitor model for the short \ac{GRB}
\cite{nakar:2007,Shibata:2007zm}. It is also possible that
these mergers will be observable electromagnetically as orphan
afterglows \cite{nakar:2007}, optical \cite{Metzger:2010sy} or radio
transients \cite{Predoi:2009af}.  Since \ac{GRB}s are well localized
both in time and on the sky by \ac{EM} observations, the corresponding
\ac{GW} search can be simplified
by reducing the volume of parameter space relative to an all-sky,
all-time search.  Targeted searches for \ac{CBC} waveforms associated
to short \ac{GRB}s were performed using data from \ac{S5} and \ac{VSR1}
\cite{Abbott:2007rh, Abadie:2010uf}. 

In this paper, we introduce a targeted, coherent search algorithm for
detecting \ac{GW} from \ac{CBC}.  This targeted search is designed as a
follow-up to \ac{EM} transients, and in particular \ac{GRB}s.
Previous searches for this source have made
use of a coincidence requirement --- namely that a signal with
consistent parameters is observed in two or more detectors in the
network.  The analysis introduced here makes use of the data from all
operational detectors, and combines the data in a coherent manner before
matched filtering against \ac{CBC} template waveforms.  In a
coherent analysis, it is straightforward to restrict the signal
model to only two independent polarizations. This allows for the
rejection of incoherent background noise, and consequently increases the
sensitivity of the search if more than two detectors are operating.

The data output by gravitational wave interferometers is neither
stationary nor Gaussian, but is contaminated by noise transients of
instrumental and environmental origin.  This makes the task of doing
data analysis a complex one, and matched filtering alone is not
sufficient to distinguish signal from noise.  Regardless of whether a
coincident or coherent search is performed, the most significant events
by \ac{SNR} would always be dominated by non-Gaussian transients, or
``glitches'', in the data.  A significant effort goes into understanding
the cause of these glitches \cite{Blackburn:2008ah} and removing times
of poor data quality from the analysis.  While these efforts greatly
reduce the number of glitches they cannot remove them entirely.
Therefore the analysis must also employ methods to distinguish signal
from noise transients.  In previous \ac{CBC} searches, signal
consistency tests \cite{Allen:2004gu, Hanna:2008} have proved very
effective at removing the non-Gaussian background.  We extend these
tests to the coherent analysis described in this paper and demonstrate
their continued effectiveness.  In addition, coherent analyses naturally
lend themselves to multi-detector consistency tests, such as the null
stream \cite{Guersel:1989th}.  We describe a number of such consistency
tests for this templated \ac{CBC} search and again demonstrate their
efficacy.  Finally, we are able to show that the various signal
consistency tests are sufficient to remove the majority of non-Gaussian
transients and render the search almost as sensitive as if the data were
Gaussian and stationary.

The layout of this paper is as follows. In section
\ref{sec:coh_matched_filter}, we describe the formulation of a targeted
coherent triggered search for \ac{CBC} signals.  In section
\ref{sec:snr_cont_consistency} we discuss an implementation of the null
stream formalism and other multi-detector consistency tests.  In section
\ref{sec:chi2tests} we describe a number of $\chi^{2}$ tests than can be
applied in a coherent search to try to separate and veto glitches.
Finally, in \ref{sec:s4data} we outline an implementation of a
targetted, coherent search for \ac{CBC} and present results on both
simulated, Gaussian data and real detector data taken from \ac{S4}.

\section{Coherent Matched filtering}
\label{sec:coh_matched_filter} 

In this section, we describe the coherent
matched-filtering search for a gravitational wave signal from a
coalescing binary in data from a network of detectors.  We restrict attention to
binaries where the component spin can be neglected.  The description is
primarily tailored towards searches where the sky location of the
gravitational wave event is known a priori, as is appropriate when
performing a followup of an \ac{EM} transient such as a \ac{GRB}
\cite{nakar:2007, Abadie:2010uf}.  Finally, since all previously
published \ac{CBC} search results \cite{Abadie:2010yba, Abbott:2009qj,
Abbott:2009tt} have used a coincidence search between detectors, we
compare the coherent analysis with the multi-detector coincident
analysis.

The coherent analysis for coalescing binary
systems has been derived previously using a similar method
in \cite{Pai:2000zt, Bose:1999pj,Bose:1999bp,Bose:2011}.
Our presentation makes use of the $\mathcal{F}$-statistic formalism,
introduced in \cite{Jaranowski:1998qm}. This was originally defined as a
method for performing searches for continuous wave searches and has been
regularly used for this task (see for example \cite{Abbott:2009nc}).
It was noted in \cite{Cutler:2005hc} that the $\mathcal{F}$-statistic and
the multiple detector inspiral statistic derived in \cite{Bose:1999pj}
are similar and the $\mathcal{F}$-statistic was adapted to searches 
for \ac{CBC} signals in \cite{Cornish:2006ms}. 

\subsection{The binary coalescence waveform}
\label{ssec:binary_waveform}

The generic binary coalescence waveform depends upon as many as
seventeen parameters.  However, we restrict attention to binaries on
circular orbits with non-spinning components.  This reduces the
parameter space to nine dimensions: the two component masses $M_{1}$ and
$M_{2}$;  the sky location of the signal ($\theta$, $\phi$); the
distance, D, to the signal; the coalescence time of the signal, $t_{o}$;
the orientation of the binary, given by the inclination $\iota$, the
polarization angle $\psi$ and the coalescence phase $\phi_{o}$.
We also assume that the sky location ($\theta$, $\phi$) of
the signal is known, thereby reducing the number of unkown parameters to
seven.  

In the radiation frame, where the gravitational wave propagates in the
$\mathbf{e}_{z}^{R}$-direction, the gravitational waveform is given by
\begin{equation}
  \mathbf{h} = h_{+} \mathbf{e}_{+} + h_{\times} \mathbf{e}_{\times}
\end{equation}
where
\begin{eqnarray}
  \mathbf{e}_{+} &=& \mathbf{e}^{R}_{x} \otimes \mathbf{e}^{R}_{x} - 
    \mathbf{e}^{R}_{y} \otimes \mathbf{e}^{R}_{y} \, , \nonumber \\
  \mathbf{e}_{\times} &=& \mathbf{e}^{R}_{x} \otimes \mathbf{e}^{R}_{y} + 
    \mathbf{e}^{R}_{y} \otimes \mathbf{e}^{R}_{x} \, ,
\end{eqnarray}
and the waveforms $h_{+, \times}$ depend upon
seven parameters ($M_{1}, M_{2}, t_{o}, D, \iota, \psi, \phi_{o}$).  The
three remaining angles $(\iota, \psi, \phi_{o})$ give the relationship
between the radiation frame and the source frame (in which
$\mathbf{e}^{S}_{z}$ lies in the direction of the binary's angular
momentum and $\mathbf{e}^{S}_{x}$ along the separation between the
binary components at $t_{o}$).  Even for a known sky location, it is
necessary to search a seven dimensional parameter space of signals.
Naively covering this space with a grid of templates would be
prohibitively costly \cite{OwenSathyaprakash98}.  However, the analysis
is greatly simplified by the observation that the last four parameters
enter only as amplitude parameters which can be analytically maximized
over at minimal cost.%
\footnote{This was observed for the inspiral signal in
\cite{Bose:1999pj} and independently for continuous wave signals in
\cite{Jaranowski:1998qm}.}  
Specifically, the two polarizations of the waveform can be expressed as 
\begin{eqnarray}\label{eq:h_plus_cross}
h_{+}(t) &=&  \mathcal{A}^{1} h_{0}(t) + \mathcal{A}^{3}
h_{\frac{\pi}{2}}(t)
\nonumber \\
h_{\times}(t) &=& \mathcal{A}^{2} h_{0}(t) + 
\mathcal{A}^{4} h_{\frac{\pi}{2}}(t) \, .
\end{eqnarray}
The two phases of the waveform are written as $h_{0}$ and
$h_{\frac{\pi}{2}}$.  These depend upon the physical parameters of the
system (in this case just the masses) as well as the coalescence time
$t_{o}$.%
\footnote{This decomposition is actually valid for all binaries in which
the plane of the orbit does not precess.  Thus, binary coalescence
waveforms in which the spins are aligned with the orbital angular
momentum can also be expressed in this form.  However, for generic spin
configurations, the orbit will precess and this simple decomposition is
no longer applicable}
 $\mathcal{A}^{i}$ are constant amplitude terms and are given
explicitly as \cite{Cornish:2006ms,Bose:2011}
\begin{eqnarray} \label{eq:amplitude_def}
\mathcal{A}^{1} &=& 
  A_{+} \cos 2 \phi_{o} \cos 2 \psi - A_{\times}\sin 2 \phi_{o} \sin 2 \psi 
  \\
\mathcal{A}^{2} &=& 
 A_{+} \cos 2 \phi_{o} \sin 2 \psi + A_{\times} \sin 2 \phi_{o} \cos 2 \psi 
  \nonumber \\
\mathcal{A}^{3} &=& 
 - A_{+} \sin 2 \phi_{o} \cos 2 \psi - A_{\times} \cos 2 \phi_{o} \sin 2 \psi 
  \nonumber \\
\mathcal{A}^{4} &=& 
 - A_{+} \sin 2 \phi_{o} \sin 2 \psi + A_{\times} \cos 2 \phi_{o} 
  \cos 2 \psi \, , \nonumber 
\end{eqnarray}
where
\begin{eqnarray}\label{eq:aplus_across}
  A_{+} &=&  \frac{D_{o}}{D}\frac{(1 + \cos^{2} \iota)}{2} \nonumber \\
  A_{\times} &=&  \frac{D_{o}}{D}\cos \iota \, ,
\end{eqnarray}
and $D_{o}$ is a fiducial distance which is used to scale
the amplitudes $\mathcal{A}^{i}$ and waveforms $h_{0, \frac{\pi}{2}}$.
Thus, the amplitudes $\mathcal{A}^{i}$ depend upon the distance to the
source and the binary orientation as encoded in the three angles
($\iota$, $\psi$, $\phi_{0}$).  For \textit{any} set of
values $\mathcal{A}^{i}$, the expressions (\ref{eq:amplitude_def}) can
be inverted to obtain the physical parameters, unique up to reflection
symmetry of the system \cite{Cornish:2006ms}.

The gravitational waveform observed in a detector $X$ is 
\begin{equation}
  h^{X} = h^{ij} D^{X}_{ij}
\end{equation}
where $D^{X}_{ij}$ denotes the detector response tensor.  For an
interferometric detector, the response tensor is given by
\begin{equation}
  \mathbf{D}^{X} = (\mathbf{e}^{X}_{x} \otimes \mathbf{e}^{X}_{x} 
    - \mathbf{e}^{X}_{y} \otimes \mathbf{e}^{X}_{y})
\end{equation}
where the basis vectors $\mathbf{e}^{X}_{x}$ and $\mathbf{e}^{X}_{y}$
point in the directions of the arms of the detector.  It is often
convenient to re-express the gravitational wave signal observed in a
given detector as
\begin{equation}
  h^{X}(t) = F_{+}(\theta^{X}, \phi^{X}, \chi^{X}) h_{+}(t) 
  + F_{\times}(\theta^{X}, \phi^{X}, \chi^{X}) h_{\times}(t) \, ,
\end{equation}
where the detector response to the two polarizations of the
gravitational wave is encoded in the functions
\begin{eqnarray}\label{eq:det_response}
  F_{+}(\theta, \phi, \chi) = 
            &-& \frac{1}{2} ( 1 + \cos^{2} \theta) \cos 2\phi \cos 2\chi \nonumber \\
            &-& \cos \theta \sin 2\phi \sin 2\chi \, ,  \\
  F_{\times}(\theta, \phi, \chi) =&& 
            \frac{1}{2} ( 1 + \cos^{2} \theta) \cos 2\phi \sin 2\chi \nonumber  \\
            &-& \cos \theta \sin 2\phi \cos 2\chi.
\end{eqnarray}
These response functions depend upon the three angles ($\theta^{X},
\phi^{X}, \chi^{X}$) which relate the detector frame to the radiation
frame: $\theta^{X}$ and $\phi^{X}$ give the sky location relative to the
detector, while $\chi^{X}$ is the polarization angle between the
detector and the radiation frames.  We have, somewhat unconventionally,
allowed for a polarization angle in transforming from source to
radiation \textit{and} radiation to detector coordinates.  In what
follows, we will often find it convenient to fix the angle $\chi^{X}$ by
explicitly tying it to the detector (or geocentric) frame; for example,
by maximizing the detector (or network) sensitivity to the $+$
polarization.  The angle $\psi$ then describes the orientation of the
source with respect to this preferred radiation frame.  

Since we are considering \ac{CBC} observed in ground-based detectors,
the time that a potential signal would spend in the sensitivity band of
any detector will be short (less than 60s for the initial detectors).
Thus the change in the source's sky location over the observation time
may be neglected, and the angles ($\theta^{X}, \phi^{X}, \chi^{X}$)
can be treated as constants.  When working with a network
of detectors, it is often useful to work in the geocentric frame.  The
location of the source $(\theta, \phi, \chi)$ is measured relative to
this frame, and coalescence time is measured at the Earth's centre.  In
this case, the location and orientation of the detector $X$ are specified
by three angles, which we denote $\vec{\alpha}^{X}$, and the detector
response will depend upon six angles $(\vec{\alpha}^{X}, \theta, \phi,
\chi)$.  Then, the observed signal in a given detector is%
\footnote{We do not give the explicit formula for the response function
dependent on the six angles $(\vec{\alpha}^{X}, \theta, \phi, \chi)$, as
the expression is somewhat lengthy.  It can be obtained by performing
six successive rotations to the detector response tensor to transform
from the detector frame, via the geocentric frame, to the radiation
frame.  The calculation is detailed in \cite{livas87a}.}
\begin{equation}\label{eq:hx_t}
  h^X(t) = F_{+}(\vec{\alpha}^{X},\theta, \phi, \chi) 
  h_{+}(t^{X} ) 
  + F_{\times}(\vec{\alpha}^{X},\theta, \phi, \chi) 
  h_{\times}(t^{X}) \, ,
\end{equation}
where $t^{X}$ is the time of arrival of the signal at detector X,
\begin{equation}
  t^X = t - dt(\vec{\alpha}^{X},\theta,\phi,\chi)
\end{equation}
and $dt$ gives the difference in arrival time of the signal between the
geocenter and detector, for the given sky position.

Combining the final expressions for the binary coalescence waveform
(\ref{eq:h_plus_cross}) and the detector response (\ref{eq:hx_t}), we
can express the gravitational waveform observed in a given detector as 
\begin{equation}\label{eq:inspiral_wave}
  h^{X}(t) = \sum_{\mu=1}^{4} \mathcal{A}^{\mu}(D, \psi, \phi_{o}, \iota)
h^{X}_{\mu}(t)
\end{equation}
where the $\mathcal{A}^{\mu}$ are defined in (\ref{eq:amplitude_def})
and $h^{X}_{\mu}$ are given by
\begin{eqnarray} \label{eq:fourh_def}
h^{X}_{1}(t) &=& F_{+}^{X} h_{0}(t^{X}) \nonumber\\
h^{X}_{2}(t) &=& F_{\times}^{X} h_{0}(t^{X}) \nonumber\\
h^{X}_{3}(t) &=& F_{+}^{X} h_{\frac{\pi}{2}}(t^{X}) \nonumber \\
h^{X}_{4}(t) &=& F_{\times}^{X} h_{\frac{\pi}{2}}(t^{X}) \, . \label{eq:hmu} 
\end{eqnarray}

\subsection{Multi detector binary coalescence search}
\label{ssec:inspiral_fstat}

Matched filtering theory \cite{Wainstein} provides a method for
determining whether the signal $h(t, \xi)$, parametrized by the time
and other parameters $\xi$, is present in a noisy data stream.  The 
data output by a detector is 
\begin{equation}
  s^{X}(t) = n^{X}(t) + h^{X}(t, \xi)
\end{equation}
where $n^{X}(t)$ is the noise, taken to be Gaussian and stationary.  The
noise $n^{X}(t)$ of the detectors is characterized by the noise \ac{PSD}
$S^{X}_{h}(f)$ as
\begin{equation}
  \langle \tilde{n}^{X}(f) [\tilde{n}^{X} (f^{\prime})]^{\star} \rangle = 
  \delta(f - f^{\prime}) S^{X}_{h}(f) \, .
\end{equation}
With this, we define the single detector inner product between two time
series $a(t)$ and $b(t)$ as
\begin{equation}\label{eq:inner_product}
  (a^{X} | b^{X}) = 4 \, \mathrm{Re} \int_{0}^{\infty}
  \, \frac{\tilde{a}^{X}(f) [\tilde{b}^{X}(f)]^{\star}}{ S^{X}_{h}(f) } \, .
\end{equation}
Then, the likelihood ratio of there being a signal $h$ present in the
data is given by:
\begin{equation}\label{eq:likelihood}
  \Lambda(h) = \frac{P(s|h)}{P(s|0)} = \frac{e^{- (s^{X} - h^{X} | s^{X}
- h^{X})
/2}}{e^{-(s^{X}| s^{X}) /2}} \, ,
\end{equation}
and the log-likelihood can be written as
\begin{equation}
  \log \Lambda = (s|h) - \frac{1}{2} (h|h) \, .
\end{equation}

The likelihood ratio for multiple detectors is a straightforward
generalization of the single detector expression (\ref{eq:likelihood}).
Assuming that the noise in different detectors are independent, in the
sense that
\begin{equation}
  \langle \tilde{n}^{X}(f) [\tilde{n}^{Y}(f^{\prime}]^{\star}) \rangle = 
  \delta^{XY} \delta(f - f^{\prime}) S^{X}_{h}(f) \, ,
\end{equation}
the multi-detector inner product is simply given by the sum of the
single detector contributions
\begin{equation}\label{eq:multi_inner}
  (\mathbf{a} | \mathbf{b} ) := \sum_{X} ( a^{X} | b^{X} ) \, .
\end{equation}
The multi-detector likelihood is given by 
\begin{equation}\label{eq:multi_log_lambda}
  \ln \Lambda = (\mathbf{s} | \mathbf{h} ) - \frac{1}{2}(\mathbf{h} |
\mathbf{h}) \, . 
\end{equation}
This is the optimal statistic for signal detection in Gaussian noise: the larger
the value of $\ln \Lambda$, the more likely that a signal is present.
It is not, however, an optimal statistic if the noise is not Gaussian, as we will
explore in more detail in sections \ref{sec:snr_cont_consistency} and \ref{sec:chi2tests}.

Specializing to the case of binary coalescence, we can substitute the
known waveform parametrization (\ref{eq:inspiral_wave}) into the general
matched filter likelihood (\ref{eq:multi_log_lambda}).  The
multi-detector likelihood becomes
\begin{equation}\label{eq:multi_det_lambda}
  \ln \Lambda = \mathcal{A}^{\mu} (\mathbf{s} | \mathbf{h}_{\mu} ) 
  - \frac{1}{2} \mathcal{A}^{\mu} \mathcal{M}_{\mu \nu} \mathcal{A}^{\nu}
\end{equation}
where the matrix $\mathcal{M}_{\mu \nu}$ is defined as
\begin{equation}
  \mathcal{M}_{\mu \nu} := ( \mathbf{h}_{\mu} | \mathbf{h}_{\nu} ) \, .
\end{equation}
The derivative of (\ref{eq:multi_det_lambda}) with respect to
$\mathcal{A}^{\mu}$ provides the values of $\mathcal{A}^{\mu}$ which
maximize the likelihood as
\begin{equation}
  \hat{A}^{\mu}  = \mathcal{M}^{\mu \nu} (\mathbf{s} | \mathbf{h}_{\nu} ) \, ,
\end{equation}
where, following \cite{Prix:2007rb}, we take $\mathcal{M}^{\mu \nu }$ to
be the inverse of $\mathcal{M}_{\mu \nu}$.  We then define the coherent \ac{SNR}
via the maximum likelihood as
\begin{equation}\label{eq:multi_det_max_lambda}
  \rho^2_{\mathrm{coh}} = 2 \ln \Lambda |_{\rm{max}} =
  (\mathbf{s} | \mathbf{h}_{\mu} ) \mathcal{M}^{\mu \nu} 
  (\mathbf{s} | \mathbf{h}_{\nu} ) \, .
\end{equation}

It is not difficult to show that $\rho^2_{\mathrm{coh}}$ follows a
$\chi^{2}$ distribution with four degrees of freedom in the absence of a
signal, and a non-central $\chi^{2}$ distribution (again with 4 degrees
of freedom) when a signal is present.  See, for example,  \cite{Prix:2007rb} for
more details.  Furthermore, $\rho^2_{\mathrm{coh}}$ is now a function of
only the waveform components $h_{\mu}$ and no longer the
$\mathcal{A}^{\mu}$ parameters.  Thus four of the original seven
waveform parameters have been analytically maximized, leaving three to
be searched over.

Calculating the maximized likelihood, as well as estimating the
parameters $\hat{\mathcal{A}}^{\mu}$ requires an inversion of the matrix
$\mathcal{M}_{\mu \nu}$.  \ac{CBC} signals will spend a large number of
cycles in the sensitive band of the detector and consequently the $0$
and $\frac{\pi}{2}$ phases will be (close to) orthogonal.  Since the
frequency evolves slowly, the amplitudes of the two phases
will be close to equal,%
\footnote{Indeed, several \ac{CBC} waveforms are generated directly in
the frequency domain \cite{Damour:2000zb}, making these equalities exact.}
 i.e.
\begin{eqnarray}
  ( h^{X}_{0} | h^{X}_{\frac{\pi}{2}} ) &\approx& 0 \, \\
  ( h^{X}_{\frac{\pi}{2}} | h^{X}_{\frac{\pi}{2}} ) &\approx& 
  ( h^{X}_{0} | h^{X}_{0} ) =: (\sigma^{X})^{2}\, .
\end{eqnarray}
Therefore, the matrix $\mathcal{M}$ simplifies to
\begin{equation}
\mathcal{M}_{\mu \nu} = \left(  \begin{array}{cccc}
A & C & 0 & 0 \\
C & B & 0 & 0 \\
0 & 0 & A & C \\
0 & 0 & C & B \end{array}
\right)
\end{equation}
where
\begin{equation}\label{eq:ab}
\begin{array}{l}
A = \sum_{X} (\sigma^{X} F^{X}_{+})^2 \\
B = \sum_{X} (\sigma^{X} F^{X}_{\times})^2 \\
C = \sum_{X} (\sigma^{X} F^{X}_{+})(\sigma^{X} F^{X}_{\times}) \, .
\end{array}
\end{equation}
In this form the detection statistic is almost identical to the $\mathcal{F}$-statistic
for detecting rotating neutron stars, as described in
\cite{Jaranowski:1998qm}, in the case where the neutron star has a small wobble angle.

\subsubsection*{Dominant Polarization}

Since we have included a polarization angle in both the transformation
between geocentric and radiation frame ($\chi$) and between radiation
and source frame ($\psi$), we have the freedom to specify one of these
without placing any physical restriction on the signal.  The coherent
\ac{SNR} is further simplified by introducing a dominant polarization
frame which renders $\mathcal{M}_{\mu \nu}$ diagonal.  

Under a rotation of the radiation frame by an angle $\chi^{\mathrm{DP}}$, the
detector response functions transform as 
\begin{eqnarray}
F^{X}_{+} &\rightarrow& F^{\mathrm{DP}, X}_{+} = 
  F^{X}_{+} \cos 2 \chi^{\mathrm{DP}} + 
  F^{X}_{\times} \sin 2 \chi^{\mathrm{DP}}  \\ \nonumber
F^{X}_{\times} &\rightarrow& F^{\mathrm{DP}, X}_{\times} = 
  - F^{X}_{+} \sin 2 \chi^{\mathrm{DP}} + 
  F^{X}_{\times} \cos 2 \chi^{\mathrm{DP}} \label{eq:f_transformation}.
\end{eqnarray}
The rotation through $\chi^{\mathrm{DP}}$ will have an identical effect
on all detectors.  Thus, there exists a polarization angle
$\chi^{\mathrm{DP}}$ which satisfies
\begin{equation}
  C^{\mathrm{DP}} = 
  \sum_{X} (\sigma^{X} F^{\mathrm{DP},X}_{+})
  (\sigma^{X} F^{\mathrm{DP},X}_{\times}) = 0 \, .
\end{equation}  
This can be solved to give $\chi^{\mathrm{DP}}$ as
\begin{equation}
 \tan 4 \chi^{\mathrm{DP}} = \frac{2 \sum_{X} 
 (\sigma^{X} F^{X}_{+})(\sigma^{X} F^{X}_{\times}) 
 }{\sum_{X} \left[ (\sigma^{X} F^{X}_{+})^{2} - 
 (\sigma^{X} F^{X}_{\times})^{2} \right]} \, .
\end{equation}
This choice serves to diagonalize the matrix
$\mathcal{M}$. To uniquely determine $\chi^{\mathrm{DP}}$, we impose an additional
requirement that the network be more sensitive to the $+$
polarization than to the $\times$ polarization.
The value of $\chi^{\mathrm{DP}}$ is a function of the
detector network, the source location and waveform; in particular it
depends upon $F_{+, \times}^{X}$ and $\sigma^{X}$.  From now on, we
assume that we are working in the dominant polarization frame and drop
the DP superscript from our expressions. 

The concept of the dominant polarization frame has been introduced previously in
un-modelled burst searches \cite{KlMoRaMi:05, KlMoRaMi:06,
1367-2630-12-5-053034}.  While the idea is very similar, the actual
implementation is somewhat different.

In the case that both $A$ and $B$ are non-zero, i.e.~that the detector
has some sensitivity to both polarizations, the coherent \ac{SNR}
can be written, in the dominant polarization, as
\begin{eqnarray}\label{eq:fstat_dp}
  \rho_{\mathrm{coh}}^{2} &=& 
  \frac{\mathbf{(s|F_{+} h_{0})}^{2} 
    + \mathbf{(s|F_{+} h_{\frac{\pi}{2}})}^{2}}{ 
    \mathbf{ (F_{+} h_{0} | F_{+} h_{0}) }} + \\
  && \, \frac{\mathbf{(s|F_{\times} h_{0})}^{2} +
  \mathbf{(s|F_{\times} h_{\frac{\pi}{2}})}^{2}}{ 
  \mathbf{ (F_{\times} h_{0} | F_{\times} h_{0}) } } . \nonumber
\end{eqnarray}
The coherent \ac{SNR} can then be seen to arise as the quadrature sum of
the power in the two phases of the waveform ($0$ and $\frac{\pi}{2}$) in
the two gravitational wave polarizations ($+$ and $\times$).

\subsubsection*{Network Degeneracy}

In many cases, a detector network is much more
sensitive to one gravitational wave polarization than the other.  In the
extreme limit (e.g. co-located and co-aligned detectors such as those at
the Hanford site) the network is entirely insensitive to the second
polarization.  In the dominant polarization frame, the network becomes
degenerate as $B \rightarrow 0$ or equivalently
\begin{equation}
  \sum_{X} (\sigma^{X} F^{X}_{\times})^2 \rightarrow 0 \, . 
\end{equation}
Thus the network will only be degenerate if $F^{X}_{\times} = 0$ for all
detectors X. If the network is degenerate then it is easy to see that the
detection statistic will be degenerate as well. In this case it is logical
to remove the $\cross$ terms from the detection statistic reducing it to
\begin{equation}\label{eq:fstat_degenerate}
  \rho_{\mathrm{coh}}^{2} =
  \frac{\mathbf{(s|F_{+} h_{0})}^{2} + \mathbf{(s|F_{+} h_{\frac{\pi}{2}})}^{2}}{ 
  \mathbf{ (F_{+} h_{0} | F_{+} h_{0}) }},
\end{equation}
which is $\chi^{2}$ distributed with two degrees of freedom.

In this formalism the coherent \ac{SNR} changes abruptly from (\ref{eq:fstat_degenerate})
to (\ref{eq:fstat_dp}). If there is any sensitivity, no matter how
small, to the $\times$ polarization, there is an entirely different
detection statistic.  This arises due to maximization over the
$\mathcal{A}^{\mu}$ parameters, allowing them to take any value.  Thus,
even though a network may have very little sensitivity to the $\times$
polarization, and consequently there be little chance of observing the
waveform in the $\times$ polarization, this is not taken into account in
the derivation.  A possible modification is to place an astrophysical
prior on the parameters $(D, \iota, \psi, \phi_{0})$ and propagate this
to the distribution of the $\mathcal{A}^{\mu}$ \cite{Prix:2009tq}.  This
would provide a smooth transition from the degenerate to non-degenerate
search.

\subsection{Comparison with Coincident Search}
\label{sec:coinc_vs_coherent}

The single detector search is a special case of the degenerate network
(\ref{eq:fstat_degenerate}) and can be written as
\begin{equation} \label{eq:sngl_det_snr}
  \rho_{X}^{2} = \frac{ (s^{X}| h_{0})^2 + (s^{X} | h_{\frac{\pi}{2}})^2}{
  (\sigma^{X})^{2}} \, .
\end{equation}
A coincidence search requires a signal to be observed in two or more
detectors, without requiring consistency of the measured waveform
amplitudes in the different detectors.  In many cases, coincidence
searches have made use of different template banks in the different
detectors \cite{Abadie:2010uf, Abadie:2010yba, Abbott:2009qj} and
required coincidence between the recovered mass parameters
\cite{Robinson:2008}.  A comparison with the coherent analysis discussed
above is facilitated if we consider a coincident search where an
identical template is used in all detectors, as was done in an analysis
of early LIGO data \cite{Abbott:2005pe}.   In this case, the
multi-detector coincident \ac{SNR} is given by
\begin{equation} \label{eq:coinc_snr}
  \rho^{2}_{\mathrm{coinc}} = \sum_{X} \rho_X^2 =\sum_{X}
  \frac{ (s^{X} | h_{0})^2 + (s^{X} | h_{\frac{\pi}{2}})^2}{(\sigma^{X})^{2}} 
  \, .
\end{equation}
This is not immediately comparable to the coherent \ac{SNR} given in
(\ref{eq:fstat_dp}).  However, both can be re-cast into similar forms by
writing the coincident \ac{SNR} as
\begin{equation} \label{eq:coinc_xy}
  \rho^{2}_{\mathrm{coinc}} = \sum_{X,Y} \sum_{i = 0, \frac{\pi}{2}}
  \Big( s^{X} \Big| \frac{h_{i}}{\sigma^{X}}\Big) 
  \left[ \delta^{XY} \right]
  \Big( s^{Y} \Big| \frac{h_{i}}{\sigma^{Y}}\Big)
\end{equation}
Similarly, the coherent \ac{SNR} can be written as
\begin{equation} \label{eq:coherent_xy}
  \rho^{2}_{\mathrm{coh}} = \sum_{X,Y} \sum_{i = 0, \frac{\pi}{2}}
  \Big( s^{X} \Big| \frac{h_{i}}{\sigma^{X}}\Big) 
  \left[ f_{+}^{X} f_{+}^{Y} + f_{\times}^{X} f_{\times}^{Y} \right]
  \Big( s^{Y} \Big| \frac{h_{i}}{\sigma^{Y}}\Big)
\end{equation}
where we have defined the orthogonal unit vectors (in detector space)
$f_{+}^{X}, f_{\times}^{X}$ as
\begin{equation}
  f_{+,\times}^{X} = \frac{\sigma^{X} F_{+,\times}^{X}}{\sqrt{\sum_{Y}
(\sigma^{Y} F_{+,\times}^{Y})^{2}}} \, .
\end{equation}

The \ac{SNR} of the coincident search (\ref{eq:coinc_xy}) is simply the
sum of all power consistent with the template waveform in each detector.
The coherent \ac{SNR} (\ref{eq:coherent_xy}) makes use of the fact that
gravitational waves have only two polarizations to restrict the
accumulated \ac{SNR} to the physical subspace spanned by $f^{+}$ and
$f^{\times}$.  For a signal, the power will lie entirely in this
subspace, while noise in the detectors  will contribute to all
components of the coincident \ac{SNR}.  Thus, the coherent analysis
obtains precisely the same signal \ac{SNR} but reduces the noise
background.  Specifically, the coherent \ac{SNR} acquires contributions
from four noise degrees of freedom, while the coincident \ac{SNR} has
$2N$ noise degrees of freedom, where $N$ indicates the number of active
detectors.  For a non-degenerate two detector
network, the coincident and coherent \ac{SNR}s are equal as in this case
$f_{+}^{X} f_{+}^{Y} + f_{\times}^{X} f_{\times}^{Y} = \delta^{XY}$.
In the case where a network is sensitive to only one
polarization, the coherent \ac{SNR} is constructed solely from the
$f^{X}_{+}$ direction and coherent \ac{SNR} is $\chi^{2}$ distributed with 2
degrees of freedom.

Finally, we note that restricting to the coherent \ac{SNR} can help to
separate transients from gravitational wave signals as those transients
which do not contribute power to the signal space will be ignored.
However, many noise transients will contribute to the coherent \ac{SNR}
and more active methods of removing them are required.  These methods
are the focus of sections \ref{sec:snr_cont_consistency} and
\ref{sec:chi2tests}.

\subsection{Synthetic + and $\times$ detectors}
\label{sec:plus_cross_dets}

In the dominant polarization the coherent \ac{SNR} is comprised of
separate $+$ and $\times$ components, with no cross terms.
 We can go one step further and interpret the coherent \ac{SNR} as
arising from two synthetic detectors, one 
sensitive to only the $+$ polarization and one sensitive to only the
$\times$ polarization.  These synthetic detectors are most easily formed
by combining the ``overwhitened'' data streams $o^{X}$ from the various
detectors, where 
\begin{equation}
  o^X(f) = \frac{s^X(f)}{S_h^X(f)} \, .
\end{equation}
The overwhitened synthetic data streams are simply
\begin{equation}
  o_{+,\times}(f) = \sum F^{X}_{+,\times} o^X(f) \, ;
\end{equation}
and the power spectra for these overwhitened data streams are
\begin{equation} \label{eq:psds}
  S_{+,\times} = \left(\sum_X
\frac{(F_{+,\times}^X)^2}{S_h^X(f)}\right)^{-1} \, .
\end{equation}
Using this, the un-whitened synthetic data streams are given as%
\footnote{There is some ambiguity in fixing the overall normalization of
the synthetic detectors.  We require that our synthetic detectors have
the same sensitivity to the two polarizations as the original network
did by requiring $(h_{0}^{+,\cross} | h_{0}^{+,\cross} )_{+, \times}
= \sum_{X} (F^{X}_{+,\cross} \sigma^{X})^{2}$.}
\begin{equation}\label{eq:splus_cross_def}
  s_{+,\times}(f) = \sum_X \frac{F_{+,\times}^X s^X(f)}{S_h^X(f)}
  \left(\sum_Y \frac{(F_{+,\times}^Y)^2}{S_h^Y(f)}\right)^{-1} \, .
\end{equation}
In terms of these synthetic detectors the detection statistic becomes
\begin{equation}\label{eq:fstat_plus_cross}
  \rho_{\mathrm{coh}}^{2} =
  \frac{ (s_+|h_{0})_{+}^{2} + (s_+| h_{\frac{\pi}{2}})_{+}^{2}}{ 
   ( h_{0} | h_{0})_{+} } +
  \frac{(s_{\times}| h_{0})_{\times}^{2} +
  (s_{\times}| h_{\frac{\pi}{2}})_{\times}^{2}}{ 
   ( h_{0} | h_{0})_{\times} } \, ,
\end{equation}
where the subscripts $+, \times$ on the inner products denote the fact
that the power spectrum of the synthetic detectors is used in their
evaluation. 



\section{Signal consistency between detectors}
\label{sec:snr_cont_consistency}

As discussed in the introduction, due to the presence of non-Gaussian
noise transients, it is essential to make use of signal
consistency requirements within search algorithms to distinguish
glitches from gravitational wave signals.

Multi-detector analyses have made good use of signal consistency between
detectors (see e.g.  \cite{1367-2630-12-5-053034}).  A particularly
powerful test is the use of a ``null stream'' \cite{Guersel:1989th}
which, by construction, contains no gravitational wave signal.  Many
noise transients will contribute power to the null stream and can
therefore be eliminated as candidate events.  In addition, requiring
that the gravitational wave signal is recovered consistently between
detectors can eliminate other noise transients, in our case this is
equivalent to imposing restrictions on the recovered values of the
parameters, $\hat{\mathcal{A}}^{\mu}$.  These two methods will be
considered in turn.  For matched filtering searches, requiring
consistency between the observed signal and template waveform has also
proven very powerful \cite{Allen:2004gu}.  A full description of
waveform consistency tests is presented in the next section.

\subsection{Null Stream Consistency}
\label{sec:nullstream}

The gravitational waveform consists of two polarizations. Thus for
networks comprising three or more detectors it is possible to construct
one or more null data streams which contain no gravitational wave signal
\cite{Guersel:1989th}.  In the context of a coherent search for \ac{CBC}
signals, the null consistency tests arise quite naturally.  In section
\ref{sec:coinc_vs_coherent}, we noted that the coherent \ac{SNR} can be
thought of as a projection of the coincident multi-detector \ac{SNR}
onto a four dimensional signal subspace.  The remaining
dimensions in the coincident search do not contain any gravitational
wave signal, but will be subject to both Gaussian and non-Gaussian
noise.  Thus, we can define the null \ac{SNR} as  
\begin{eqnarray}\label{eq:null_overwhite}
  \rho_{\mathrm{N}}^{2} &=& \rho^{2}_{\mathrm{coinc}} -
  \rho^{2}_{\mathrm{coh}} \\
  &=& \sum_{X,Y} \sum_{i = 0 , \frac{\pi}{2}}
  \Big( s^{X} \Big| \frac{h_{i}}{\sigma^{X}}\Big) 
  \left[ N^{XY} \right]
  \Big( s^{Y} \Big| \frac{h_{i}}{\sigma^{Y}}\Big)
  \nonumber \, ,
\end{eqnarray}
where 
\begin{equation} \label{eq:null_cont}
  N^{XY} = \delta^{XY} - f_{+}^{X} f_{+}^{Y} - f_{\times}^{X}
f_{\times}^{Y} \, . 
\end{equation}

A gravitational wave signal matching the template $h$ will provide no
contribution to the null \ac{SNR}, so we expect that, for signals, this
statistic will be $\chi^{2}$ distributed with $(2N - 4)$ degrees of
freedom.  A noise transient that is incoherent across the data
streams, may give a large coherent \ac{SNR}, but it is likely to also give a
large null \ac{SNR}.  Thus requiring a \textit{small} null \ac{SNR} will prove
effective at distinguishing incoherent noise transients from real
gravitational wave signals.   Since the definition of the null \ac{SNR}
(\ref{eq:null_overwhite}) makes use of the template waveform,
gravitational waveforms which do not match the template $h$ can
contribute to the null \ac{SNR}.

We can go one step further and introduce synthetic null detectors in
analogy with the  synthetic $+$ and $\times$ detectors.  For
concreteness, we describe the three detector case, but this can be
extended in a straightforward manner to larger networks.   To do so, we
introduce the null direction
\begin{equation}
  n^{X} = \epsilon^{XYZ} f_{+}^{Y} f_{\times}^{Z}.
\end{equation}
such that $N^{XY} = n^{X} n^{Y}$.
Then, the over-whitened synthetic null detector is 
\begin{equation}
  o_{N}(f) = \sum \frac{n^{X}}{\sigma^{X}} o^X(f) \, .
\end{equation}
The power spectrum of the null stream is%
\footnote{In this case, there is a normalization ambiguity.  For the
synthetic plus and cross streams, it was natural to require that the
synthetic detectors have the same sensitivities as the original
network.  For the null stream this is not feasible as the network has
zero sensitivity to a signal in the null stream, so we normalize such
that $(h_{0} | h_{0})_{N} = 1$.}
\begin{equation} \label{eq:nullpsds}
  S_{N}(f) = \left(\sum_X \frac{(n^X)^2}{(\sigma^{X})^{2} S_h^X(f)}\right)^{-1}
\end{equation}
and the un-whitened null stream is%
\begin{equation}\label{eq:sn}
  s_{N}(f) = \left( \sum_X \frac{n^{X} s^X(f)}{\sigma^{X} S_h^X(f)}
\right) \cdot S_{N}(f) \, .
\end{equation}
Finally, the null \ac{SNR} can be written as
\begin{equation}
  \rho_{\mathrm{N}}^{2}  = \frac{
  \left( s_{N} | h_{0} \right)_{N}^{2} + 
  \left( s_{N} | h_{\frac{\pi}{2}} \right)_{N}^{2} }{ (h_{0} | h_{0})_{N} } \, . 
\end{equation}

The null \ac{SNR} described above differs from the multi-detector null
stream formalism introduced in \cite{Guersel:1989th} and used by several
other authors.  A null stream is constructed to be a data stream which
contains no contribution from the $h_{+}$ and $h_{\times}$ gravitational
waveforms, regardless of the details of the waveform.  To provide a
concrete comparison between the null stream and null \ac{SNR}, we
restrict attention to a three detector network.  The null stream is
explicitly constructed as
\begin{equation} \label{eq:snormalnull}
  \mathrm{s_{\mathrm{Null}}}(f) = \epsilon_{XYZ} F^{X}_{+} F^{Y}_{\times}
s^{Z}(f) \, .
\end{equation}

By comparing the null stream in (\ref{eq:snormalnull}) with the
synthetic null detector (\ref{eq:sn}), it is clear that these will
generically differ.  To get an insight into the differences, consider a
network with two co-located detectors $A$ and $B$, with power spectra
$S^{A}_{h}(f)$ and $S^{B}_{h}(f)$ respectively, and a third detector $C$
which is sensitive to the other polarization of gravitational waves.
For this network, the null stream will be a combination of only the $A$
and $B$ detector data.  The power spectrum of the null stream is 
\begin{equation}
  S_{\mathrm{Null}}(f) = S^{A}(f) + S^{B}(f).
\end{equation}
while for the synthetic null detector, it is
\begin{equation} 
  \frac{1}{S_{\mathrm{N}}(f)} = 
  \frac{1}{(\sigma^{A})^{2} S^{A}(f)} + \frac{1}{(\sigma^{B})^{2} S^{B}(f)} \, .
\end{equation}

Thus, if the power spectra of detectors $A$ and $B$ are identical, then
the two null streams are also identical.  In the extreme case that the
sensitivity bands of the two detectors do not overlap at all, then there
is no null stream ($S_{\mathrm{Null}} \rightarrow \infty$).  However,
the null \ac{SNR} need not vanish and is similar to a two bin version of
the $\chi^{2}$ test described in section \ref{ssec:standard_chisq}.
Thus, it is possible to construct scenarios in which these two null
stream formulations differ significantly.

For the most part, the power spectra of the ground based detectors are
rather comparable.  So, in general there will not be a significant
difference between these two forms.  There are advantages to both
methods.  The null stream is designed to cancel all gravitational wave
signals from the data, thus making it more robust when the signal is not
well known.  However, by making use of the template signal, there are
instances in which the null \ac{SNR} provides a more powerful
consistency test.  Furthermore, it has a computational benefit in that
it does not require the production of a null stream --- all
manipulations are performed on the single detector \ac{SNR} data
streams which are subsequently separated into coherent and null
componenets.  In practice we have found very little difference in
performance, and choose to compute the null \ac{SNR}
(\ref{eq:null_overwhite}) for computational simplicity.

Finally, we note that both null stream formalisms will perform optimally
only if the three detectors have similar sensitivities. In the case
where one detector is significantly less sensitive than the others, the
null stream will generally tend to the data of that less sensitive
detector.  Also, the null formalisms described here will only completely
cancel a gravitational wave signal provided that the calibration of the
data streams is accurate, any error in calibration will lead to a signal
surviving in the null stream data.

\subsection{Amplitude Consistency}
\label{sec:ampcons}

The four amplitude parameters $\mathcal{A}^{i}$, encoding the distance
to and orientation of the binary system, can take any values.  Indeed,
any set of $\mathcal{A}^{i}$ corresponds to unique  values of the
distance, inclination angle, coalescence phase and polarization angle,
up to symmetries of the system.  However, some of these values will be
significantly more likely to occur astrophysically than others.
For example, the number of binary coalescence events is expected to be
approximately proportional to star formation rate \cite{Temp:2010cfa} and
consequently should be roughly uniform in volume.  Thus, events are more
likely to occur at a greater distance.  Similarly, the gravitational
wave amplitude, at a fixed distance, is greater for face on signals than
edge on ones, as is clear from (\ref{eq:aplus_across}).  Thus,
approximately face on signals at a large distance are more likely
signals than nearby, edge on ones. 
  
The distribution of noise events will follow its own characteristic
distribution.  For Gaussian noise, this distribution can be evaluated
and the signal and noise distributions incorporated into the ranking
statistic \cite{Prix:2009tq}.  Non-stationarities in the data will again
produce a different distribution of amplitudes.  Specifically, the
majority of transients are caused by a disturbance or glitch in a single
detector with little or no signal in the other detectors.  For networks
with three or more detectors, this will typically be inconsistent with a
coherent signal across the network, leading to a large value of the null
\ac{SNR}.  In certain scenarios, most notably for two detector networks,
there will be a consistent set of values for the $\mathcal{A}^{i}$.
However, these values carry the characteristic signature of a glitch.
Specifically, the \ac{SNR} contributions will  typically be consistent
with a nearby, close to edge on system ($A_{\times} \approx 0$), with a
very specific orientation to provide essentially no response in all but
one detector.  Thus, the glitch distribution of the $\mathcal{A}^{i}$
parameters, will be significantly different from the distribution
expected for gravitational wave signals.  In the remainder of this
section, we explore the possibility of making use of the extracted
$\mathcal{A}^{i}$ parameters to distinguish between glitches and
signals.  Unlike the null stream, amplitude consistency tests are
available for  two detector networks.  They should be especially useful
in the case of the two 4km LIGO instruments, which have similar
sensitivities to the majority of points on the sky. 

\begin{figure}
  \begin{minipage}[t]{0.95\linewidth}
    \includegraphics[width=\linewidth]{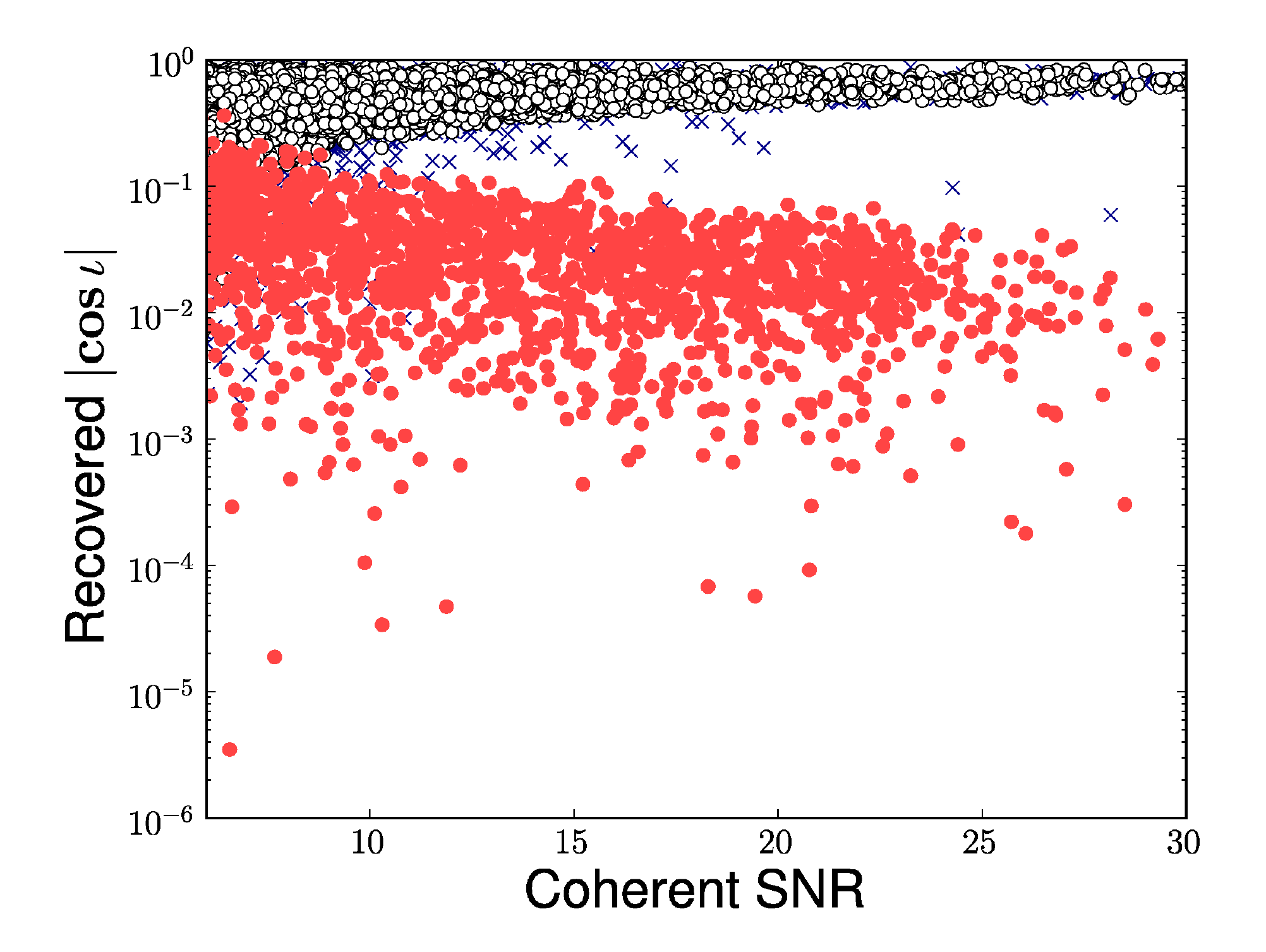}
  \end{minipage}
  \begin{minipage}[t]{0.95\linewidth}
    \includegraphics[width=\linewidth]{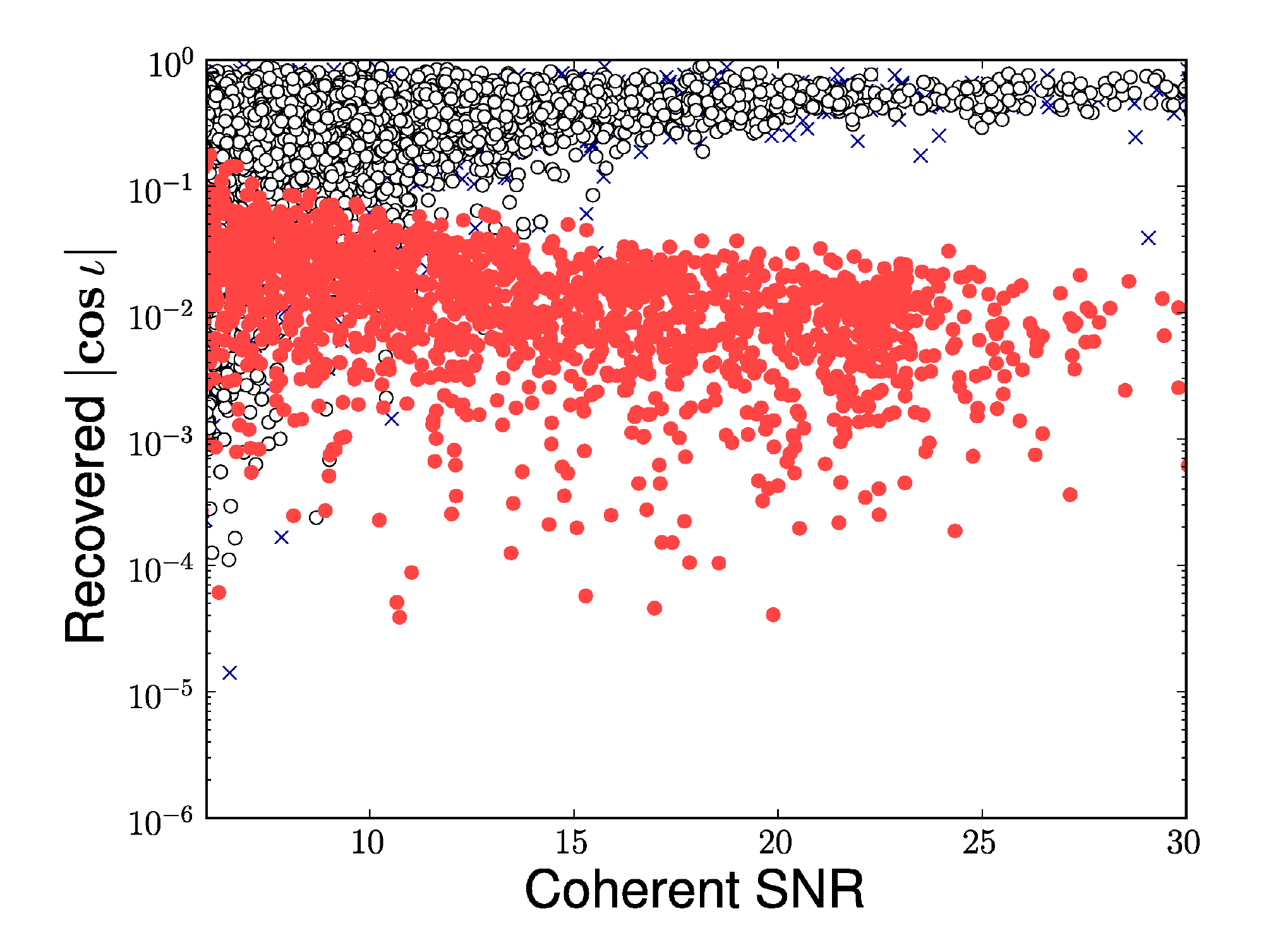}
  \end{minipage}
\caption{\label{fig:sim_inc_injs} 
The distribution of the recovered inclination angle plotted against
coherent \ac{SNR} for optimally oriented signals (unfilled circles),
uniformly distributed orientations (blue crosses) and simulated glitches
(red circles). The top figure shows a network configuration where we are
equally sensitive to both gravitational wave polarizations. The bottom
figure shows a configuration where we are 5 times more sensitive to the
$+$ polarization than to the $\cross$.}
\end{figure}

We have argued that the majority of gravitational wave signals will
originate from (close to) face on binaries while the majority of noise
transients will mimic (close to) edge on binaries.  The recovered value
of the inclination angle $\iota$ should then serve to separate signals
from noise.   To investigate this, we simulated a large number of
simulated \ac{CBC} signals and a large number of noise glitches; added
Gaussian noise and plotted the recovered inclination angle in Figure
\ref{fig:sim_inc_injs}.  The glitches were generated as events with a
large \ac{SNR} in one detector coincident with Gaussian noise in a
second detector.  The signals were separated into two groups: the first
with only face on binaries ($|\cos \iota| = 1$) and the second a uniform
distribution over the two sphere (uniform in $\cos \iota$ and $\psi$) of
the binary orientation.  In both cases, they were distributed uniformly
in volume and orbital phase.  We also consider different network
configurations, both containing two equally sensitive detectors.  In the
first case one detector is sensitive to $+$ and the other to $\times$
polarization ; in the second case both detectors have strong and equal
sensitivity to the $+$ polarization and weak but opposite sensitivity to
the $\cross$ polarization --- rather typical for the Hanford, Livingston
network.  For both sets of signals and choices of network, there is a
clear distinction between signal and glitch distribution.  However,
there is a clear downwards bias on the recovered values of $\iota$.
This can be understood by looking at the expressions for  $A_{+}$ and
$A_{\times}$.  For face on binaries, these will be equal but, in the
presence of noise, $A_{\times}$ will be reconstructed to be somewhat
smaller than $A_{+}$.  A relative difference of only $5\%$ leads to a
recovered inclination of $45^{\circ}$, so even for loud signals there
can be large discrepancy between the actual and recovered inclination
angle.

Despite the difference in distribution between signal and noise, there
is also a significant overlap of the populations at low SNRs.  Consequently, any
threshold imposed on the recovered inclination angle is liable to
either reject a fraction of signals or pass a fraction of glitches.  It
is, however, quite possible that knowledge of these expected
distributions could be folded into the detection statistic in a Bayesian
manner.

\begin{figure}
  \begin{minipage}[t]{0.95\linewidth}
    \includegraphics[width=\linewidth]{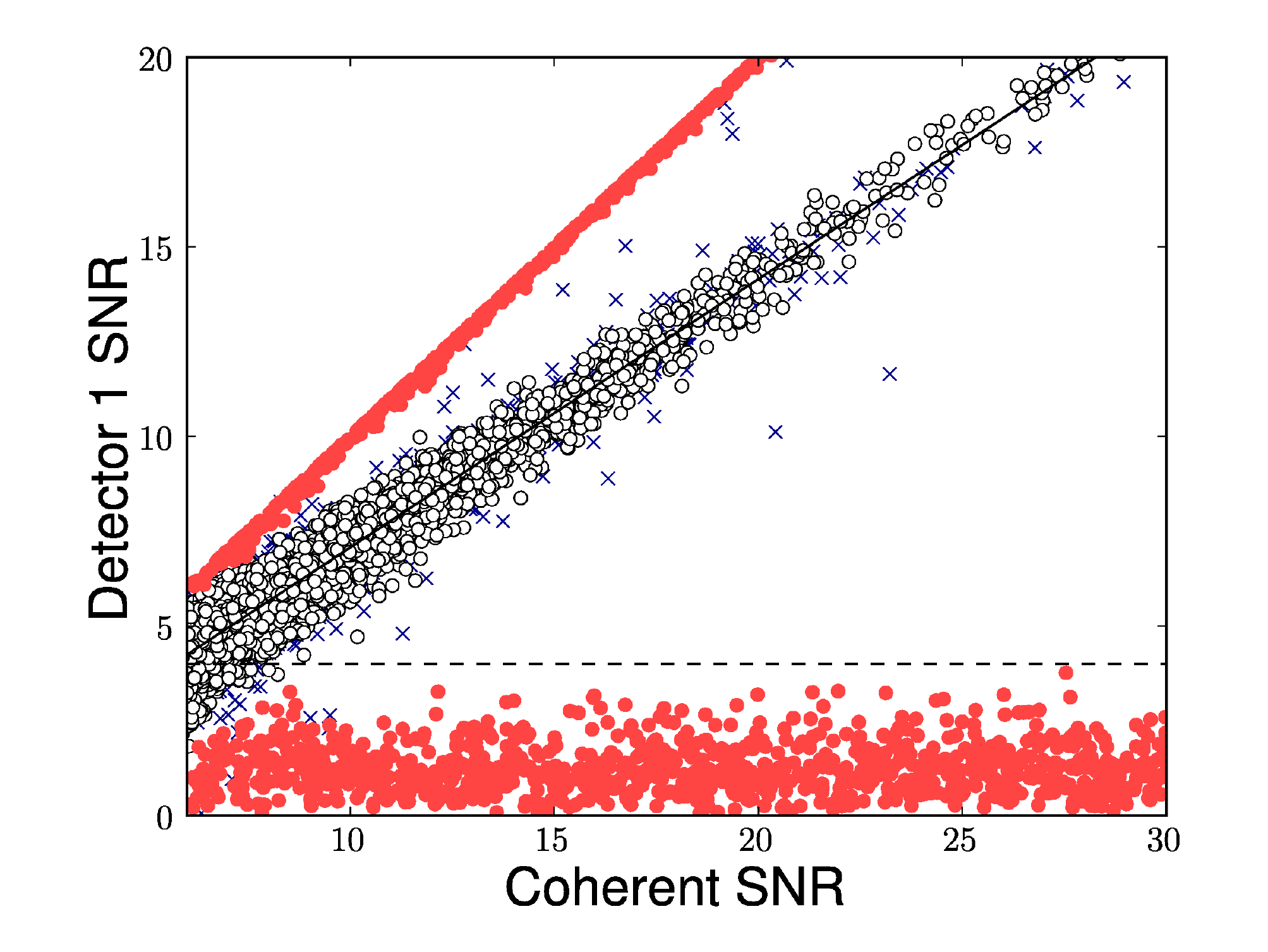}
  \end{minipage}
  \begin{minipage}[t]{0.95\linewidth}
    \includegraphics[width=\linewidth]{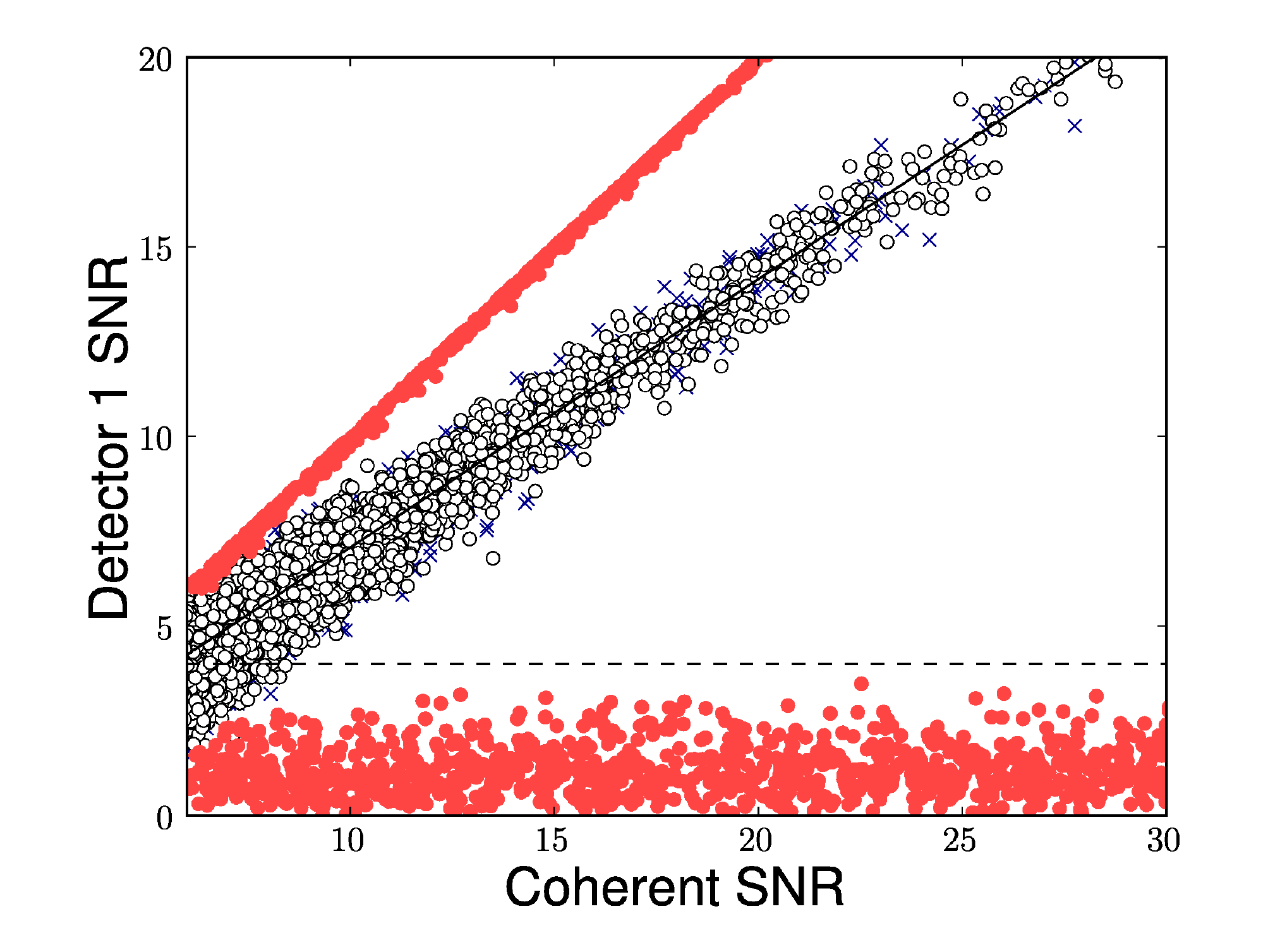}
  \end{minipage}
  \caption{\label{fig:sim_snr_injs} 
The distribution of single detector \ac{SNR} plotted against coherent
\ac{SNR} for optimally oriented signals (unfilled circles), uniformly
oriented signals (blue crosses) and glitches (red circles). The top
figure shows a two equally sensitive detector network configuration
where one detector sees only the $+$ polarization and the second
detector sees only the $\cross$ polarization. The bottom figure shows a
two detector network configuration where both detectors have strong and
equal sensitivity to the $+$ polarization and weak but opposite
sensitivity to the $\cross$ polarization. This is meant to emulate a
typical instance for the Hanford, Livingston network. The diagonal solid
black line shows the expected \ac{SNR} for the optimally oriented
signals. The horizontal dashed black line indicates an \ac{SNR} of 4.}
\end{figure}

We have found that using the observed \ac{SNR} in the individual
instruments to be a more effective discriminator of signal and noise.
To demonstrate the efficacy of such an approach, in Figure
\ref{fig:sim_snr_injs} we plot the single detector \ac{SNR} as a
function of the coherent \ac{SNR} for the same population of glitches
and the two classes of signals (face on and uniformly distributed
orientation) described above.  The glitches fall into two groups
depending upon which detector suffered the glitch.  Since our model
detectors are equally sensitive, then on average one expects each
detector to accrue $1/\sqrt{2}$ of the coherent \ac{SNR}.  Even allowing
for non-optimally oriented signals and the addition of Gaussian noise,
the signals follow this expectation.  Only a small number of signals 
are found with \ac{SNR}s inconsistent with the expected values, these
are ones that have very specific orientations. Overall, the
signal and glitch populations are very well separated, at least until
the coherent \ac{SNR} becomes rather small.

The most effective strategy we have found is to require that all events
have an \ac{SNR} above 4 in the two most sensitive detectors in the
network.  The cut is illustrated in Figure \ref{fig:sim_snr_injs}.  This
strategy removes the majority of glitch signals while having a negligible
effect on the signal population at large \ac{SNR}.  For lower \ac{SNR}
the signals which are lost due to this cut would be unlikely to be
detection candidates as Gaussian noise alone produces similar events.

\section{Coherent $\chi^2$ tests}
\label{sec:chi2tests}

Data from gravitational wave detectors contain numerous
non-stationarities due to both instrumental and environmental causes.
These non-stationarities, or glitches, typically do not
match well with the \ac{CBC} waveform. However, they often contain enough power
that, even though the match with the template is poor, a large \ac{SNR}
is observed.  In the previous section, we have seen how the use of
various coherent consistency tests can mitigate this problem.
Additionally, a number of other signal consistency tests have been
implemented \cite{Allen:2004gu, Hanna:2008} and used in searches for CBC
signals \cite{Abadie:2010yba, Abbott:2009qj, Abbott:2009tt}.  These
tests are all designed to eliminate glitches which have a different
signal morphology than the template waveform.  This is essentially done
by testing whether the detector data orthogonal to the signal is well
described as Gaussian and stationary ---  for a glitch, there will be
residual power which does not match the template waveform.  These tests
are commonly known as ``$\chi^2$ tests'' as they construct a statistic
which is $\chi^{2}$ distributed in the presence of Gaussian noise plus a
signal matching the template waveform.  If the data contains a glitch,
the $\chi^{2}$ statistic will generally have a large value, thereby
allowing for differentiation of signal from non-stationary noise.  In
this section, we briefly review the general formulation of $\chi^{2}$
tests before presenting a detailed description of three such tests which
have been implemented for the coherent search described in section
\ref{sec:coh_matched_filter}. 

\subsection{A general framework for $\chi^{2}$ tests}
\label{sec:gen_chi}

Consider the data from a gravitational wave detector at a time $t$ which
has produced a large \ac{SNR} when filtered against a template $h(t)$.
Generically, the data $s(t)$ can be decomposed as
\begin{equation}\label{eq:signal_model}
  s(t) = n(t) + A h(t) + B g(t) 
\end{equation}
where $n(t)$ represents a Gaussian noise component, $h(t)$ is the
template waveform, $g(t)$ is an additional non-Gaussian noise
contribution to the data stream and $A$ and $B$ are amplitude factors.
The glitch contribution $g(t)$ is taken to be the power orthogonal to
$h(t)$ and both $g(t)$ and $h(t)$ are normalized,
so that
\begin{equation}
  ( g | g ) = 1 \, , (h | h) = 1 \, ,  ( g | h ) = 0 \, .
\end{equation}

In order to construct a $\chi^{2}$ test, we must introduce an additional
set of waveforms $T^{i}$.  These waveforms are required to be
orthonormal and orthogonal to $h$, 
\begin{equation}
  (h|T^i) = 0 \, , (T^i | T^j) = \delta^{ij} \, . 
\end{equation}
Furthermore, for the $\chi^{2}$ test to be effective, the $T^{i}$ must
have a good overlap with the glitch waveform $g(t)$.

The $\chi^{2}$ discriminator is constructed as 
\begin{equation}\label{eq:chi2} 
  \chi^2 = \sum_{i=1}^{N} (T^i | s)^2 \, .
\end{equation}
When the data comprises only signal plus Gaussian noise, i.e. $B=0$ in
equation (\ref{eq:signal_model}), 
\begin{equation} 
  \chi^2 = \sum_{i=1}^{N} (T^i | n)^2.
\end{equation}
and the statistic is the sum of squares of independent Gaussian
variables with zero mean and unit variance.  Thus the test is $\chi^{2}$
distributed with $N$ degrees of freedom, with a mean and variance of 
\begin{equation} 
  \langle\chi^2\rangle = N \, ,
  \quad
  {\rm Var}(\chi^2) = 2N  \, .
\end{equation}
This is true for \emph{any} set of waveforms $T^i$ given the above
assumptions. 

In the case where the data are not an exact match to the signal, we take
both $A$ and $B$ non-zero, i.e. any signal or glitch can be decomposed
into a part $A h(t)$ proportional to the template under consideration
plus a second orthogonal contribution $B g(t)$.  Clearly, for different
glitches, the waveform $g(t)$ as well as the amplitude factor $B$ will
be different.  In this case the $\chi^{2}$ test takes the form
\begin{equation} 
  \chi^2 = \sum_{i=1}^N \left[ (T^i | n)^2 + 2 B (T^i | n)(T^i | g) +
  B^{2} (T^i|g)^2 \right] \, .  
\end{equation}
This has a mean 
\begin{equation} 
  \langle\chi^2\rangle = N + B^{2} \sum_i^N (T^i | g)^2 
\end{equation}
and a variance 
\begin{equation} 
  {\rm Var}(\chi^2) = 2N + 4 B^{2} \sum_i(T^i|g)^{2} \, .
\end{equation}
The $\chi^{2}$ test is distributed as a non-central $\chi^{2}$
distribution with $N$ degrees of freedom and a non-centrality parameter
\cite{Allen:2004gu}
\begin{equation}
  \lambda =  B^{2} \sum_{i=1}^N (T^i | g)^2 \, .
\end{equation}
The challenge in constructing a $\chi^{2}$ test is to select the basis
waveforms $T^{i}$ such they have large overlaps with the observed
glitches in the data.  If this is done successfully, then any glitch
producing a large \ac{SNR} will also give a large value of $\chi^{2}$,
inconsistent with a signal in Gaussian noise.

In many cases, there is some uncertainty in the template waveform.  For
example, the post-Newtonian (PN) expansion used in generating \ac{CBC}
waveforms is truncated at a finite (typically 3 or 3.5 PN
\cite{Bliving}) order and there will be differences between this
analytically calculated waveform and the one provided by nature.  There
are similar uncertainties in waveforms obtained from numerical
relativity simulations \cite{Hannam:2009hh}.  Additionally, to search
the full parameter space of coalescing binaries, a discrete template
bank is used which allows for some mismatch between the templates and
any potential signal within the parameter space
\cite{OwenSathyaprakash98}.  Normally the template bank is created so
that the mismatch is no larger than $3\%$ at any point in the parameter
space.  Finally, there are uncertainties in instrumental calibration
\cite{Abadie:2010px} which will affect the match between signal and
template.

We model these effects by parametrizing the signal as
\begin{equation}\label{eq:mismatch_sig}
  H(t) = A ( \sqrt{1 - \epsilon^{2}}\,  h(t) + \epsilon \, m(t) ) \, ,
\end{equation}
where $m(t)$ is the component of $H$ that is orthogonal to $h$
[$(m|h) = 0$] and $\epsilon$ encodes the mismatch between signal and
template in the sense that
\begin{equation}\label{eq:mismatch}
  1 - \frac{(H | h)}{\sqrt{(H | H) (h |h)}} 
  = 1 - \sqrt{1 - \epsilon^{2}} \approx \epsilon\, .
\end{equation}
In most cases, it is reasonable to assume a mismatch of less than $5\%$.
The obvious counter-example is when searching for highly spinning
systems using non-spinning waveforms, see e.g.
\cite{Vandenbroeck:2009cv,Fazi:2009}.  

Since (\ref{eq:mismatch_sig}) is a special case of
(\ref{eq:signal_model}) it follows directly that the mean and variance
of the $\chi^{2}$ test in the presence of a mis-matched signal are
\begin{eqnarray} 
  \langle\chi^2\rangle &=& N +  A^2 \epsilon^2 \sum_{i=1}^N (T^i | m)^2
  \nonumber \\
  {\rm Var}(\chi^2) &=& 2N + 4A^2\epsilon^2\sum_{i=1}^{N} (T^i | m)^2. 
\end{eqnarray}
Since the \ac{SNR} of the signal is proportional to $A$, the expected
$\chi^{2}$ value for a mis-matched signal increases with the strength of
the signal.  However, for mismatched signals $\chi^{2} \propto
\epsilon^{2} A^2$ while for glitches $\chi^{2} \propto B^{2}$ and
provided $\epsilon A \ll B$ the two can be separated. See
\cite{Allen:2004gu} for a more detailed discussion.   

When introducing the $\chi^{2}$ test, we assumed that the $T^{i}$ were
orthonormal and orthogonal to the template waveform $h$.  In practice,
this can be difficult to guarantee.  The signal consistency tests
discussed in the remainder of this section are constructed from
gravitational waveforms.  If one picks a set of gravitational waveforms,
$t^i$, there is no guarantee that they will be either orthonormal or
orthogonal to $h$.  We can, at least, construct waveforms which are
orthogonal to $h$ by introducing
\begin{equation} 
  T^i = \frac{t^i - (t^i | h) h }{\sqrt{ 1 - (t^i | h)^2} }.
\end{equation}
While this ensures $(h|T^i)=0$ it does not guarantee
orthonormality of the $T^{i}$, $(T^i|T^j) = \delta^{ij}$.  Thus this
method will not produce a $\chi^{2}$ distribution, and will instead form
a generalized $\chi^{2}$ distribution.  The mean of the distribution
remains $N$ but the variance is increased, 
\begin{equation}
  {\rm Var}(\chi^2) = 2N + 2 \sum_{i \neq j}  
  (T^{i} | T^{j})^{2}. 
\end{equation}
It has been found, however, that this does not present a significant
obstacle to using these tests, especially as the thresholds are
tuned empirically \cite{LIGOS3S4Tuning}.

\subsubsection*{Multi-detector $\chi^{2}$ tests}
\label{sec:cohchi2test}

In section \ref{sec:coh_matched_filter}, we derived a coherent
multi-detector search for coalescing binaries.  The search involves
filtering four waveform components $h_{\mu}$ against the multi-detector
data stream.  Our initial discussion of $\chi^{2}$ tests was
limited to the description of a single phase template waveform $h$ and
test waveforms $T^{i}$.  The extension to a two phase waveform has been
described previously \cite{Allen:2004gu} and here we extend that to a
four component waveform across multiple detectors, as is appropriate for
this search.  We begin by noting that the four waveform components
$h_{\mu}$ are orthogonal in the dominant polarization basis.  They are,
however, not generally normalized, as
\begin{equation}
  ( \mathbf{h}_{\mu} | \mathbf{h}_{\nu} ) = \mathcal{M}_{\mu \nu} =
\mathrm{diag} (A, B, A, B) \, ,
\end{equation}
where $A$ and $B$ are defined in (\ref{eq:ab}).  Thus, we first
normalize so that
\begin{equation}
  ( \hat{\mathbf{h}}_{\mu} | \hat{\mathbf{h}}_{\nu} ) = \delta_{\mu \nu}.
\end{equation}

To construct a network $\chi^{2}$ test, we require a set of
(4-component) normalized, test waveforms $\hat{t}_{\mu}^{i}$. The
components
\begin{equation} \label{eq:genchisq} 
T^{i}_{\mu} = \frac{\hat{t}^i_{\mu} 
  - \sum_{\nu} (\hat{\mathbf{t}}^i_{\mu}|\hat{\mathbf{h}}_{\nu}) \hat{h}_{\nu}}
{\sqrt{1 - \sum_{\sigma}
(\hat{\mathbf{t}}^i_{\sigma}|\hat{\mathbf{h}}_{\sigma})^2}} \, ,
\end{equation}
constructed to be orthogonal to $h_{\mu}$, are used in the $\chi^{2}$
test.  Thus, the coherent, multi-detector $\chi^2$ test is 
\begin{equation} \label{eq:cohchi2} 
  \chi^2 = \sum_{\mu=1}^4 \sum_{i=1}^N (\mathbf{T}^i_{\mu} | \mathbf{s})^2 \, .
\end{equation}
Provided the test waveforms are orthonormal, in the sense that
\begin{equation}\label{eq:template_orth}
  (\mathbf{T}^{i}_{\mu} | \mathbf{T}^{j}_{\nu}) = \delta^{ij}
\delta_{\mu\nu} \, ,
\end{equation}
the distribution for a signal matching $h_{\mu}$ plus Gaussian noise
will be $\chi^{2}$ distributed with $4N$ degrees of freedom.  As for the
single phase filter, we cannot always guarantee (\ref{eq:template_orth})
is satisfied, although it is relatively simple to ensure the four
components of a given template \textit{are} orthogonal.  This means that
the statistic will not, in general, be $\chi^{2}$ distributed: The mean
remains $4N$ but the variance increases to
\begin{equation} 
  {\rm Var}(\chi^2) = 8N + 
  2 \sum_{i,j=1}^N \sum_{\mu,\nu=1}^{4} 
  \left[ (T^{i}_{\mu} | T^{j}_{\nu})^{2} - \delta^{ij}\delta_{\mu\nu} \right]. 
\end{equation}

\subsection{The coherent bank $\chi^{2}$ test}
\label{sec:bank_chi}

The bank $\chi^{2}$ test was designed to test the consistency of the
observed \ac{SNR} across different templates in the bank at the time of
a candidate signal. It was first described in \cite{Hanna:2008} for the
case of a single detector.  A glitch will typically cause a high
\ac{SNR} in many templates across the bank, while a real signal will
give a well prescribed distribution of \ac{SNR} across the template
bank.

\begin{figure}
  \begin{minipage}[t]{0.95\linewidth}
    \includegraphics[width=\linewidth]{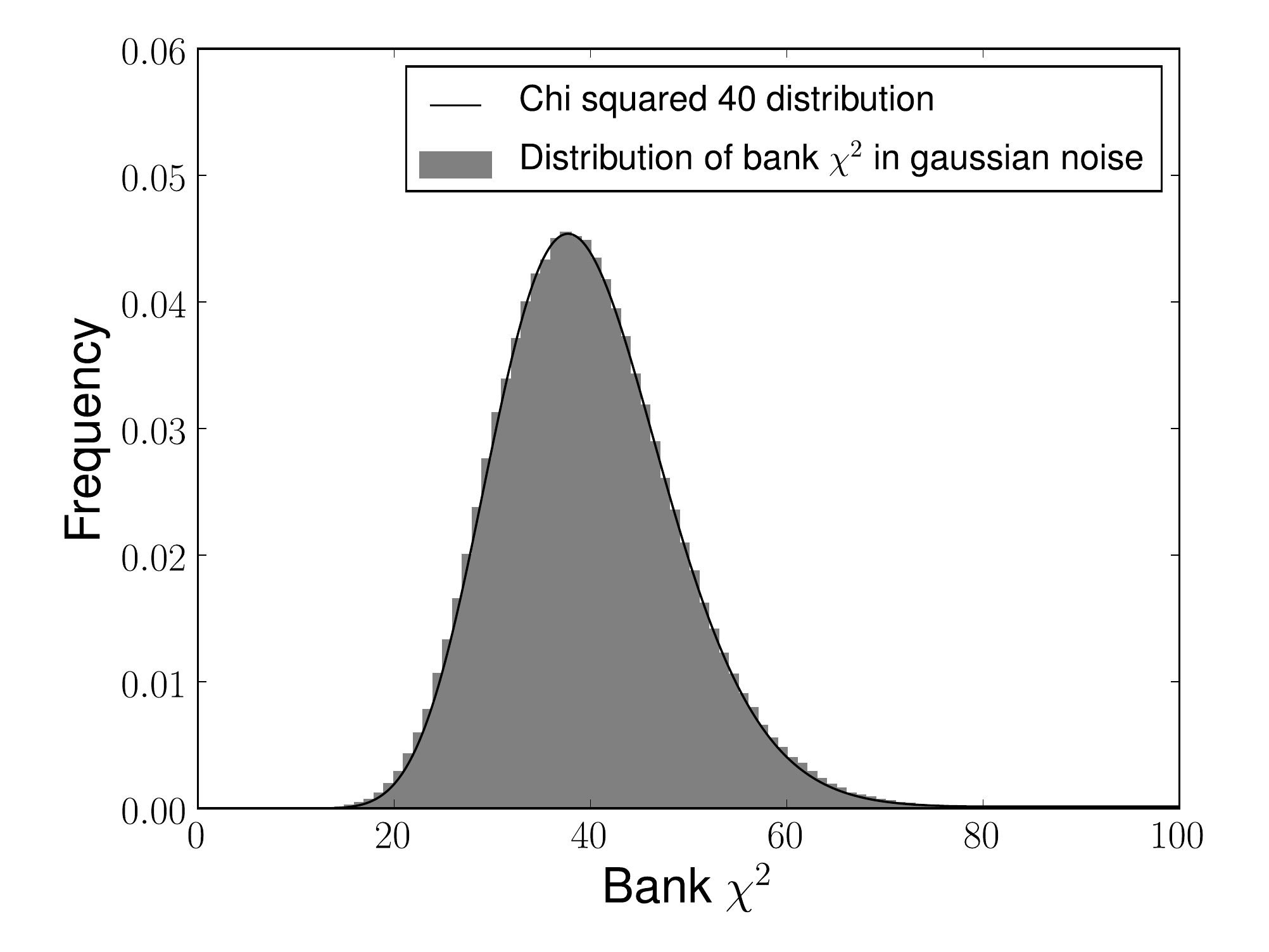}
  \end{minipage}
  \caption{\label{fig:bank_hist} 
The distribution of the bank $\chi^{2}$ test for a single template $h$,
with a bank of size $10$.  The plot shows the distribution of the bank
veto calculated for every time sample in $128 s$ of simulated Gaussian
data (with no signal present).  In the case that the ten bank templates
are orthogonal, the expected distribution is $\chi^{2}$ with 40 degrees
of freedom (shown as the solid black line).  As can be seen, the actual
distribution follows the expected one closely.} 
\end{figure}

The bank $\chi^{2}$ makes use of other CBC templates as the waveforms
$t^{i}$ to construct the $\chi^{2}$ test.  These $N$ templates are taken
from different points across the mass space.  In implementing the bank
$\chi^{2}$, we choose a \textit{fixed} set of template waveforms $t^{i}$
which remain the same for every template $h$ in the search template
bank.  The bank $\chi^{2}$ statistic is then constructed following
(\ref{eq:genchisq}) and (\ref{eq:cohchi2}).  The test is most effective
when the set of $T_{\mu}^{i}$ is close to orthogonal \cite{Hanna:2008}
so we select templates which are well distributed across the mass space,
ensuring the overlaps $(T_{\mu}^{i} | T_{\nu}^{j})$ are small for $i
\neq j$.  Figure \ref{fig:bank_hist} shows the distribution of the bank
$\chi^{2}$ for a single template filtered against Gaussian noise.  The
set of fixed bank waveforms consisted of ten waveforms distributed over
the full mass parameter space.  Using these waveforms, the deviation
from  a $\chi^{2}$ distribution is negligible.  

For the bank $\chi^{2}$ to be effective, glitches in the data must have
a good overlap with a reasonable fraction of the templates $t^{i}$.
While, in general, it is difficult to predict the composition of
glitches in the data, it seems reasonable to assume that glitches which
produce a large \ac{SNR} for the template $h$ will also have a good
overlap with other waveforms in the template space.  Thus, the set of
templates which is spread across the parameter space is suitable.

\subsection{The coherent autocorrelation $\chi^{2}$ test}
\label{sec:auto_chi}

\begin{figure}
  \begin{minipage}[t]{0.95\linewidth}
    \includegraphics[width=\linewidth]{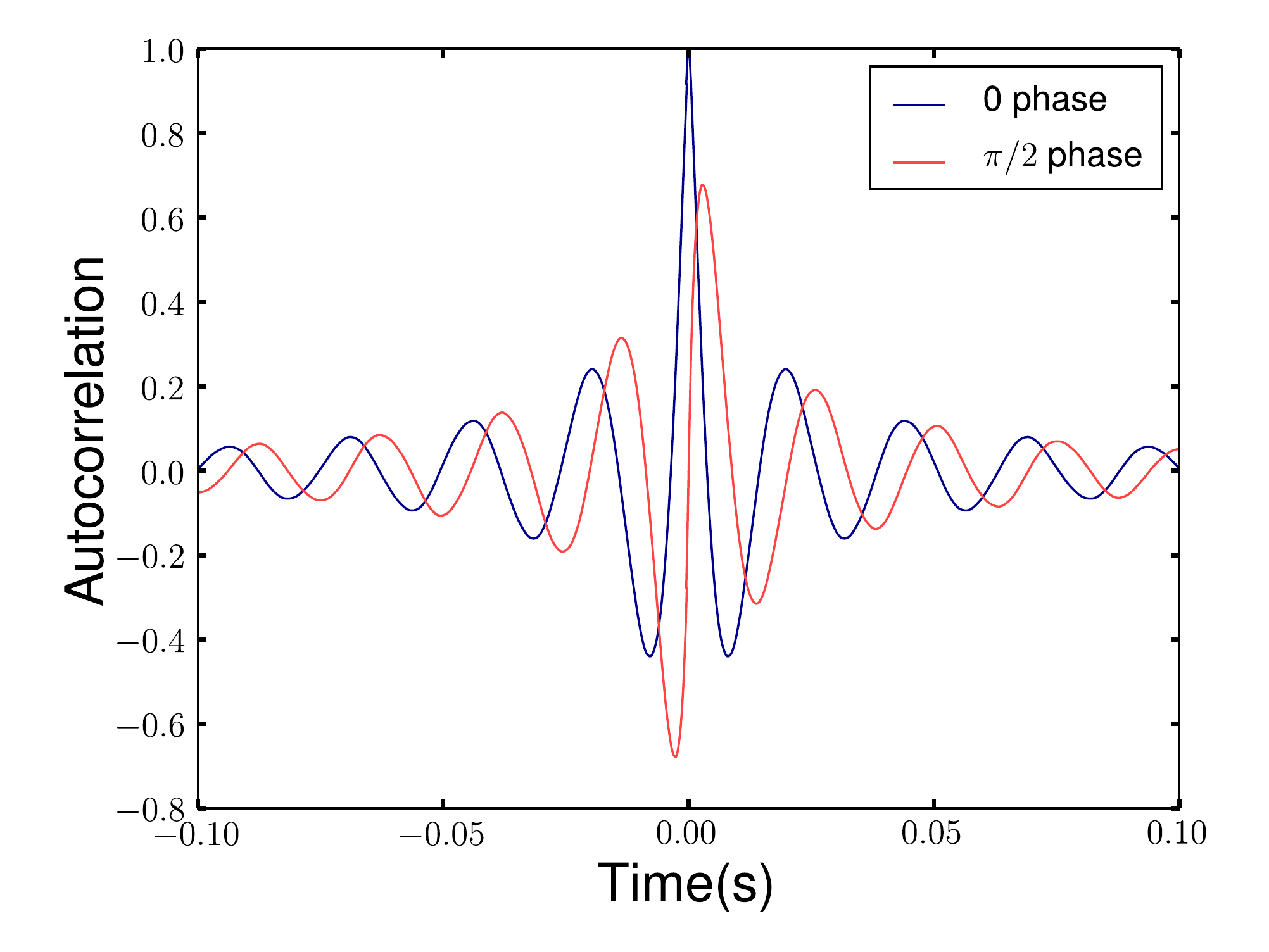}
  \end{minipage}
  \caption{\label{fig:auto_corr} 
The single detector auto-correlation of a gravitational wave inspiral signal from a
1.4,1.4 solar mass binary neutron star.  Both phases of the waveform are
shown. }
\end{figure}

Filtering a gravitational wave template against data containing a
matching signal produces a peak in the \ac{SNR} at the time of
the signal.  Furthermore, there is a characteristic shape of this peak
which depends upon the template waveform and also the noise power
spectrum of the data.  An example of this ``autocorrelation''
for a binary neutron star template is shown in Figure
\ref{fig:auto_corr}.  A noise transient in the data will produce a peak
in the \ac{SNR} but it will typically lack the characteristic shape
produced by a genuine \ac{CBC} signal. 

The ``auto'' $\chi^{2}$ test was designed to test the consistency of the
\ac{SNR} peak \cite{Hanna:2008}.  It is a similar test to the bank
$\chi^{2}$, but where the bank $\chi^{2}$ investigates consistency in
\ac{SNR} across the mass space, the auto $\chi^{2}$ tests for
consistency of the \ac{SNR} time series.  The set of templates $t^{i}$
are chosen to be the original template $h$ with time shifts $\delta
t^{i}$ applied. The values of $\delta t^{i}$ are all unique and chosen
to be of the same time-scale as the auto-correlation of the template
waveform (typically $0.1s$ or less) and the duration of
non-stationarities in the data, which is similar. 

\begin{figure}
  \begin{minipage}[t]{0.95\linewidth}
    \includegraphics[width=\linewidth]{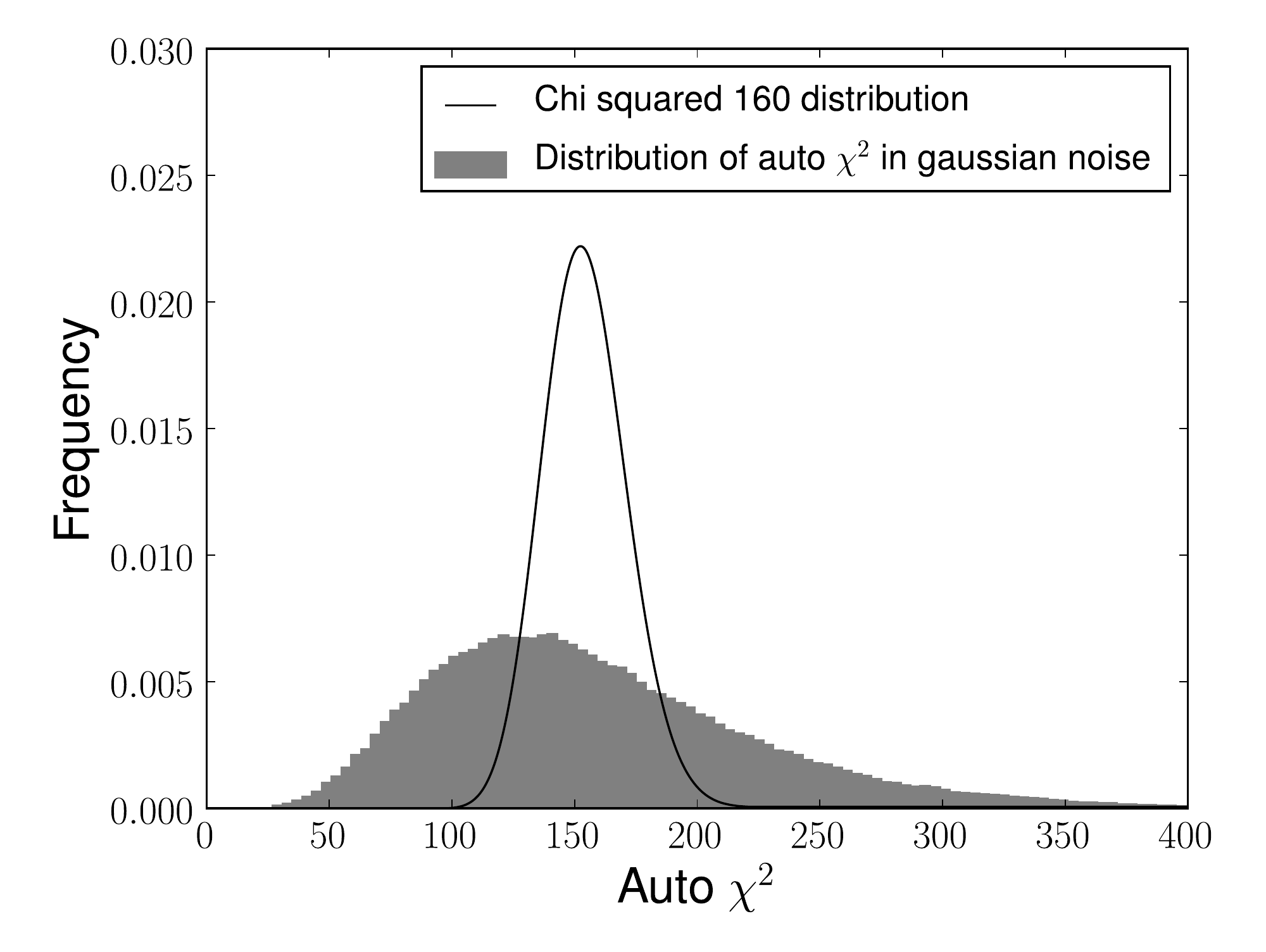}
  \end{minipage}
  \caption{\label{fig:auto_hist} 
The distribution of the auto $\chi^{2}$ test for a single template $h$,
generated with 40 time shifted templates, with shifts between $0.001$ and 
$0.04$ seconds.  The plot shows the distribution of the auto veto
calculated for every time sample in $128 s$ of simulated Gaussian data
(with no signal present).  In the case that the forty time shifted
templates are orthogonal, the expected distribution is $\chi^{2}$ with
160 degrees of freedom (shown in black).  As can be seen, the actual
distribution differs significantly from this due to the
non-orthogonality of the $t^{i}$ waveforms.} 
\end{figure}

In Figure \ref{fig:auto_hist}, we show the distribution of the auto
$\chi^{2}$ for a single template waveform filtered in Gaussian data.
For this result, forty waveforms $t^{i}$ were used, equally spaced with
a $1$ ms spacing, and all with coalescence times prior to that of
$h$.  Thus, the auto $\chi^{2}$ is testing the consistency of the \ac{SNR}
time series for $0.04$ seconds prior to the \ac{SNR} peak.  The overlap
$(t^{i}| t^{j})$ depends only upon the difference $\delta t^{i} - \delta
t^{j}$ and Figure \ref{fig:auto_corr} shows clearly that a significant
fraction of the overlaps are far from zero.  Consequently, the auto
$\chi^{2}$ test has a distribution with a large deviation from a
$\chi^{2}$ distribution with $4N$ degrees of freedom.  

\subsection{The coherent $\chi^2$ test}
\label{ssec:standard_chisq}

The ``standard'' $\chi^{2}$ test originally proposed in
\cite{Allen:2004gu} has been used as a discriminator in many
gravitational wave searches for CBCs. Given the template waveforms and
the detector sensitivity, it is possible to predict the accumulation of
\ac{SNR} as a function of frequency.  By calculating the observed
\ac{SNR} contribution from a number of frequency bins, and comparing to
the predicted value, one can construct a $\chi^{2}$ consistency test.

Formally, given a template $h$ which produced a candidate signal with an
\ac{SNR} of $\rho$, calculate $N$ non-overlapping frequency windows such
that the expected \ac{SNR} is $\rho/N$ in each.  Then, calculate the
actual \ac{SNR} $\rho^{i}$ in each of these frequency bins and compare
with the expected value by calculating 

\begin{equation}\label{eq:standard_chisq}
  \chi^2 = N \sum_i^N (\rho^i - \rho/N)^2 \, .
\end{equation}
For a gravitational wave signal matching the template $h$ plus Gaussian
noise, this statistic will be $\chi^2$ distributed with $N-1$ degrees of
freedom.  Written in the form (\ref{eq:standard_chisq}) it appears
different from the general case we discussed earlier. In
\cite{Allen:2004gu} it was shown that it can be re-expressed in the form
(\ref{eq:chi2}).

This $\chi^{2}$ test can be extended to coherent, multi-detector
searches.  Indeed, in \cite{Itoh:2004zd}, the construction was
applied to a coherent search for continuous 
gravitational waves. Here, we present the extension to a coherent CBC
search.  First, define
\begin{equation}
 \rho_{\mu}^i = 
  \frac{\mathbf{(s | h_{\mu}^i})}{\mathbf{\sqrt{(h_{\mu} | h_{\mu})}}} 
\end{equation}
to be the \ac{SNR} contribution in the $i$th frequency bin to the
\ac{SNR}.%
\footnote{Strictly speaking the frequency bins for the $F_+$ and
$F_{\cross}$ components will be different because, as we have noted in
equation (\ref{eq:psds}), the \ac{PSD}s for the synthetic $+$ and $\cross$
detectors are not equal. However, usually the difference between the two
is small enough that it can be safely ignored to avoid computing twice
the number of filters. Alternatively, in \cite{Allen:2004gu} a method
was presented for calculating the standard $\chi^2$ test using unequal
frequency bins, that method could easily be incorporated into a coherent
search.}
The coherent $\chi^{2}$ statistic is then
constructed as
\begin{equation}
  \chi^2 = N \sum_{i=1}^N \sum_{\mu = 1}^4 (\rho_{\mu}^i - \rho_{\mu}/N)^2. 
\end{equation}

\begin{figure}
  \begin{minipage}[t]{0.95\linewidth}
    \includegraphics[width=\linewidth]{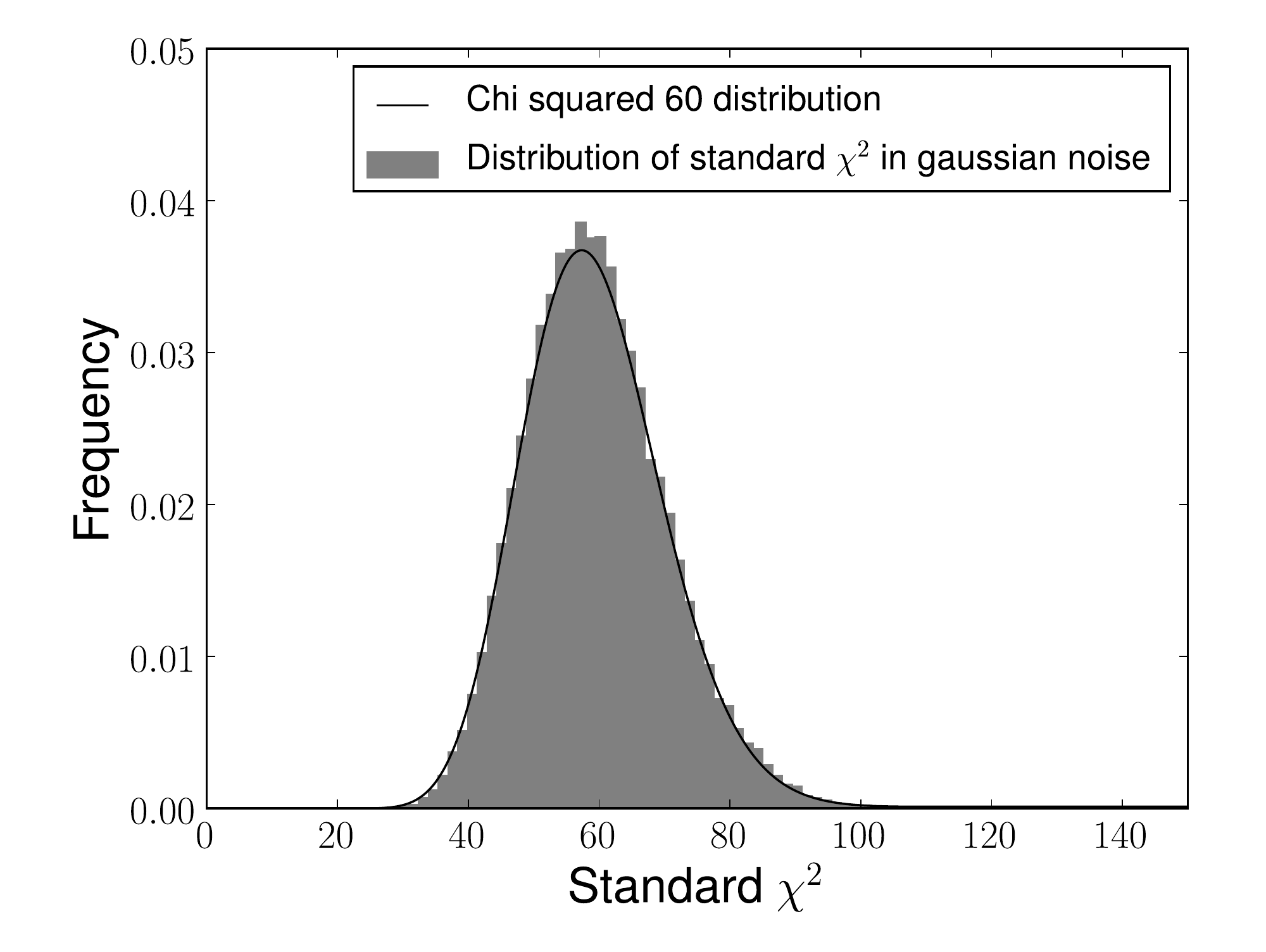}
  \end{minipage}
  \caption{\label{fig:chi_hist} 
The distribution of the $\chi^{2}$ test for a single template $h$, split
into 16 non-overlapping frequency bins.  The plot shows the distribution
of the $\chi^{2}$ test calculated for every time sample in $128 s$ of
simulated Gaussian data (with no signal present).  The observed
distribution of values shows (shown in grey) is an excellent match with the
expected $\chi^{2}$ distribution with sixty degrees of freedom (shown in
black). }
\end{figure}

As all the components are orthogonal it is easy to see that this
statistic will be exactly $\chi^{2}$ distributed with $4N - 4$ degrees
of freedom.  One can interpret this as the sum of the single detector
$\chi^{2}$ values for the $h_{0}$ and $h_{\frac{\pi}{2}}$ waveforms in the
synthetic $+$ and $\cross$ detectors.  Figure \ref{fig:chi_hist} shows
the distribution of the ``standard''
$\chi^{2}$, using 16 frequency bins.  The distribution matches the
expected $\chi^{2}$ with 60 degrees of freedom.

An alternative approach to applying the $\chi^2$ test to a coherent search
was proposed in \cite{Pai:2000zt}. This approach involves calculating the
$\chi^2$ values for each of the active detectors and using these values
to veto glitches.

\section{Implementation and Performance of a Coherent Search}
\label{sec:s4data}

In the previous sections we have presented a method for detecting
gravitational waves from CBCs in a network of detectors.  The coherent
\ac{SNR} described in \ref{sec:coh_matched_filter} is ideal for
distinguishing signals in Gaussian data and, in sections
\ref{sec:snr_cont_consistency} and \ref{sec:chi2tests}, we have
introduced a number of strategies for discriminating between signal and
noise in non-Gaussian, non-stationary data. Here, we describe an
implementation of the targeted, coherent search for gravitational waves
from \ac{CBC}.  In addition, we demonstrate its  efficacy by performing
test analyses of simulated data and real data taken from \ac{S4}.

\subsection{Implementation of a coherent triggered search for CBCs}
\label{sec:searchimplementation}

Here, we describe the main steps by which the algorithms described in
sections \ref{sec:coh_matched_filter}, \ref{sec:snr_cont_consistency}
and \ref{sec:chi2tests} have been implemented.  The analysis is
available in the  \ac{LIGO} Scientific Collaboration Applications Library
(LAL) suite \cite{lalsuite}, and makes use of a large number of tools
and methods previously implemented in that library.

\subsubsection{Analysis setup}
\label{sec:searchsetup}

A targeted, coincident search for gravitational waves from \ac{CBC}
associated to \ac{GRB}s has been implemented, and used in a search of
\ac{S5} and \ac{VSR1} data \cite{Abadie:2010uf}.  The
coherent search pipeline uses many of the same definitions, and much of
the same architecture to determine the analysis details.  Specifically,
``onsource'' time is [-5,+1) seconds around the reported time of the
GRB; this is when a gravitational wave signal would be expected
\cite{Shibata:2007zm,DavLevKing04}.  The noise background is estimated using 1,944 seconds of
``offsource'' data split into 324 trials of 6 second length each.  These
are used to calculated the significance of any event occuring in the
onsource. As in Ref.~\cite{Abadie:2010uf} we impose a 48s buffer zone
between the onsource and offsource regions.  To obtain an accurate
estimate of the detectors' power spectra, we only analyse data from a
detector if it has taken at least 2190s of continuous data around the
time of the GRB.  Modulo this restriction, the coherent analysis is
designed to make use of data from all detectors that were on at the time
of the GRB.

\subsubsection{Template bank generation}
\label{sec:templategeneration}

The problem of placing a template bank for a single detector has been
extensively studied \cite{Owen96, OwenSathyaprakash98, Bank06,
Cokelaer:2007kx}.  However, less thought has been given to the problem
of placing an appropriate bank for a coherent analysis.  Our current
method is to use a template bank generated for one of the detectors in
the network. In the results presented later, we have made use of a bank
generated with the initial \ac{LIGO} design spectrum, with a maximum
total mass of 40$M_{\odot}$ and a minimum component mass of
1$M_{\odot}$, these are the same values as used in \cite{Abadie:2010uf}.
This method enables us to demonstrate the efficiency of this coherent
search it is not the optimal solution.  A simple improvement would
involve placing a template bank appropriate for the (maximally
sensitive) synthetic $+$ detector defined in equation (\ref{eq:psds}).
In many cases, a network is significantly more sensitive to one GW
polarization and, in these cases, this template bank would perform well.

\subsubsection{Coherent SNR and Null Streams}
\label{sec:cohsnrproduction}

The data are first read in and conditioned using the methods and
algorithms developed for the \ac{S4} search for post-merger ringdowns from
CBCs \cite{Abbott:2009km, Goggin:2008dz}.  The data are downsampled to a
frequency of 4096Hz and split into overlapping 256 second segments for
analysis. The noise \ac{PSD}s are calculated using the same method as in
\cite{Abbott:2009km}.

Each template of the bank is filtered against the data from each
detector to generate the single detector filters
$(s^{X}|h^{X}_{0,\frac{\pi}{2}})$ and sensitivities $\sigma^X$.  The
algorithms used are taken from the LAL FindChirp library
\cite{Allen:2005fk}, specifically those written to perform a search for
spinning waveforms \cite{Fazi:2009} using the physical template family
(PTF) waveforms \cite{PBCV04}.%
\footnote{This choice stems from the desire to extend this search to
incorporate a single spin.  This is particularly appropriate for
\ac{NSBH} binaries where the spin of the \ac{NS} can be safely
neglected.}
 The waveform templates are generated using the TaylorT4 post--Newtonian
approximant \cite{Damour:2000zb}.  The single detector filter outputs are
shifted in time to account for the relative delays from the given
\ac{GRB} sky location.  They are then combined to form the coherent and
the null \ac{SNR}s as described by (\ref{eq:coherent_xy}) and
(\ref{eq:null_overwhite}).  A ``trigger'' is recorded at any time the
coherent \ac{SNR} is greater than 6, and no louder event occurred in
any template in the bank within $0.1$ seconds.

\subsubsection{Calculating the $\chi^{2}$ tests}
\label{sec:sbvproduction}

The analysis calculates signal based vetoes in the same manner that it
does the coherent \ac{SNR}: The necessary single detector filters are
constructed and then these are combined together to create the
$\chi^{2}$ tests as described in the earlier sections.

Calculating the ``standard'' $\chi^2$ test is computationally expensive.
Therefore this veto is only calculated for a segment if there is at
least one event within that segment with \ac{SNR} larger than the
threshold and with values of bank $\chi^2$, auto $\chi^2$ and null
\ac{SNR} that do not immediately lead to it being dismissed as a glitch.

\subsubsection{Simulated Signals}
\label{sec:softwareinjs}

The sensitivity of the analysis to gravitational wave signals from
\ac{CBC} is assessed by adding simulated signals to the data stream
before performing the search.  The analysis uses the architecture from
\cite{Abbott:2009km} to perform these simulations.  A simulation is
deemed to have been recovered by the analysis is there is an event
within $0.1$ seconds of the signal time; no attempt is made to guarantee
a good match between simulated and recovered parameters.  While this
does mean that, in principle, signals may be found due to a nearby
glitch, the effect is minimal, particularly when considering detection
candidates which are louder than all background.

The results shown in the remainder of the section are based on a set of
simulated signals comprised of binary neutron stars (component masses
limited to be between 1 and 3 $M_{\odot}$) at appropriate distances to
be observable by the detector network.  All of these binaries are
oriented face on to the detectors.  These choices are motivated by the
fact that the current implementation is designed as a search for
gravitational waves associated with \ac{GRB}s.

\subsection{Analysis of simulated data}
\label{sec:gaussiannoise}

The analysis was first run on simulated data for the initial \ac{LIGO}
network (H1, H2, L1).  Data were simulated to be Gaussian and
stationary, with no noise transients.  Specifically, the coherent
analysis pipeline was run on 2190 sections of
Gaussian data as if a GRB had occurred in the middle of the data
stretch.  This provides a benchmark with which to compare results when
running on real data. 

\begin{figure}
  \begin{minipage}[t]{0.95\linewidth}
  \includegraphics[width=\linewidth]{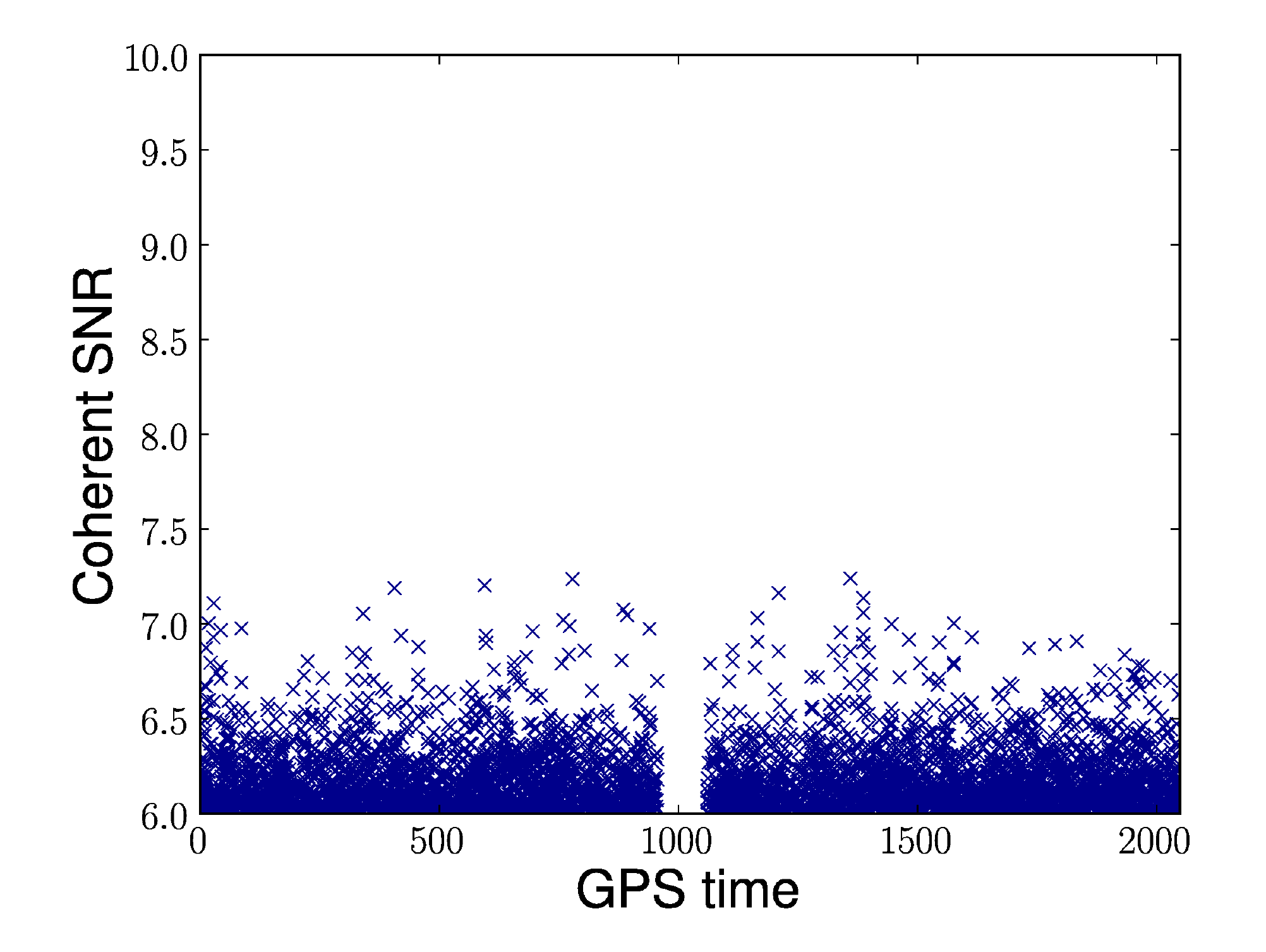}
  \end{minipage}
  \caption{\label{fig:S4gaussiansnr} 
The distribution of \ac{SNR} triggers in the offsource region plotted
against time for an analysis of simulated Gaussian noise in the initial
\ac{LIGO} (H1, H2, L1) network. }
\end{figure}

Figure \ref{fig:S4gaussiansnr} shows all the triggers produced by the
pipeline in the off source time. The loudest event in the approximately
2000 seconds of off-source data has an \ac{SNR} of 7.24.
Ideally, the various signal consistency tests described previously will
reduce the amplitude of the loudest surviving event in real data to
something similar to this.

\subsection{Analysis of real data}
\label{sec:realdata}

\begin{figure}
  \begin{minipage}[t]{0.95\linewidth}
  \includegraphics[width=\linewidth]{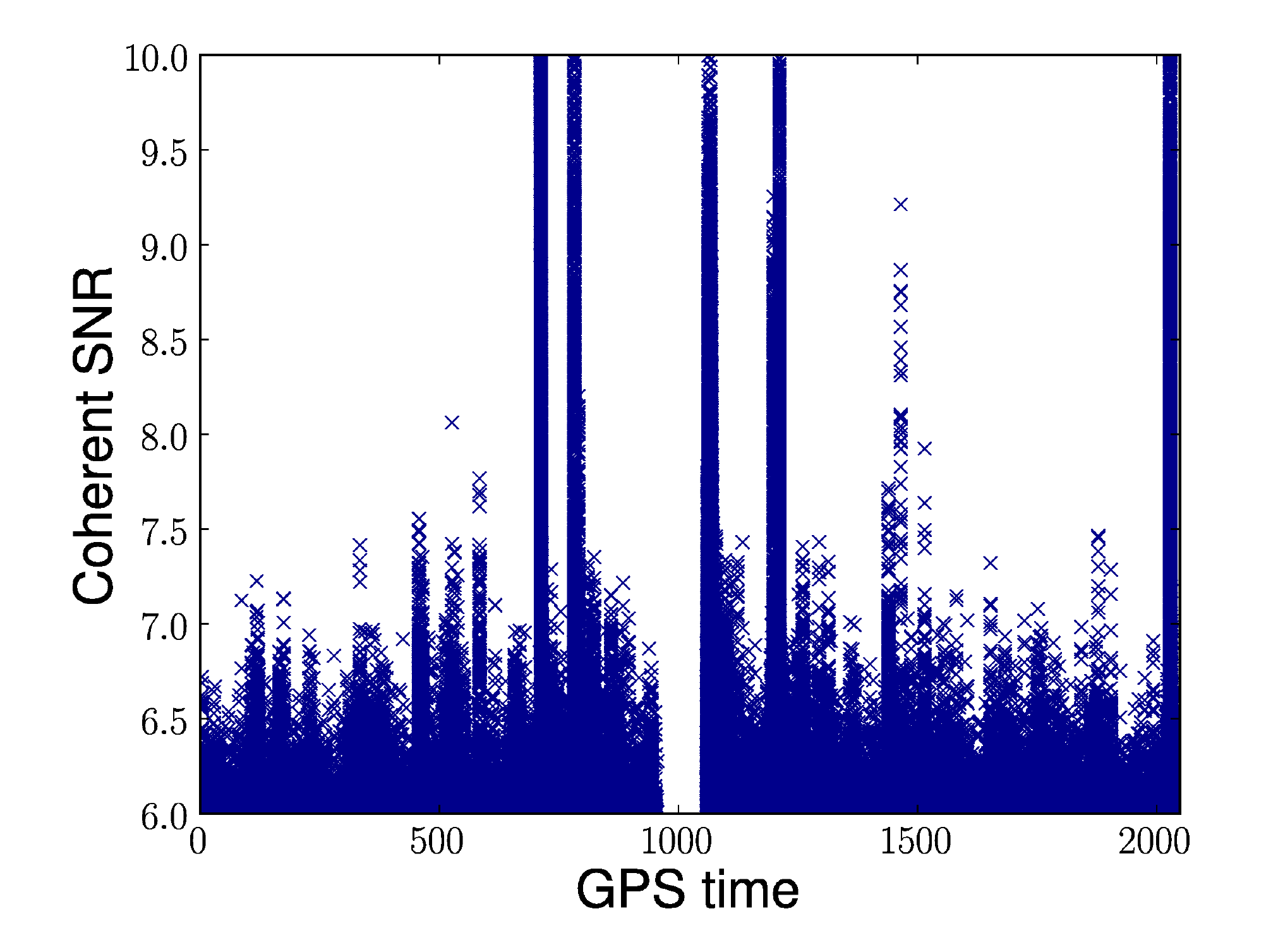}
  \end{minipage}
  \caption{\label{fig:S4snrtime} 
The distribution of \ac{SNR} triggers in the offsource region plotted against
time for an analysis of a mock S4 GRB.  The axes on the plot are chosen to be
identical to those for figure \ref{fig:S4gaussiansnr} to make the plots easier
to compare.  The S4 data has a large number of non-Gaussian features.  The
largest of these peaks extends to a coherent \ac{SNR} of 40, although
non-Gaussian structure is visible at \ac{SNR}s as low as 7.}
\end{figure}

A test analysis was performed on real data taken from \ac{S4}.  We chose
an arbitrary block of 2190 seconds of data for which all three of the
\ac{LIGO} detectors were operating and ran the analysis as if a GRB had
occurred during this time.  The simulated sky location of the GRB was
(184.623\textdegree,42.294\textdegree) in right ascension and
declination respectively. For this chosen time and sky location the 
sensitivity of the H1 and L1 detectors were roughly equal and the H2
detector was half as sensitive as the other two.

\subsubsection*{Coherent SNR}
\label{sec:coh_snr}

Figure \ref{fig:S4snrtime} shows the coherent \ac{SNR} of triggers
produced during the analysis of the S4 data.  It clearly demonstrates
that this data is not well characterized by Gaussian noise alone. A
number of loud transients are present in the data which show up as short
duration peaks of large \ac{SNR}. The loudest of these has an
\ac{SNR} of almost 40.  If events were only ranked on \ac{SNR}, a signal
would have to be very loud to show against this non-Gaussian background.
In addition to the loud peaks there are also a large number of
quieter peaks that show up at all times in the data. If the search
is to begin to approach the efficiency of the Gaussian case then the
signal consistency tests must allow us to reject the quieter noise
transients, as well as the loud ones.

In the remainder of this subsection we demonstrate the performance  of
the signal consistency tests introduced in sections
\ref{sec:snr_cont_consistency} and \ref{sec:chi2tests}.  Finally, we
formulate a detection statistic, which arises as a combination of the
coherent \ac{SNR} and signal consistency tests, and demonstrate that
with it the majority of the non-Gaussian background can be removed.

\subsubsection*{Null SNR}
\label{sec:s4null}

\begin{figure}
  \begin{minipage}[t]{0.95\linewidth}
    \includegraphics[width=\linewidth]{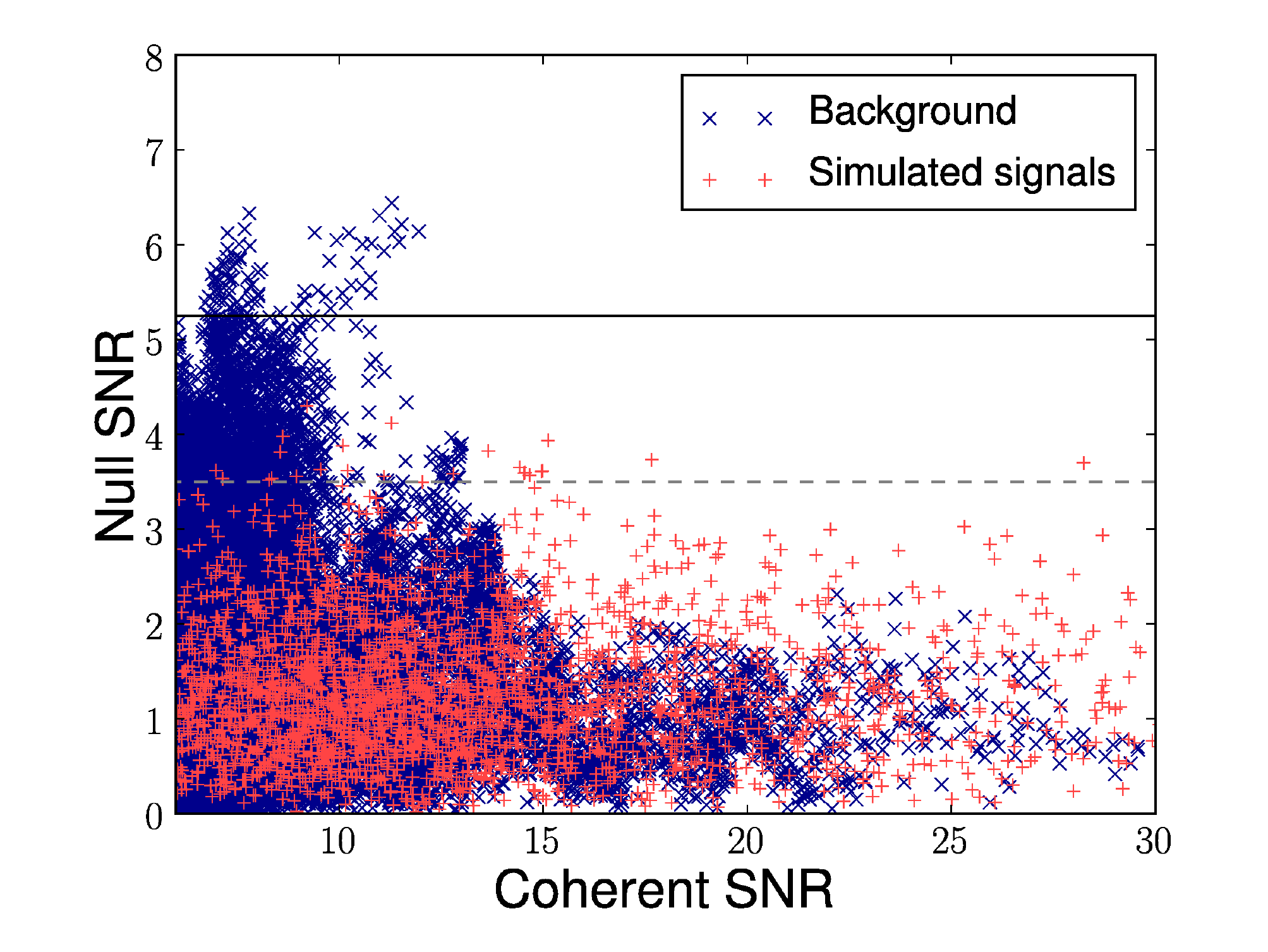}
  \end{minipage}
  \caption{\label{fig:S4null} 
The distribution of the null \ac{SNR} plotted against coherent \ac{SNR}.
The solid line at null \ac{SNR} of 5.5 is the line above which
triggers are vetoed. The dashed line at 3.5 is the line above which
triggers are downweighted (see section \ref{sec:detstats4}). }
\end{figure}

Figure \ref{fig:S4null} shows the performance of the null stream for
both simulated signals and background noise.  The ability of the null
\ac{SNR} to distinguish signal from noise is relatively poor in this
example.  The mock GRB analysis uses data from the two Hanford detectors
and the detector at Livingston.  Thus, the null stream is
derived from a combination of the H1 and H2 detectors; the Livingston
detector does not contribute.  The loudest glitches during the time
of this analysis originated in L1, and therefore do not have significant
null \ac{SNR}.  However, quieter glitches in the Hanford detectors at an
\ac{SNR} around 10 do produce a large null \ac{SNR}.  Any trigger with a
null \ac{SNR} greater than 5 is eliminated from the analysis.

\subsubsection*{Single detector SNR}
\label{sec:S4snglsnrs}

\begin{figure}
  \begin{minipage}[t]{0.95\linewidth}
    \includegraphics[width=\linewidth]{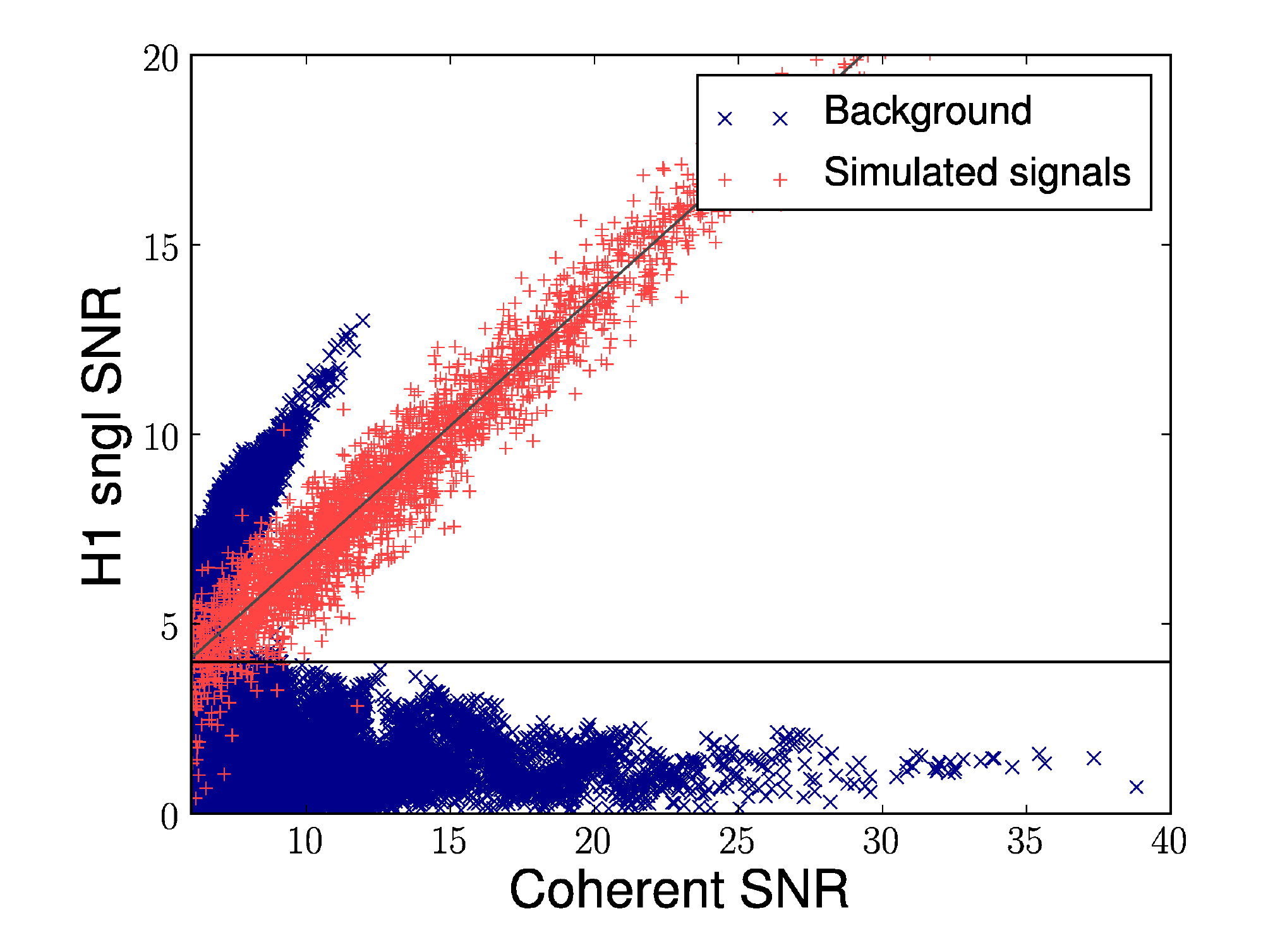}
  \end{minipage}
  \begin{minipage}[t]{0.95\linewidth}
    \includegraphics[width=\linewidth]{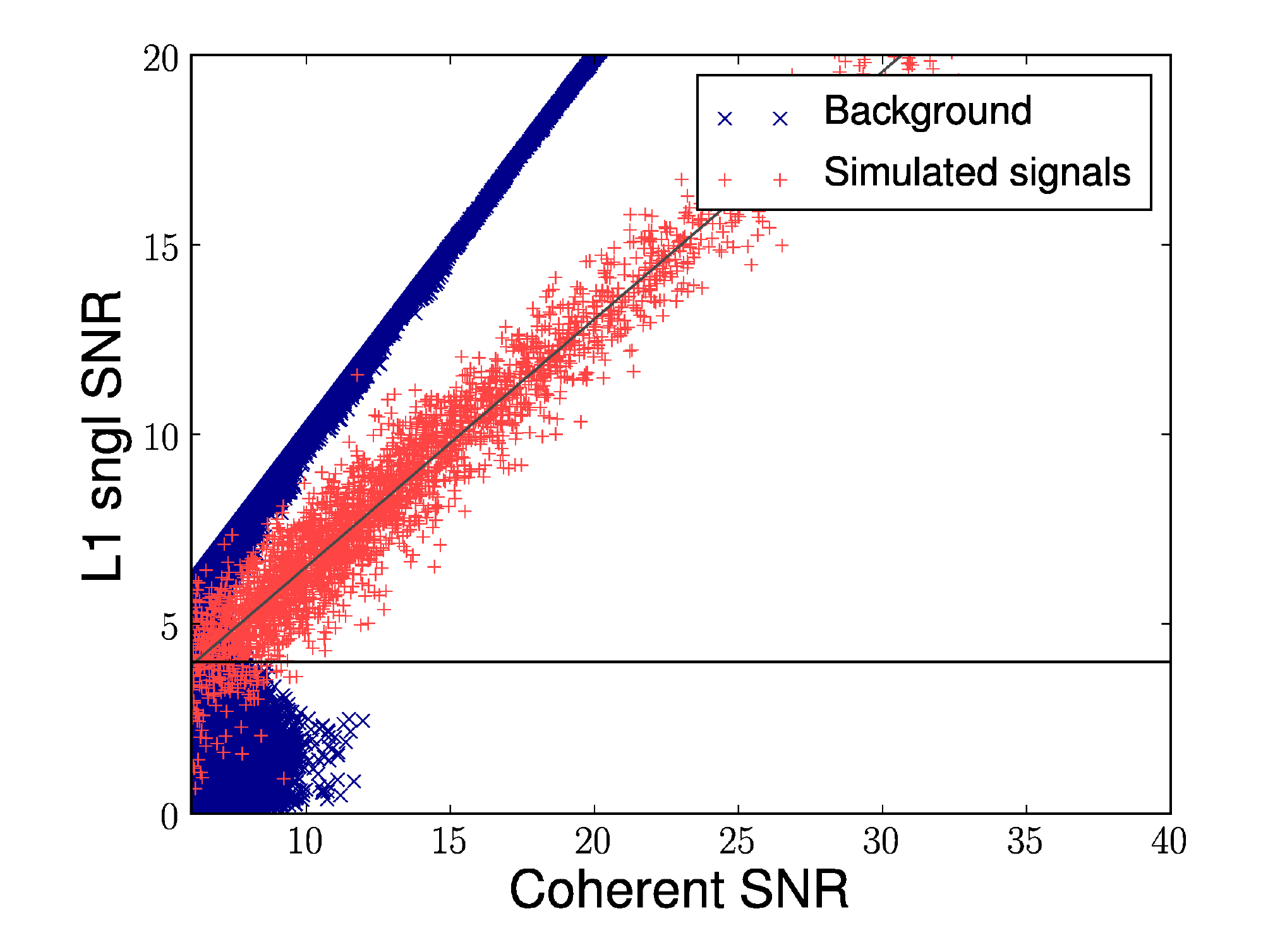}
  \end{minipage}
\caption{\label{fig:S4h1l1snr} 
The distribution of single detector \ac{SNR} for the more sensitive H1
and L1 detectors, plotted against coherent \ac{SNR}. The top figure
shows the H1 \ac{SNR}, the bottom figure shows the L1 \ac{SNR}. The
horizontal line indicates \ac{SNR}=4. Below this line triggers
will be vetoed. The inclined dark gray line indicates the expected
\ac{SNR} of these face on simulated signals.}
\end{figure}

The most straightforward, and most effective amplitude consistency test
we have found is the requirement of a single detector \ac{SNR} greater
than 4 in the two most sensitive detectors;  In this analysis, the L1
and H1 detectors.  Figure \ref{fig:S4h1l1snr} demonstrates that this is
a particularly effective strategy for removing noise glitches.  Triggers
arising due to glitches in the L1 detector have large coherent \ac{SNR}
but a negligible contribution from H1 and are consequently discarded.

\subsubsection*{$\chi^2$ tests}
\label{sec:s4chi2}

\begin{figure}
  \begin{minipage}[t]{0.95\linewidth}
    \includegraphics[width=\linewidth]{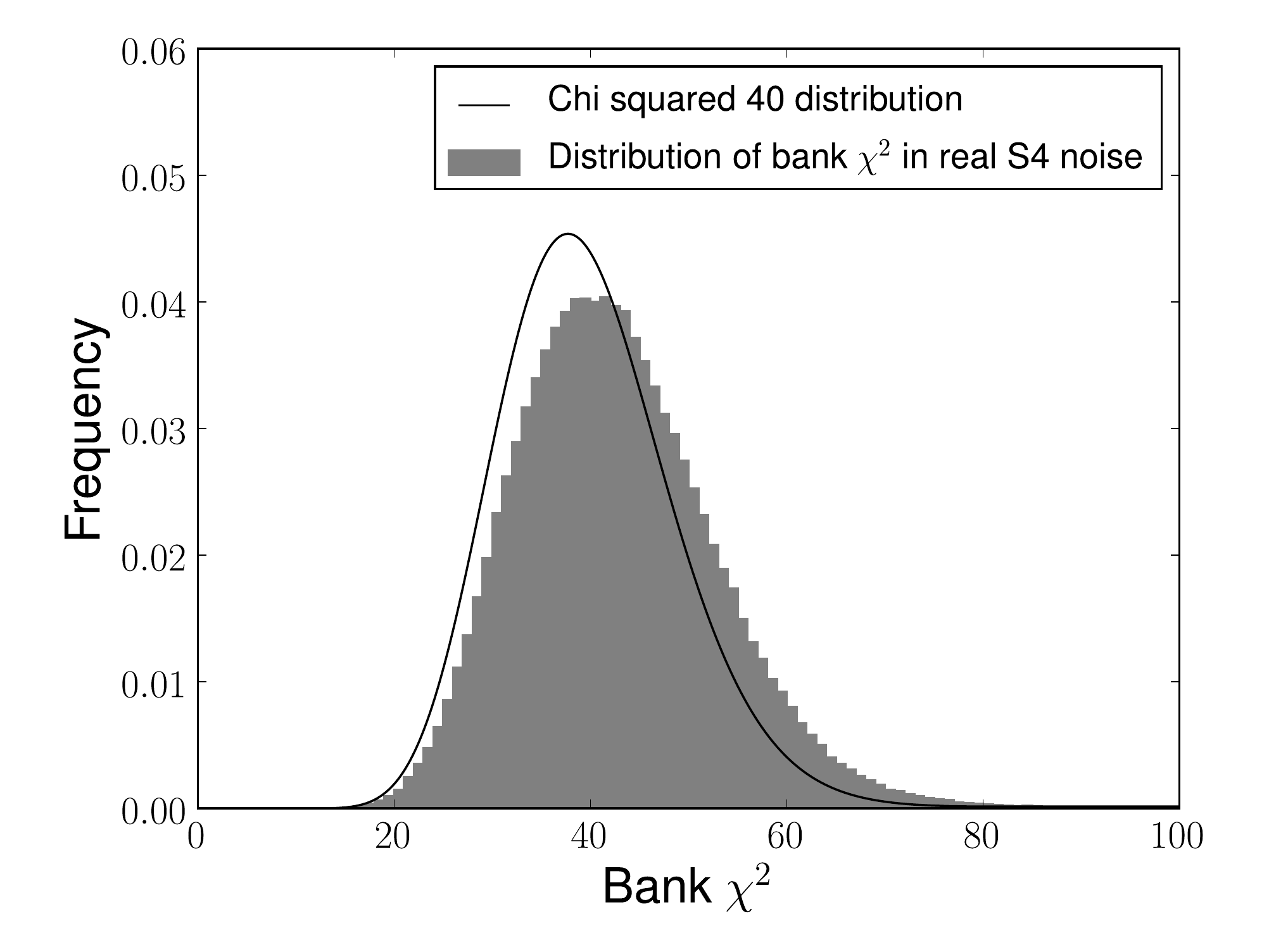}
  \end{minipage}
  \caption{\label{fig:S4bank_hist} 
The distribution of the bank $\chi^{2}$ test for a single template $h$,
with a bank of size $10$.  The plot shows the distribution of the bank
veto calculated for every time sample in $128 s$ of data.  The observed
distribution is inconsistent with the expected result in Gaussian noise
(the black curve).}
\end{figure}

In section \ref{sec:chi2tests} we introduced three $\chi^2$ tests
designed to separate signals from noise glitches in the data.  Figure
\ref{fig:S4bank_hist} shows the distribution of the bank $\chi^2$ for
every time sample for a single template.  This is directly comparable to
Figure \ref{fig:bank_hist} which shows the same for Gaussian data.  The
deviation from the predicted $\chi^{2}$ distribution is due to the
non-Gaussianity of the data.

\begin{figure}
  \begin{minipage}[t]{0.95\linewidth}
    \includegraphics[width=\linewidth]{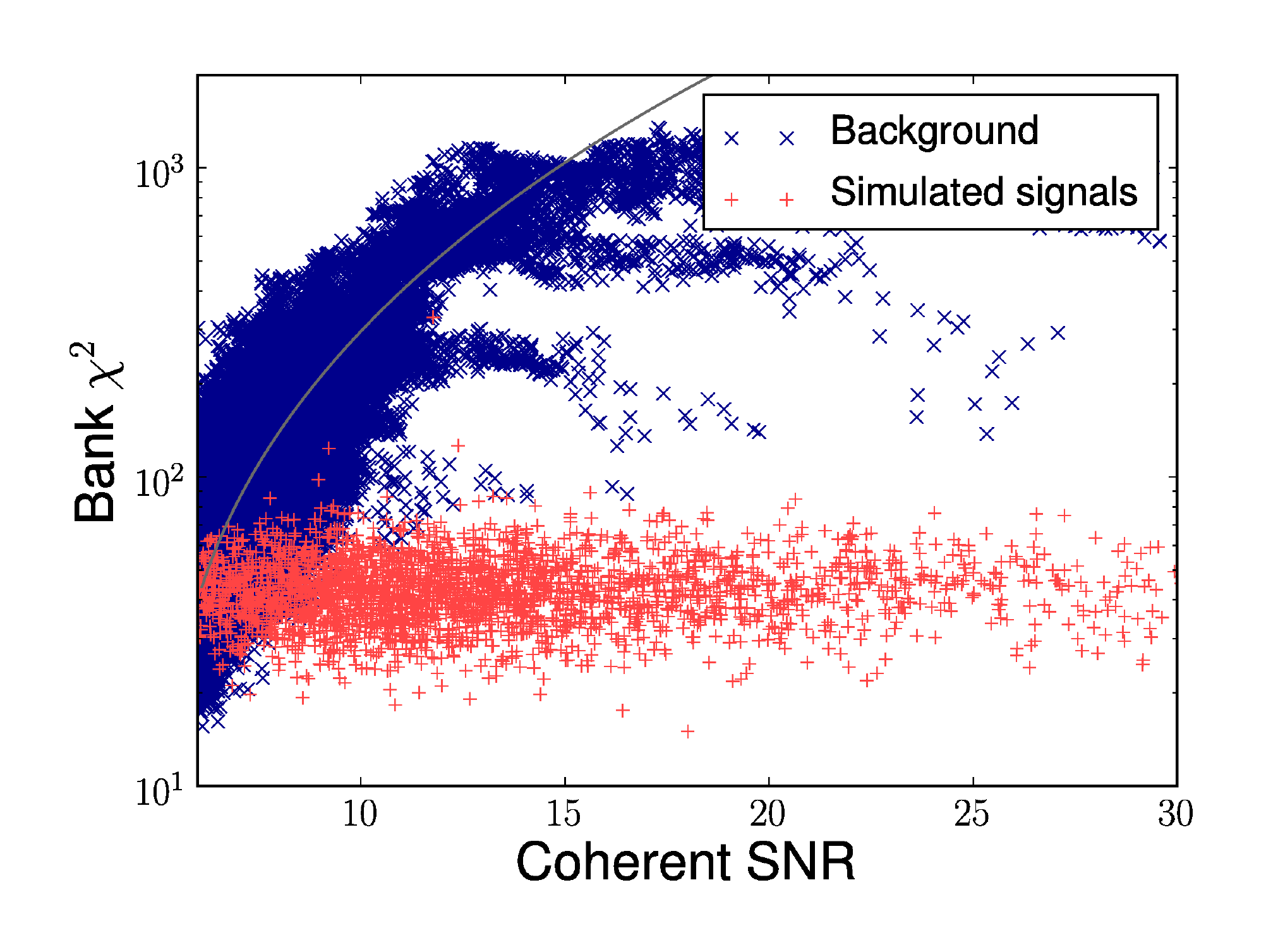}
  \end{minipage}
  \begin{minipage}[t]{0.95\linewidth}
    \includegraphics[width=\linewidth]{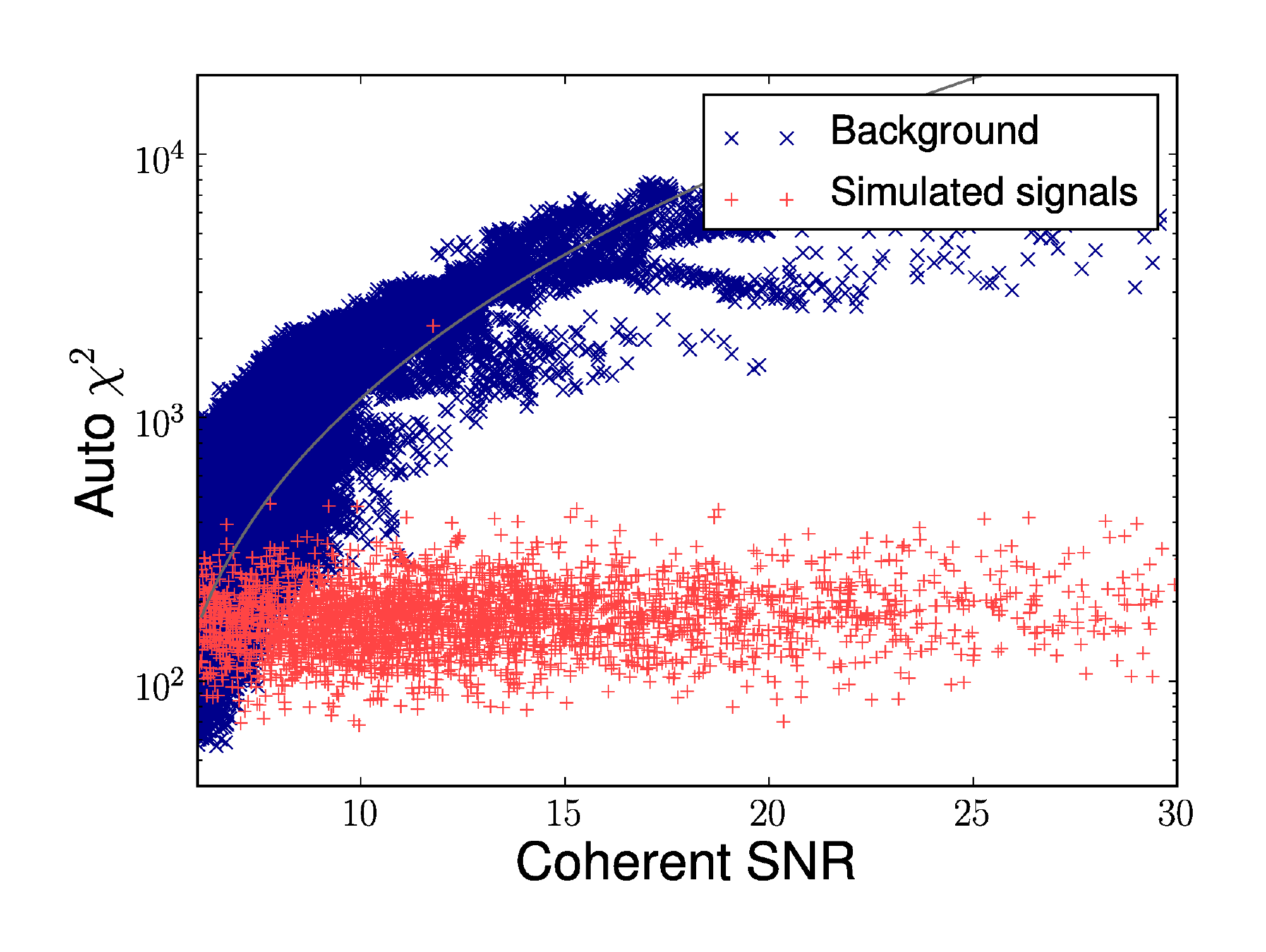}
  \end{minipage}
  \caption{\label{fig:S4bank_auto trig} 
The distribution of bank and auto $\chi^{2}$ test plotted
against \ac{SNR}.  The results for the off-source data triggers are
plotted in blue, with simulated signals in red.
The solid line shows the line of \textit{newSNR} = 6,
defined in section \ref{sec:detstats4}.
Triggers with \textit{newSNR} $ < 6$ are vetoed.}
\end{figure}

The distribution of bank and auto $\chi^{2}$ for both simulated signals
and noise triggers is shown in Figure \ref{fig:S4bank_auto trig}.  Both
of these tests are effective at separating the simulated signals from
noise transients.  In order to quantify this, we make use of the
\textit{newSNR} formalism that is being used in the latest
coincident searches for CBCs \cite{s6allsky,s6grb}.  The idea is to
downweight the significance of noise triggers with large $\chi^{2}$
values relative to signals.  This is achieved by introducing the ``\textit{newSNR}'':
\begin{equation}
 \rho_{\mathrm{new}} = \left\{
\begin{array}{cl}
\displaystyle \rho, & \chi^2 \le n_{\mathrm{dof}} \\[0.1in]
\displaystyle \frac{\rho}{\left[\left(1 +
\frac{\chi^2}{n_{dof}}^{4/3}\right)/2\right]^{1/4}}, & \chi^2 >
n_{\mathrm{dof}} 
\end{array} 
\right.
\end{equation}
where $n_{\mathrm{dof}}$ is the number of degrees of freedom of the
$\chi^{2}$ test.
For a signal, the mean $\chi^{2}$ value is one per degree of freedom and
consequently the \textit{newSNR} will be similar to the SNR.  Noise
transients with a large $\chi^{2}$ value are significantly downweighted.
The \textit{newSNR} is calculated using both the auto and bank $\chi^{2}$
values.  Any trigger with \textit{either} an auto \textit{newSNR} or bank \textit{newSNR}
less than 6 is discarded.  The curves on Figure \ref{fig:S4bank_auto
trig} show this \textit{newSNR} threshold for the two $\chi^{2}$ tests.

\begin{figure}
  \begin{minipage}[t]{0.95\linewidth}
    \includegraphics[width=\linewidth]{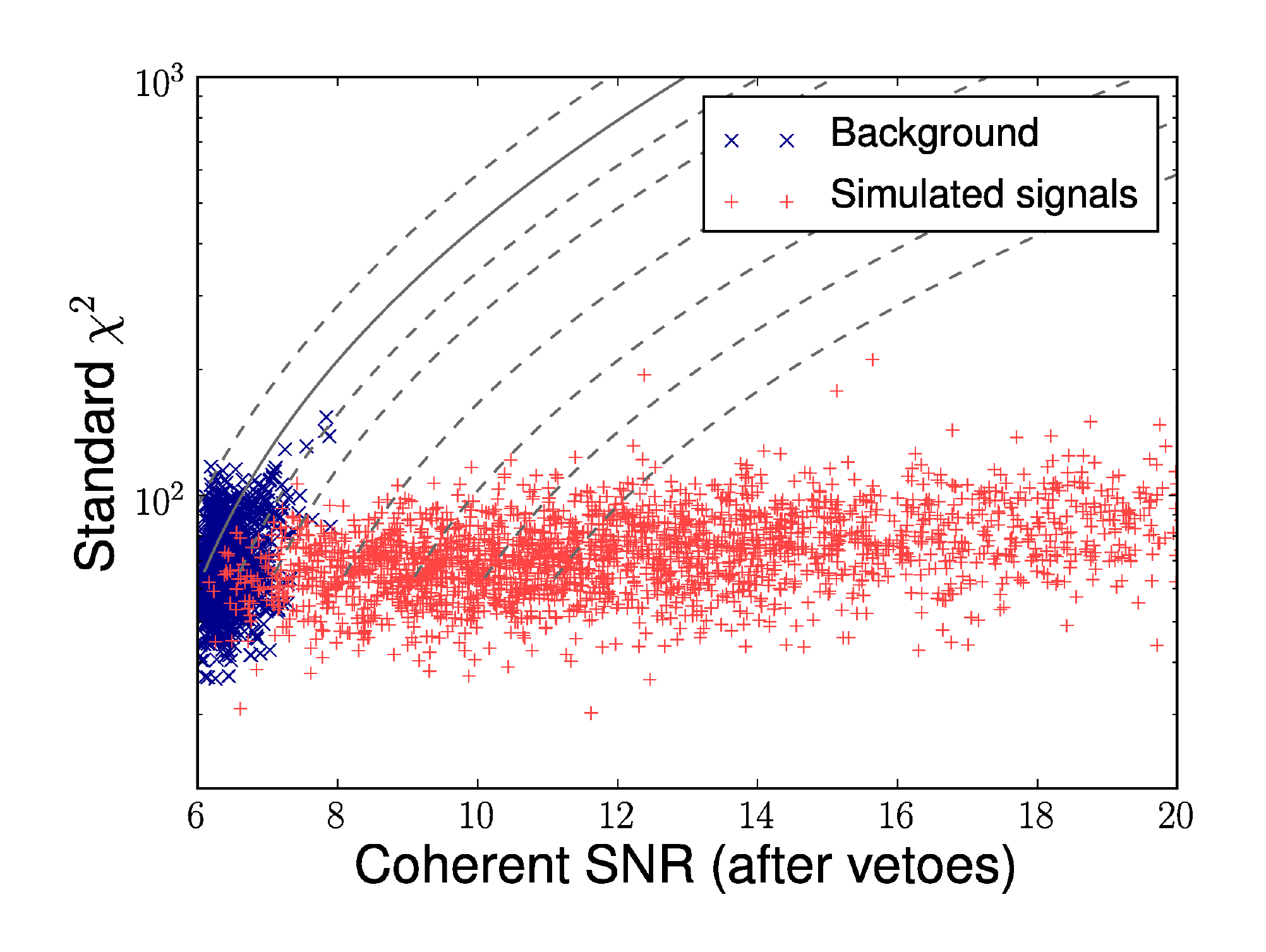}
  \end{minipage}
  \caption{\label{fig:S4chisq}
The distribution of standard $\chi^{2}$ test plotted
against \ac{SNR}.  The results for the off-source data triggers are
plotted in blue, with simulated signals in red. The dashed lines show
contours of \textit{newSNR}, defined in section
\ref{sec:detstats4}, the solid line shows the line of \textit{newSNR} = 6.
Triggers with \textit{newSNR} $< 6$ are vetoed.} 
\end{figure}

Finally, we turn to the standard $\chi^{2}$ test.  As this is rather
costly to compute, we only do so for triggers which have passed all of
the previously described thresholds (on coherent, null and single
detector \ac{SNR}, the bank and auto $\chi^{2}$).  Figure
\ref{fig:S4chisq} shows the distribution of the standard $\chi^{2}$ test
for simulated signals and noise.  The preceding tests have succeeded in
removing the vast majority of non-Gaussian triggers from the data.  A
threshold of 6 on \textit{newSNR} serves to eliminate a few more.  We have
found that the standard $\chi^{2}$ is the most effective at separating
signal from background so we also make use of it in the final ranking of
events.  Figure \ref{fig:S4chisq}, shows contours of constant \textit{newSNR}
which will be used in the final ranking.

\subsubsection*{Detection statistic}
\label{sec:detstats4}

In the preceeding discussion, we have imposed a number of cuts on the
initial candidate events produced by the analysis pipeline.  Let us
briefly recap those cuts:

\begin{itemize}

\item Generate a trigger at any time for which $\rho > 6$.  Only keep
the loudest one in each 0.1 seconds.

\item Discard any triggers with $\rho_{N} > 5$.

\item Discard any triggers for which $\rho_{H1} < 4$ or $\rho_{L1} < 4$.

\item Discard any triggers for which $\rho_{new} < 6$ for the bank or
auto $\chi^{2}$.

\end{itemize}

Finally, we rank the remaining triggers based upon the \textit{newSNR}
calculated using the standard $\chi^{2}$ as well as the null SNR as:

\begin{itemize}

\item Rank remaining triggers using a detection statistic $\rho_{det}$
given by
\begin{equation}\label{eq:rho_det}
\rho_{\mathrm{det}} =  \left\{
\begin{array}{ll}
 \displaystyle \rho_{\mathrm{new}} , & \rho_{N} \le 3.5 \\[0.1in]
 \displaystyle \frac{\rho_{\mathrm{new}}}{\rho_{\mathrm{Null}} - 2.5} , 
 & 3.5 < \rho_{N} < 5.0
\end{array}
\right.
\end{equation}

\end{itemize}

The length of time a \ac{CBC} spends in the sensitive band of the
detector varies greatly with the mass, and it has been found that the
shorter, high mass templates are more susceptible to occuring with large
\ac{SNR} at the time of glitches \cite{abbott:122001}.  Also, the
various signal consistency tests are less effective for these short
templates.  Therefore, we follow Ref.~\cite{abbott:122001} and split
the template bank into three regions based on the chirp mass of the
template.  The false alarm probability for a given trigger is based on a
comparison of the detection statistic to off source triggers \textit{in
the same mass bin}.

\subsubsection*{Performance of search}

\begin{figure}
  \begin{minipage}[t]{0.95\linewidth}
    \includegraphics[width=\linewidth]{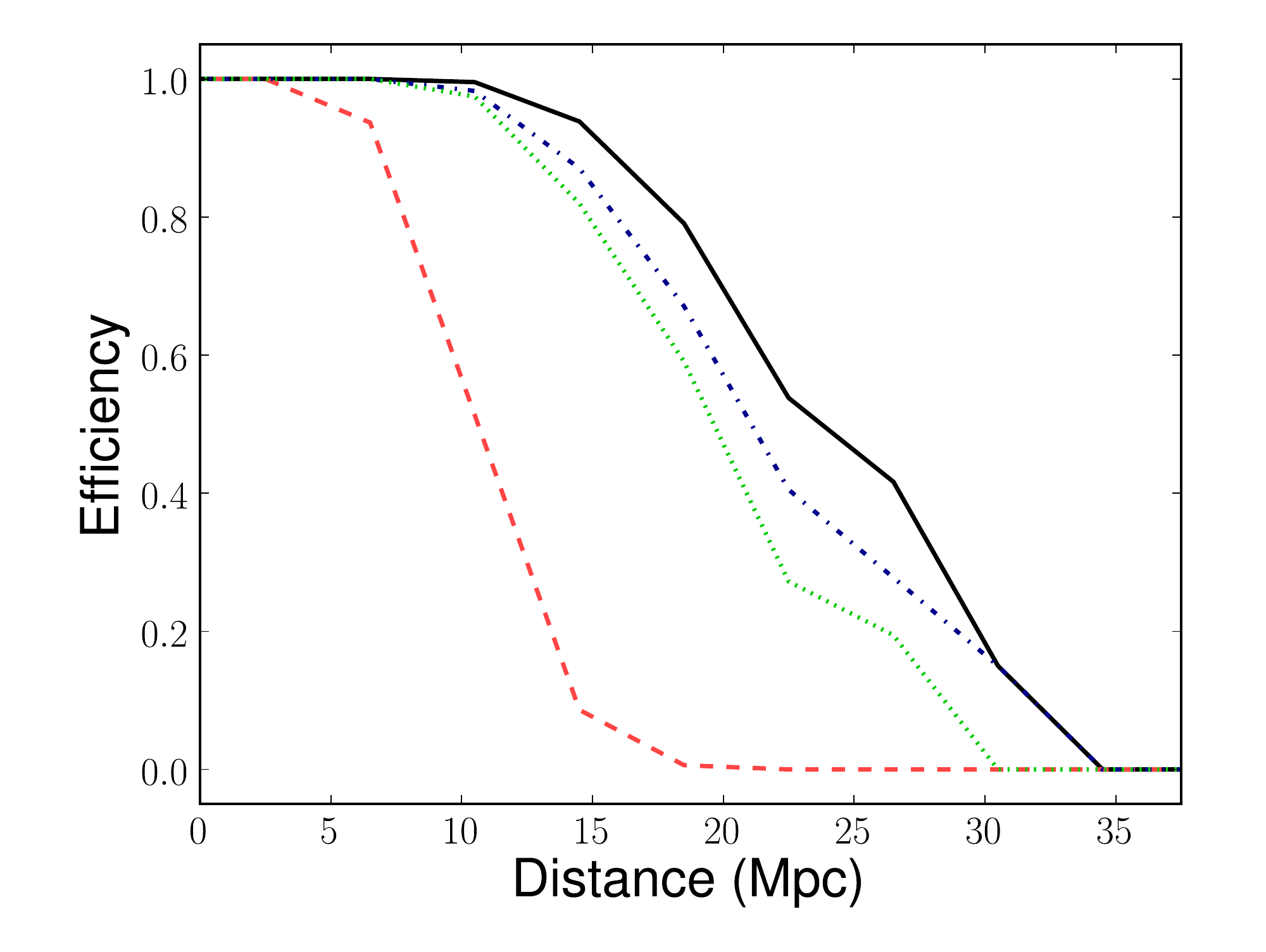}
  \end{minipage}
\caption{\label{fig:S4injrec} 
Efficiency of recovery of simulated signals in real data.  The
efficiency is shown for four different cases: i) signals found above an
\ac{SNR} of 7.24, the loudest background trigger in Gaussian noise
(black solid line); ii) signals found above an \ac{SNR} of 12.88, the
loudest background trigger in real noise (red dashed line); iii) signals
found with a value of the detection statistic (\ref{eq:rho_det}) above
7.41, the loudest background event; iv) signals found
louder than all background in an H1-L1 search (green dotted line).}
\end{figure}

The sensitivity of the analysis can be assessed by examining the
performance of recovering simulated signals. The signal population was
\ac{BNS} with optimal orientation originating from the location of the
fake GRB.  For these purposes we will only consider an signal found
if its associated trigger is louder than the loudest event that occured
in the low mass bin during the offsource time --- this ensures that the
false alarm probability is less than one in 324 (the number of offsource
trials).  At this level, a candidate event starts to be interesting but
realisitically a false alarm probability closer to $10^{-4}$ or
$10^{-5}$ would be required for a detection candidate.  

Figure \ref{fig:S4injrec} shows the efficiency with which simulated
signals are recovered as a function of the distance.  Recall that the
largest \ac{SNR} recorded in the analysis of simulated, Gaussian data
was $7.24$.  We show the efficiency of the search at finding simulations
with an \ac{SNR} greater than this.  If the data were Gaussian, or the
signal consistency tests were able to remove all non-Gaussianities, then
the real search would match this sensitivity.  By using \textit{only}
\ac{SNR} to rank events, the sensitivity of the search is substantially
worse than in Gaussian data.  Although the offsource contained events
with an \ac{SNR} up to 39, the loudest low mass trigger had a coherent
\ac{SNR} of 12.88.  This is almost double the loudest Gaussian noise and
thus the sensitivity of the search is reduced by about a factor of two
--- illustrated by the 50\% efficiency moving down from 20 Mpc to 10
Mpc.  The signal consistency tests --- null stream, $\chi^{2}$ tests and
single detector \ac{SNR} threshold --- are designed to bring the
sensitivity closer to the Gaussian case.   The simulated signals are
considered found if they are recovered with a detection \ac{SNR} greater
than $7.41$ (the loudest off-source event) and all signal consistency
tests.  The efficiency of the detection search is only about 10\% less
than it would be in Gaussian noise.  The 10\% loss in
sensitivity can be attributed to having a slightly louder offsource
event ($7.41$ rather than $7.24$) and a \textit{small} loss of
efficiency due to the various signal consistency checks. 

To illustrate the benefits of a coherent search, we would like to
compare the performance with a coincidence search. For the initial LIGO
network, signals close to detection threshold would be unlikely to have
passed any SNR threshold placed on the H2 detector.  Specifically, for a
signal with a coherent \ac{SNR} of 7.5, the expected \ac{SNR} in H2
would be around 2.5 which is a prohibitively low threshold.  Therefore
it is reasonable to compare the coherent search to a two detector, H1-L1
coincidence search.  For a two detector search, the coherent and
coincident \ac{SNR}s are equal, and the null stream test is not
available.  Consequently, the performance of two detector coherent and
concident searches should be comparable.  Therefore, we present the
results of an H1-L1 detector coherent search to give an indication of
the performance of a coincidence search. The efficiency of this two
detector search is also shown in Figure \ref{fig:S4injrec}.  The
sensitivity of the two detector search is about a factor of 10\% lower
than the three detector coherent search.  For a network of three
approximately equally sensitive detectors, we would expect an even
greater sensitivity improvement from employing coherent techniques.

\section{Discussion}

We have presented a formulation of a targeted coherent search for
compact binary coalescences.  For Gaussian noise, the
coherent \ac{SNR} would be ideal for separating signals from the noise
background. However, since data from gravitational wave interferometers
is neither Gaussian nor stationary, we have also discussed a number of
methods of separating the non-stationary noise background from the
signal population.  These tests include various $\chi^{2}$ tests, which
were originally designed for use in single detectors.  We have extended
them to the network analysis and demonstrated their continued efficacy.
Additionally, the coherent analysis allows for some additional tests
which are not readily available in the coincidence case.  The most
significant of these is the null \ac{SNR} which can be used to reject
events which are not consistent with two gravitational wave
polarizations.  Additionally, we explored consistency tests between the
recovered amplitudes of the gravitational wave and found that a simple
\ac{SNR} threshold on the two most sensitive detectors gave excellent
results.  

The analysis described in this paper has been implemented and in the
final section we showed results of a test run.  This made use of the
\ac{S4} data from the LIGO detectors.  Although the data was far from
Gaussian, after the application of all of the signal consistency tests
the results were remarkably close to what would be expected in Gaussian
noise.  This analysis is available to be used in searches for GW inspiral
signals associated with GRBs in more recent LIGO and Virgo data,
such as S6 and VSR2 and VSR3.

There are a number of ways in which this analysis could be enhanced to
broaden its use and increase its sensitivity.  First, a number of
\ac{GRB}s, particularly those observed by Fermi \cite{fermiwebsite} and
IPN \cite{ipnwebsite} are not localized sufficiently accurately that the
error box can be treated as a point on the sky.  Thus, it would be nice
to extend this analysis to allow for a region of the sky to be covered.
This would require looping over the relevant sky points; incorporating
the correct detector sensitivities $F_{+,\times}$ and time delays.  In
principle, this would not greatly slow down the analysis as the majority
of time is taken in performing the single detector filters and these
would \textit{not} need to be re-calculated.  As well as looking at a
patch on the sky, the analysis could be extended to cover the whole sky,
as appropriate for an un-triggered search.  This brings in a host of new
complications which have been met and dealt with by other coherent
search methods \cite{1367-2630-12-5-053034, Klimenko:2008fu}.  In order
to obtain a good estimate of the background for an all sky, un-triggered
search we would need to implement background estimation and time
shifting the data would likely be the best way to do this.  

Since \ac{GRB}s are thought to be rather tightly beamed, it is
reasonable to take them as being face on, or close to.  In this case,
the gravitational waves are circularly polarized and there is, in
effect, only a single polarization.  This opens the possibility of
limiting the signal space to just this one polarization and adding an
extra ``null'' test.  Alternatively, it should be possible to perform a
Bayesian marginalization over the astrophysically expected distributions
of the various parameters.  
 
The progenitors of short \ac{GRB}s are thought to be \ac{BNS} or
\ac{NSBH}.  The search we have described is ideal for the \ac{BNS} case
as the spins of the neutron stars are unlikely to have a significant
effect on the waveform.  However, when one of the components of the
binary is a black hole, the spin could be large.  Furthermore, the mass
ratio is likely to be relatively large.  In this case, the spin of the
black hole can have a significant effect on the observed waveform
\cite{PBCV04}.  Consequently, we would like to extend this search to
incorporate spin effects.  The infrastructure described in this paper
can already accept spinning waveforms, but the implementation of signal
based vetoes proves somewhat more complex.  Work is underway on this
\cite{HFspin}. 

\section*{Acknowledgements}

The authors would like to thank Sukanta Bose, Duncan Brown, James Clark, Alex Dietz,
Tom Dent, Nick Fotopolous, Phil Hall, Ioannis Kamaretsos, Duncan Macleod, Laura Nuttall,
Valeriu Predoi, Patrick Sutton, Bangalore Sathyaprakash, John Veitch and Alan
Weinstein for useful discussion and helpful comments about this work.
In this work
IWH was supported by the Science and Technology Facilities Council, UK,
studentship ST/F005954/1, SF was supported by the Royal Society.
This paper has been given the internal \ac{LSC} document number
P1000104.
\\
\\
\\
\\

\bibliography{references}

\begin{thebibliography}{70}%
\makeatletter
\providecommand \@ifxundefined [1]{%
 \@ifx{#1\undefined}
}%
\providecommand \@ifnum [1]{%
 \ifnum #1\expandafter \@firstoftwo
 \else \expandafter \@secondoftwo
 \fi
}%
\providecommand \@ifx [1]{%
 \ifx #1\expandafter \@firstoftwo
 \else \expandafter \@secondoftwo
 \fi
}%
\providecommand \natexlab [1]{#1}%
\providecommand \enquote  [1]{``#1''}%
\providecommand \bibnamefont  [1]{#1}%
\providecommand \bibfnamefont [1]{#1}%
\providecommand \citenamefont [1]{#1}%
\providecommand \href@noop [0]{\@secondoftwo}%
\providecommand \href [0]{\begingroup \@sanitize@url \@href}%
\providecommand \@href[1]{\@@startlink{#1}\@@href}%
\providecommand \@@href[1]{\endgroup#1\@@endlink}%
\providecommand \@sanitize@url [0]{\catcode `\\12\catcode `\$12\catcode
  `\&12\catcode `\#12\catcode `\^12\catcode `\_12\catcode `\%12\relax}%
\providecommand \@@startlink[1]{}%
\providecommand \@@endlink[0]{}%
\providecommand \url  [0]{\begingroup\@sanitize@url \@url }%
\providecommand \@url [1]{\endgroup\@href {#1}{\urlprefix }}%
\providecommand \urlprefix  [0]{URL }%
\providecommand \Eprint [0]{\href }%
\@ifxundefined \urlstyle {%
  \providecommand \doi  [0]{\begingroup \@sanitize@url \@doi}%
  \providecommand \@doi [1]{\endgroup \@@startlink {\doibase
  #1}doi:\discretionary {}{}{}#1\@@endlink }%
}{%
  \providecommand \doi  [0]{doi:\discretionary{}{}{}\begingroup
  \urlstyle{rm}\Url }%
}%
\providecommand \doibase [0]{http://dx.doi.org/}%
\providecommand \Doi [0]{\begingroup \@sanitize@url \@Doi }%
\providecommand \@Doi  [1]{\endgroup\@@startlink{\doibase#1}\@@Doi}%
\providecommand \@@Doi [1]{#1\@@endlink}%
\providecommand \selectlanguage [0]{\@gobble}%
\providecommand \bibinfo  [0]{\@secondoftwo}%
\providecommand \bibfield  [0]{\@secondoftwo}%
\providecommand \translation [1]{[#1]}%
\providecommand \BibitemOpen [0]{}%
\providecommand \bibitemStop [0]{}%
\providecommand \bibitemNoStop [0]{.\EOS\space}%
\providecommand \EOS [0]{\spacefactor3000\relax}%
\providecommand \BibitemShut  [1]{\csname bibitem#1\endcsname}%
\bibitem [{\citenamefont {Abbott}\ \emph
  {et~al.}(2009){\natexlab{a}}\citenamefont {Abbott} \emph
  {et~al.}}]{Abbott:2009li}%
  \BibitemOpen
  \bibfield  {author} {\bibinfo {author} {\bibfnamefont {B.}~\bibnamefont
  {Abbott}} \emph {et~al.} (\bibinfo {collaboration} {LIGO Scientific
  Collaboration}),\ }\Doi {10.1088/0034-4885/72/7/076901} {\bibfield  {journal}
  {\bibinfo  {journal} {Reports on Progress in Physics},\ }\textbf {\bibinfo
  {volume} {72}},\ \bibinfo {pages} {076901} (\bibinfo {year}
  {2009}{\natexlab{a}})},\ \Eprint {http://arxiv.org/abs/0711.3041}
  {arXiv:0711.3041 [gr-qc]} \BibitemShut {NoStop}%
\bibitem [{\citenamefont {Acernese}\ \emph {et~al.}(2006)\citenamefont
  {Acernese} \emph {et~al.}}]{Acernese:2006bj}%
  \BibitemOpen
  \bibfield  {author} {\bibinfo {author} {\bibfnamefont {F.}~\bibnamefont
  {Acernese}} \emph {et~al.} (\bibinfo {collaboration} {Virgo Scientific
  Collaboration}),\ }\href@noop {} {\bibfield  {journal} {\bibinfo  {journal}
  {Class. Quant. Grav.},\ }\textbf {\bibinfo {volume} {23}},\ \bibinfo {pages}
  {S635} (\bibinfo {year} {2006})}\BibitemShut {NoStop}%
\bibitem [{\citenamefont {Willke}(2007)}]{Willke:2007zz}%
  \BibitemOpen
  \bibfield  {author} {\bibinfo {author} {\bibfnamefont {B.}~\bibnamefont
  {Willke}} (\bibinfo {collaboration} {LIGO Scientific Collaboration}),\ }\Doi
  {10.1088/0264-9381/24/19/S02} {\bibfield  {journal} {\bibinfo  {journal}
  {Class. Quant. Grav.},\ }\textbf {\bibinfo {volume} {24}},\ \bibinfo {pages}
  {S389} (\bibinfo {year} {2007})}\BibitemShut {NoStop}%
\bibitem [{\citenamefont {Abadie}\ \emph
  {et~al.}(2010){\natexlab{a}}\citenamefont {Abadie} \emph
  {et~al.}}]{Abadie:2010yba}%
  \BibitemOpen
  \bibfield  {author} {\bibinfo {author} {\bibfnamefont {J.}~\bibnamefont
  {Abadie}} \emph {et~al.} (\bibinfo {collaboration} {LIGO and Virgo Scientific
  Collaborations}),\ }\Doi {10.1103/PhysRevD.82.102001} {\bibfield  {journal}
  {\bibinfo  {journal} {Phys. Rev.},\ }\textbf {\bibinfo {volume} {D82}},\
  \bibinfo {pages} {102001} (\bibinfo {year} {2010}{\natexlab{a}})},\ \Eprint
  {http://arxiv.org/abs/1005.4655} {arXiv:1005.4655 [gr-qc]} \BibitemShut
  {NoStop}%
\bibitem [{\citenamefont {Abbott}\ \emph
  {et~al.}(2009){\natexlab{b}}\citenamefont {Abbott} \emph
  {et~al.}}]{Abbott:2009ws}%
  \BibitemOpen
  \bibfield  {author} {\bibinfo {author} {\bibfnamefont {B.~P.}\ \bibnamefont
  {Abbott}} \emph {et~al.} (\bibinfo {collaboration} {LIGO and Virgo Scientific
  Collaborations}),\ }\Doi {10.1038/nature08278} {\bibfield  {journal}
  {\bibinfo  {journal} {Nature},\ }\textbf {\bibinfo {volume} {460}},\ \bibinfo
  {pages} {990} (\bibinfo {year} {2009}{\natexlab{b}})},\ \Eprint
  {http://arxiv.org/abs/0910.5772} {arXiv:0910.5772 [astro-ph.CO]} \BibitemShut
  {NoStop}%
\bibitem [{\citenamefont {Abadie}\ \emph
  {et~al.}(2010){\natexlab{b}}\citenamefont {Abadie} \emph
  {et~al.}}]{Abadie:2010mt}%
  \BibitemOpen
  \bibfield  {author} {\bibinfo {author} {\bibfnamefont {J.}~\bibnamefont
  {Abadie}} \emph {et~al.} (\bibinfo {collaboration} {LIGO and Virgo Scientific
  Collaborations}),\ }\Doi {10.1103/PhysRevD.81.102001} {\bibfield  {journal}
  {\bibinfo  {journal} {Phys. Rev.},\ }\textbf {\bibinfo {volume} {D81}},\
  \bibinfo {pages} {102001} (\bibinfo {year} {2010}{\natexlab{b}})},\ \Eprint
  {http://arxiv.org/abs/1002.1036} {arXiv:1002.1036 [gr-qc]} \BibitemShut
  {NoStop}%
\bibitem [{\citenamefont {Abott}\ \emph {et~al.}(2010)\citenamefont {Abott}
  \emph {et~al.}}]{Collaboration:2009rfa}%
  \BibitemOpen
  \bibfield  {author} {\bibinfo {author} {\bibfnamefont {B.}~\bibnamefont
  {Abott}} \emph {et~al.} (\bibinfo {collaboration} {LIGO and Virgo Scientific
  Collaborations}),\ }\Doi {10.1088/0004-637X/713/1/671} {\bibfield  {journal}
  {\bibinfo  {journal} {Astrophys. J.},\ }\textbf {\bibinfo {volume} {713}},\
  \bibinfo {pages} {671} (\bibinfo {year} {2010})},\ \Eprint
  {http://arxiv.org/abs/0909.3583} {arXiv:0909.3583 [astro-ph.HE]} \BibitemShut
  {NoStop}%
\bibitem [{\citenamefont {Harry}(2010)}]{Harry:2010zz}%
  \BibitemOpen
  \bibfield  {author} {\bibinfo {author} {\bibfnamefont {G.~M.}\ \bibnamefont
  {Harry}} (\bibinfo {collaboration} {LIGO Scientific}),\ }\Doi
  {10.1088/0264-9381/27/8/084006} {\bibfield  {journal} {\bibinfo  {journal}
  {Class. Quant. Grav.},\ }\textbf {\bibinfo {volume} {27}},\ \bibinfo {pages}
  {084006} (\bibinfo {year} {2010})}\BibitemShut {NoStop}%
\bibitem [{avl()}]{avlligowebsite}%
  \BibitemOpen
  \href@noop {} {\enquote {\bibinfo {title} {{Advanced LIGO}},}\ }\bibinfo
  {howpublished} {\url{http://www.ligo.caltech.edu/advLIGO/}}\BibitemShut
  {NoStop}%
\bibitem [{adv()}]{advvirgowebsite}%
  \BibitemOpen
  \href@noop {} {\enquote {\bibinfo {title} {{Advanced Virgo}},}\ }\bibinfo
  {howpublished} {\url{http://wwwcascina.virgo.infn.it/advirgo/}}\BibitemShut
  {NoStop}%
\bibitem [{\citenamefont {Abadie}\ \emph
  {et~al.}(2010){\natexlab{c}}\citenamefont {Abadie} \emph
  {et~al.}}]{Temp:2010cfa}%
  \BibitemOpen
  \bibfield  {author} {\bibinfo {author} {\bibfnamefont {J.}~\bibnamefont
  {Abadie}} \emph {et~al.} (\bibinfo {collaboration} {LIGO and Virgo Scientific
  Collaborations}),\ }\Doi {10.1088/0264-9381/27/17/173001} {\bibfield
  {journal} {\bibinfo  {journal} {Class. Quant. Grav.},\ }\textbf {\bibinfo
  {volume} {27}},\ \bibinfo {pages} {173001} (\bibinfo {year}
  {2010}{\natexlab{c}})},\ \Eprint {http://arxiv.org/abs/1003.2480}
  {arXiv:1003.2480 [astro-ph.HE]} \BibitemShut {NoStop}%
\bibitem [{lcg()}]{lcgtwebsite}%
  \BibitemOpen
  \href@noop {} {\enquote {\bibinfo {title} {{LCGT: Large Scale Cryogenic
  Gravitational Wave Telescope}},}\ }\bibinfo {howpublished}
  {\url{http://gw.icrr.u-tokyo.ac.jp/lcgt/}}\BibitemShut {NoStop}%
\bibitem [{aig()}]{aigowebsite}%
  \BibitemOpen
  \href@noop {} {\enquote {\bibinfo {title} {{AIGO: Australian International
  Gravitational Observatory}},}\ }\bibinfo {howpublished}
  {\url{http://www.aigo.org.au/}}\BibitemShut {NoStop}%
\bibitem [{\citenamefont {Hild}\ \emph {et~al.}(2008)\citenamefont {Hild},
  \citenamefont {Chelkowski},\ and\ \citenamefont {Freise}}]{Hild:2008ng}%
  \BibitemOpen
  \bibfield  {author} {\bibinfo {author} {\bibfnamefont {S.}~\bibnamefont
  {Hild}}, \bibinfo {author} {\bibfnamefont {S.}~\bibnamefont {Chelkowski}}, \
  and\ \bibinfo {author} {\bibfnamefont {A.}~\bibnamefont {Freise}},\
  }\href@noop {} { (\bibinfo {year} {2008})},\ \Eprint
  {http://arxiv.org/abs/0810.0604} {arXiv:0810.0604 [gr-qc]} \BibitemShut
  {NoStop}%
\bibitem [{\citenamefont {Abbott}\ \emph {et~al.}(2008)\citenamefont {Abbott}
  \emph {et~al.}}]{Abbott:2007rh}%
  \BibitemOpen
  \bibfield  {author} {\bibinfo {author} {\bibfnamefont {B.}~\bibnamefont
  {Abbott}} \emph {et~al.} (\bibinfo {collaboration} {LIGO Scientific
  Collaboration}),\ }\Doi {10.1086/587954} {\bibfield  {journal} {\bibinfo
  {journal} {Astrophys. J.},\ }\textbf {\bibinfo {volume} {681}},\ \bibinfo
  {pages} {1419} (\bibinfo {year} {2008})},\ \Eprint
  {http://arxiv.org/abs/0711.1163} {arXiv:0711.1163 [astro-ph]} \BibitemShut
  {NoStop}%
\bibitem [{\citenamefont {Abadie}\ \emph
  {et~al.}(2010){\natexlab{d}}\citenamefont {Abadie} \emph
  {et~al.}}]{Abadie:2010uf}%
  \BibitemOpen
  \bibfield  {author} {\bibinfo {author} {\bibfnamefont {J.}~\bibnamefont
  {Abadie}} \emph {et~al.} (\bibinfo {collaboration} {LIGO and Virgo Scientific
  Collaborations}),\ }\Doi {10.1088/0004-637X/715/2/1453} {\bibfield  {journal}
  {\bibinfo  {journal} {Astrophys. J.},\ }\textbf {\bibinfo {volume} {715}},\
  \bibinfo {pages} {1453} (\bibinfo {year} {2010}{\natexlab{d}})},\ \Eprint
  {http://arxiv.org/abs/1001.0165} {arXiv:1001.0165 [astro-ph.HE]} \BibitemShut
  {NoStop}%
\bibitem [{\citenamefont {Abbott}\ \emph {et~al.}(2010)\citenamefont {Abbott}
  \emph {et~al.}}]{Collaboration:2009kk}%
  \BibitemOpen
  \bibfield  {author} {\bibinfo {author} {\bibfnamefont {B.~P.}\ \bibnamefont
  {Abbott}} \emph {et~al.} (\bibinfo {collaboration} {LIGO and Virgo Scientific
  Collaborations}),\ }\Doi {10.1088/0004-637X/715/2/1438} {\bibfield  {journal}
  {\bibinfo  {journal} {Astrophys. J.},\ }\textbf {\bibinfo {volume} {715}},\
  \bibinfo {pages} {1438} (\bibinfo {year} {2010})},\ \Eprint
  {http://arxiv.org/abs/0908.3824} {arXiv:0908.3824 [astro-ph.HE]} \BibitemShut
  {NoStop}%
\bibitem [{\citenamefont {Kanner}\ \emph {et~al.}(2008)\citenamefont {Kanner}
  \emph {et~al.}}]{Kanner:2008zh}%
  \BibitemOpen
  \bibfield  {author} {\bibinfo {author} {\bibfnamefont {J.}~\bibnamefont
  {Kanner}} \emph {et~al.},\ }\Doi {10.1088/0264-9381/25/18/184034} {\bibfield
  {journal} {\bibinfo  {journal} {Class. Quant. Grav.},\ }\textbf {\bibinfo
  {volume} {25}},\ \bibinfo {pages} {184034} (\bibinfo {year} {2008})},\
  \Eprint {http://arxiv.org/abs/0803.0312} {arXiv:0803.0312 [astro-ph]}
  \BibitemShut {NoStop}%
\bibitem [{\citenamefont {Nakar}(2007)}]{nakar:2007}%
  \BibitemOpen
  \bibfield  {author} {\bibinfo {author} {\bibfnamefont {E.}~\bibnamefont
  {Nakar}},\ }\Doi {10.1016/j.physrep.2007.02.005} {\bibfield  {journal}
  {\bibinfo  {journal} {Phys. Rept.},\ }\textbf {\bibinfo {volume} {442}},\
  \bibinfo {pages} {166} (\bibinfo {year} {2007})},\ \Eprint
  {http://arxiv.org/abs/astro-ph/0701748} {arXiv:astro-ph/0701748} \BibitemShut
  {NoStop}%
\bibitem [{\citenamefont {Shibata}\ and\ \citenamefont
  {Taniguchi}(2008)}]{Shibata:2007zm}%
  \BibitemOpen
  \bibfield  {author} {\bibinfo {author} {\bibfnamefont {M.}~\bibnamefont
  {Shibata}}\ and\ \bibinfo {author} {\bibfnamefont {K.}~\bibnamefont
  {Taniguchi}},\ }\Doi {10.1103/PhysRevD.77.084015} {\bibfield  {journal}
  {\bibinfo  {journal} {Phys. Rev.},\ }\textbf {\bibinfo {volume} {D77}},\
  \bibinfo {pages} {084015} (\bibinfo {year} {2008})},\ \Eprint
  {http://arxiv.org/abs/0711.1410} {arXiv:0711.1410 [gr-qc]} \BibitemShut
  {NoStop}%
\bibitem [{\citenamefont {Metzger}\ \emph {et~al.}(2010)\citenamefont {Metzger}
  \emph {et~al.}}]{Metzger:2010sy}%
  \BibitemOpen
  \bibfield  {author} {\bibinfo {author} {\bibfnamefont {B.~D.}\ \bibnamefont
  {Metzger}} \emph {et~al.},\ }\href@noop {} { (\bibinfo {year} {2010})},\
  \Eprint {http://arxiv.org/abs/1001.5029} {arXiv:1001.5029 [astro-ph.HE]}
  \BibitemShut {NoStop}%
\bibitem [{\citenamefont {Predoi}\ \emph {et~al.}(2010)\citenamefont {Predoi}
  \emph {et~al.}}]{Predoi:2009af}%
  \BibitemOpen
  \bibfield  {author} {\bibinfo {author} {\bibfnamefont {V.}~\bibnamefont
  {Predoi}} \emph {et~al.},\ }\Doi {10.1088/0264-9381/27/8/084018} {\bibfield
  {journal} {\bibinfo  {journal} {Class. Quant. Grav.},\ }\textbf {\bibinfo
  {volume} {27}},\ \bibinfo {pages} {084018} (\bibinfo {year} {2010})},\
  \Eprint {http://arxiv.org/abs/0912.0476} {arXiv:0912.0476 [gr-qc]}
  \BibitemShut {NoStop}%
\bibitem [{\citenamefont {Blackburn}\ \emph {et~al.}(2008)\citenamefont
  {Blackburn} \emph {et~al.}}]{Blackburn:2008ah}%
  \BibitemOpen
  \bibfield  {author} {\bibinfo {author} {\bibfnamefont {L.}~\bibnamefont
  {Blackburn}} \emph {et~al.},\ }\Doi {10.1088/0264-9381/25/18/184004}
  {\bibfield  {journal} {\bibinfo  {journal} {Class. Quant. Grav.},\ }\textbf
  {\bibinfo {volume} {25}},\ \bibinfo {pages} {184004} (\bibinfo {year}
  {2008})},\ \Eprint {http://arxiv.org/abs/0804.0800} {arXiv:0804.0800 [gr-qc]}
  \BibitemShut {NoStop}%
\bibitem [{\citenamefont {Allen}(2005)}]{Allen:2004gu}%
  \BibitemOpen
  \bibfield  {author} {\bibinfo {author} {\bibfnamefont {B.}~\bibnamefont
  {Allen}},\ }\Doi {10.1103/PhysRevD.71.062001} {\bibfield  {journal} {\bibinfo
   {journal} {Phys. Rev.},\ }\textbf {\bibinfo {volume} {D71}},\ \bibinfo
  {pages} {062001} (\bibinfo {year} {2005})},\ \Eprint
  {http://arxiv.org/abs/gr-qc/0405045} {arXiv:gr-qc/0405045} \BibitemShut
  {NoStop}%
\bibitem [{\citenamefont {Hanna}(2008)}]{Hanna:2008}%
  \BibitemOpen
  \bibfield  {author} {\bibinfo {author} {\bibfnamefont {C.}~\bibnamefont
  {Hanna}},\ }\emph {\bibinfo {title} {Searching for gravitational waves from
  binary systems in non-stationary data}},\ \href
  {http://etd.lsu.edu/docs/available/etd-03272008-092832/} {Ph.D. thesis},\
  \bibinfo  {school} {Louisiana State University} (\bibinfo {year}
  {2008})\BibitemShut {NoStop}%
\bibitem [{\citenamefont {Guersel}\ and\ \citenamefont
  {Tinto}(1989)}]{Guersel:1989th}%
  \BibitemOpen
  \bibfield  {author} {\bibinfo {author} {\bibfnamefont {Y.}~\bibnamefont
  {Guersel}}\ and\ \bibinfo {author} {\bibfnamefont {M.}~\bibnamefont
  {Tinto}},\ }\Doi {10.1103/PhysRevD.40.3884} {\bibfield  {journal} {\bibinfo
  {journal} {Phys. Rev.},\ }\textbf {\bibinfo {volume} {D40}},\ \bibinfo
  {pages} {3884} (\bibinfo {year} {1989})}\BibitemShut {NoStop}%
\bibitem [{\citenamefont {Abbott}\ \emph
  {et~al.}(2009){\natexlab{c}}\citenamefont {Abbott} \emph
  {et~al.}}]{Abbott:2009qj}%
  \BibitemOpen
  \bibfield  {author} {\bibinfo {author} {\bibfnamefont {B.~P.}\ \bibnamefont
  {Abbott}} \emph {et~al.} (\bibinfo {collaboration} {LIGO Scientific
  Collaboration}),\ }\Doi {10.1103/PhysRevD.80.047101} {\bibfield  {journal}
  {\bibinfo  {journal} {Phys. Rev.},\ }\textbf {\bibinfo {volume} {D80}},\
  \bibinfo {pages} {047101} (\bibinfo {year} {2009}{\natexlab{c}})},\ \Eprint
  {http://arxiv.org/abs/0905.3710} {arXiv:0905.3710 [gr-qc]} \BibitemShut
  {NoStop}%
\bibitem [{\citenamefont {Abbott}\ \emph
  {et~al.}(2009){\natexlab{d}}\citenamefont {Abbott} \emph
  {et~al.}}]{Abbott:2009tt}%
  \BibitemOpen
  \bibfield  {author} {\bibinfo {author} {\bibfnamefont {B.~P.}\ \bibnamefont
  {Abbott}} \emph {et~al.} (\bibinfo {collaboration} {LIGO Scientific
  Collaboration}),\ }\Doi {10.1103/PhysRevD.79.122001} {\bibfield  {journal}
  {\bibinfo  {journal} {Phys. Rev.},\ }\textbf {\bibinfo {volume} {D79}},\
  \bibinfo {pages} {122001} (\bibinfo {year} {2009}{\natexlab{d}})},\ \Eprint
  {http://arxiv.org/abs/0901.0302} {arXiv:0901.0302 [gr-qc]} \BibitemShut
  {NoStop}%
\bibitem [{\citenamefont {Pai}\ \emph {et~al.}(2001)\citenamefont {Pai},
  \citenamefont {Dhurandhar},\ and\ \citenamefont {Bose}}]{Pai:2000zt}%
  \BibitemOpen
  \bibfield  {author} {\bibinfo {author} {\bibfnamefont {A.}~\bibnamefont
  {Pai}}, \bibinfo {author} {\bibfnamefont {S.}~\bibnamefont {Dhurandhar}}, \
  and\ \bibinfo {author} {\bibfnamefont {S.}~\bibnamefont {Bose}},\ }\Doi
  {10.1103/PhysRevD.64.042004} {\bibfield  {journal} {\bibinfo  {journal}
  {Phys. Rev.},\ }\textbf {\bibinfo {volume} {D64}},\ \bibinfo {pages} {042004}
  (\bibinfo {year} {2001})},\ \Eprint {http://arxiv.org/abs/gr-qc/0009078}
  {arXiv:gr-qc/0009078} \BibitemShut {NoStop}%
\bibitem [{\citenamefont {Bose}\ \emph {et~al.}(2000)\citenamefont {Bose},
  \citenamefont {Pai},\ and\ \citenamefont {Dhurandhar}}]{Bose:1999pj}%
  \BibitemOpen
  \bibfield  {author} {\bibinfo {author} {\bibfnamefont {S.}~\bibnamefont
  {Bose}}, \bibinfo {author} {\bibfnamefont {A.}~\bibnamefont {Pai}}, \ and\
  \bibinfo {author} {\bibfnamefont {S.~V.}\ \bibnamefont {Dhurandhar}},\ }\Doi
  {10.1142/S0218271800000360} {\bibfield  {journal} {\bibinfo  {journal} {Int.
  J. Mod. Phys.},\ }\textbf {\bibinfo {volume} {D9}},\ \bibinfo {pages} {325}
  (\bibinfo {year} {2000})},\ \Eprint {http://arxiv.org/abs/gr-qc/0002010}
  {arXiv:gr-qc/0002010} \BibitemShut {NoStop}%
\bibitem [{\citenamefont {Bose}\ \emph {et~al.}(1999)\citenamefont {Bose},
  \citenamefont {Dhurandhar},\ and\ \citenamefont {Pai}}]{Bose:1999bp}%
  \BibitemOpen
  \bibfield  {author} {\bibinfo {author} {\bibfnamefont {S.}~\bibnamefont
  {Bose}}, \bibinfo {author} {\bibfnamefont {S.~V.}\ \bibnamefont
  {Dhurandhar}}, \ and\ \bibinfo {author} {\bibfnamefont {A.}~\bibnamefont
  {Pai}},\ }\Doi {10.1007/s12043-999=} {\bibfield  {journal} {\bibinfo
  {journal} {Pramana},\ }\textbf {\bibinfo {volume} {53}},\ \bibinfo {pages}
  {1125} (\bibinfo {year} {1999})},\ \Eprint
  {http://arxiv.org/abs/gr-qc/9906064} {arXiv:gr-qc/9906064} \BibitemShut
  {NoStop}%
\bibitem [{\citenamefont {Bose}\ \emph {et~al.}(2011)\citenamefont {Bose},
  \citenamefont {Dayanga}, \citenamefont {Ghosh},\ and\ \citenamefont
  {Talukder}}]{Bose:2011}%
  \BibitemOpen
  \bibfield  {author} {\bibinfo {author} {\bibfnamefont {S.}~\bibnamefont
  {Bose}}, \bibinfo {author} {\bibfnamefont {T.}~\bibnamefont {Dayanga}},
  \bibinfo {author} {\bibfnamefont {S.}~\bibnamefont {Ghosh}}, \ and\ \bibinfo
  {author} {\bibfnamefont {D.}~\bibnamefont {Talukder}},\ }\href@noop {}
  {\bibfield  {journal} {\bibinfo  {journal} {In preparation}} (\bibinfo {year}
  {2011})}\BibitemShut {NoStop}%
\bibitem [{\citenamefont {Jaranowski}\ \emph {et~al.}(1998)\citenamefont
  {Jaranowski}, \citenamefont {Krolak},\ and\ \citenamefont
  {Schutz}}]{Jaranowski:1998qm}%
  \BibitemOpen
  \bibfield  {author} {\bibinfo {author} {\bibfnamefont {P.}~\bibnamefont
  {Jaranowski}}, \bibinfo {author} {\bibfnamefont {A.}~\bibnamefont {Krolak}},
  \ and\ \bibinfo {author} {\bibfnamefont {B.~F.}\ \bibnamefont {Schutz}},\
  }\Doi {10.1103/PhysRevD.58.063001} {\bibfield  {journal} {\bibinfo  {journal}
  {Phys. Rev.},\ }\textbf {\bibinfo {volume} {D58}},\ \bibinfo {pages} {063001}
  (\bibinfo {year} {1998})},\ \Eprint {http://arxiv.org/abs/gr-qc/9804014}
  {arXiv:gr-qc/9804014} \BibitemShut {NoStop}%
\bibitem [{\citenamefont {Abbott}\ \emph
  {et~al.}(2009){\natexlab{e}}\citenamefont {Abbott} \emph
  {et~al.}}]{Abbott:2009nc}%
  \BibitemOpen
  \bibfield  {author} {\bibinfo {author} {\bibfnamefont {B.~P.}\ \bibnamefont
  {Abbott}} \emph {et~al.} (\bibinfo {collaboration} {LIGO Scientific
  Collaboration}),\ }\Doi {10.1103/PhysRevD.80.042003} {\bibfield  {journal}
  {\bibinfo  {journal} {Phys. Rev.},\ }\textbf {\bibinfo {volume} {D80}},\
  \bibinfo {pages} {042003} (\bibinfo {year} {2009}{\natexlab{e}})},\ \Eprint
  {http://arxiv.org/abs/0905.1705} {arXiv:0905.1705 [gr-qc]} \BibitemShut
  {NoStop}%
\bibitem [{\citenamefont {Cutler}\ and\ \citenamefont
  {Schutz}(2005)}]{Cutler:2005hc}%
  \BibitemOpen
  \bibfield  {author} {\bibinfo {author} {\bibfnamefont {C.}~\bibnamefont
  {Cutler}}\ and\ \bibinfo {author} {\bibfnamefont {B.~F.}\ \bibnamefont
  {Schutz}},\ }\Doi {10.1103/PhysRevD.72.063006} {\bibfield  {journal}
  {\bibinfo  {journal} {Phys. Rev.},\ }\textbf {\bibinfo {volume} {D72}},\
  \bibinfo {pages} {063006} (\bibinfo {year} {2005})},\ \Eprint
  {http://arxiv.org/abs/gr-qc/0504011} {arXiv:gr-qc/0504011} \BibitemShut
  {NoStop}%
\bibitem [{\citenamefont {Cornish}\ and\ \citenamefont
  {Porter}(2007)}]{Cornish:2006ms}%
  \BibitemOpen
  \bibfield  {author} {\bibinfo {author} {\bibfnamefont {N.~J.}\ \bibnamefont
  {Cornish}}\ and\ \bibinfo {author} {\bibfnamefont {E.~K.}\ \bibnamefont
  {Porter}},\ }\Doi {10.1088/0264-9381/24/23/001} {\bibfield  {journal}
  {\bibinfo  {journal} {Class. Quant. Grav.},\ }\textbf {\bibinfo {volume}
  {24}},\ \bibinfo {pages} {5729} (\bibinfo {year} {2007})},\ \Eprint
  {http://arxiv.org/abs/gr-qc/0612091} {arXiv:gr-qc/0612091} \BibitemShut
  {NoStop}%
\bibitem [{\citenamefont {Owen}\ and\ \citenamefont
  {Sathyaprakash}(1999)}]{OwenSathyaprakash98}%
  \BibitemOpen
  \bibfield  {author} {\bibinfo {author} {\bibfnamefont {B.~J.}\ \bibnamefont
  {Owen}}\ and\ \bibinfo {author} {\bibfnamefont {B.~S.}\ \bibnamefont
  {Sathyaprakash}},\ }\Doi {10.1103/PhysRevD.60.022002} {\bibfield  {journal}
  {\bibinfo  {journal} {Phys. Rev.},\ }\textbf {\bibinfo {volume} {D60}},\
  \bibinfo {pages} {022002} (\bibinfo {year} {1999})},\ \Eprint
  {http://arxiv.org/abs/gr-qc/9808076} {arXiv:gr-qc/9808076} \BibitemShut
  {NoStop}%
\bibitem [{\citenamefont {Livas}(1987)}]{livas87a}%
  \BibitemOpen
  \bibfield  {author} {\bibinfo {author} {\bibfnamefont {J.~C.}\ \bibnamefont
  {Livas}},\ }\emph {\bibinfo {title} {Upper limits for gravitational radiation
  from some astrophysical sources}},\ \href@noop {} {Ph.D. thesis},\ \bibinfo
  {school} {Massachussetts Institute of Technology} (\bibinfo {year}
  {1987})\BibitemShut {NoStop}%
\bibitem [{\citenamefont {Wainstein}\ and\ \citenamefont
  {Zubakov}(1962)}]{Wainstein}%
  \BibitemOpen
  \bibfield  {author} {\bibinfo {author} {\bibfnamefont {L.~A.}\ \bibnamefont
  {Wainstein}}\ and\ \bibinfo {author} {\bibfnamefont {V.~D.}\ \bibnamefont
  {Zubakov}},\ }\href@noop {} {\emph {\bibinfo {title} {Extraction of Signals
  from Noise}}}\ (\bibinfo  {publisher} {Prentice-Hall},\ \bibinfo {address}
  {Englewood Cliffs},\ \bibinfo {year} {1962})\BibitemShut {NoStop}%
\bibitem [{\citenamefont {Prix}(2007)}]{Prix:2007rb}%
  \BibitemOpen
  \bibfield  {author} {\bibinfo {author} {\bibfnamefont {R.}~\bibnamefont
  {Prix}},\ }\Doi {10.1088/0264-9381/24/19/S11} {\bibfield  {journal} {\bibinfo
   {journal} {Classical and Quantum Gravity},\ }\textbf {\bibinfo {volume}
  {24}},\ \bibinfo {pages} {S481} (\bibinfo {year} {2007})},\ \Eprint
  {http://arxiv.org/abs/0707.0428} {arXiv:0707.0428 [gr-qc]} \BibitemShut
  {NoStop}%
\bibitem [{\citenamefont {Damour}\ \emph {et~al.}(2001)\citenamefont {Damour},
  \citenamefont {Iyer},\ and\ \citenamefont {Sathyaprakash}}]{Damour:2000zb}%
  \BibitemOpen
  \bibfield  {author} {\bibinfo {author} {\bibfnamefont {T.}~\bibnamefont
  {Damour}}, \bibinfo {author} {\bibfnamefont {B.~R.}\ \bibnamefont {Iyer}}, \
  and\ \bibinfo {author} {\bibfnamefont {B.~S.}\ \bibnamefont
  {Sathyaprakash}},\ }\Doi {10.1103/PhysRevD.63.044023} {\bibfield  {journal}
  {\bibinfo  {journal} {Phys. Rev.},\ }\textbf {\bibinfo {volume} {D63}},\
  \bibinfo {pages} {044023} (\bibinfo {year} {2001})},\ \Eprint
  {http://arxiv.org/abs/gr-qc/0010009} {arXiv:gr-qc/0010009} \BibitemShut
  {NoStop}%
\bibitem [{\citenamefont {Klimenko}\ \emph {et~al.}(2005)\citenamefont
  {Klimenko}, \citenamefont {Mohanty}, \citenamefont {Rakhmanov},\ and\
  \citenamefont {Mitselmakher}}]{KlMoRaMi:05}%
  \BibitemOpen
  \bibfield  {author} {\bibinfo {author} {\bibfnamefont {S.}~\bibnamefont
  {Klimenko}}, \bibinfo {author} {\bibfnamefont {S.}~\bibnamefont {Mohanty}},
  \bibinfo {author} {\bibfnamefont {M.}~\bibnamefont {Rakhmanov}}, \ and\
  \bibinfo {author} {\bibfnamefont {G.}~\bibnamefont {Mitselmakher}},\ }\Doi
  {10.1103/PhysRevD.72.122002} {\bibfield  {journal} {\bibinfo  {journal}
  {Phys. Rev.},\ }\textbf {\bibinfo {volume} {D72}},\ \bibinfo {pages} {122002}
  (\bibinfo {year} {2005})},\ \Eprint {http://arxiv.org/abs/gr-qc/0508068}
  {arXiv:gr-qc/0508068} \BibitemShut {NoStop}%
\bibitem [{\citenamefont {Klimenko}\ \emph {et~al.}(2006)\citenamefont
  {Klimenko}, \citenamefont {Mohanty}, \citenamefont {Rakhmanov},\ and\
  \citenamefont {Mitselmakher}}]{KlMoRaMi:06}%
  \BibitemOpen
  \bibfield  {author} {\bibinfo {author} {\bibfnamefont {S.}~\bibnamefont
  {Klimenko}}, \bibinfo {author} {\bibfnamefont {S.}~\bibnamefont {Mohanty}},
  \bibinfo {author} {\bibfnamefont {M.}~\bibnamefont {Rakhmanov}}, \ and\
  \bibinfo {author} {\bibfnamefont {G.}~\bibnamefont {Mitselmakher}},\ }\Doi
  {10.1088/1742-6596/32/1/003} {\bibfield  {journal} {\bibinfo  {journal} {J.
  Phys. Conf. Ser.},\ }\textbf {\bibinfo {volume} {32}},\ \bibinfo {pages} {12}
  (\bibinfo {year} {2006})}\BibitemShut {NoStop}%
\bibitem [{\citenamefont {Sutton}\ \emph {et~al.}(2010)\citenamefont {Sutton}
  \emph {et~al.}}]{1367-2630-12-5-053034}%
  \BibitemOpen
  \bibfield  {author} {\bibinfo {author} {\bibfnamefont {P.~J.}\ \bibnamefont
  {Sutton}} \emph {et~al.},\ }\Doi {10.1088/1367-2630/12/5/053034} {\bibfield
  {journal} {\bibinfo  {journal} {New J. Phys.},\ }\textbf {\bibinfo {volume}
  {12}},\ \bibinfo {pages} {053034} (\bibinfo {year} {2010})},\ \Eprint
  {http://arxiv.org/abs/0908.3665} {arXiv:0908.3665 [gr-qc]} \BibitemShut
  {NoStop}%
\bibitem [{\citenamefont {Prix}\ and\ \citenamefont
  {Krishnan}(2009)}]{Prix:2009tq}%
  \BibitemOpen
  \bibfield  {author} {\bibinfo {author} {\bibfnamefont {R.}~\bibnamefont
  {Prix}}\ and\ \bibinfo {author} {\bibfnamefont {B.}~\bibnamefont
  {Krishnan}},\ }\Doi {10.1088/0264-9381/26/20/204013} {\bibfield  {journal}
  {\bibinfo  {journal} {Class. Quant. Grav.},\ }\textbf {\bibinfo {volume}
  {26}},\ \bibinfo {pages} {204013} (\bibinfo {year} {2009})},\ \Eprint
  {http://arxiv.org/abs/0907.2569} {arXiv:0907.2569 [gr-qc]} \BibitemShut
  {NoStop}%
\bibitem [{\citenamefont {Robinson}\ \emph {et~al.}(2008)\citenamefont
  {Robinson}, \citenamefont {Sathyaprakash},\ and\ \citenamefont
  {Sengupta}}]{Robinson:2008}%
  \BibitemOpen
  \bibfield  {author} {\bibinfo {author} {\bibfnamefont {C.~A.~K.}\
  \bibnamefont {Robinson}}, \bibinfo {author} {\bibfnamefont {B.~S.}\
  \bibnamefont {Sathyaprakash}}, \ and\ \bibinfo {author} {\bibfnamefont
  {A.~S.}\ \bibnamefont {Sengupta}},\ }\Doi {10.1103/PhysRevD.78.062002}
  {\bibfield  {journal} {\bibinfo  {journal} {Phys. Rev.},\ }\textbf {\bibinfo
  {volume} {D78}},\ \bibinfo {pages} {062002} (\bibinfo {year} {2008})},\
  \Eprint {http://arxiv.org/abs/0804.4816} {arXiv:0804.4816 [gr-qc]}
  \BibitemShut {NoStop}%
\bibitem [{\citenamefont {Abbott}\ \emph {et~al.}(2005)\citenamefont {Abbott}
  \emph {et~al.}}]{Abbott:2005pe}%
  \BibitemOpen
  \bibfield  {author} {\bibinfo {author} {\bibfnamefont {B.}~\bibnamefont
  {Abbott}} \emph {et~al.} (\bibinfo {collaboration} {LIGO Scientific
  Collaboration}),\ }\Doi {10.1103/PhysRevD.72.082001} {\bibfield  {journal}
  {\bibinfo  {journal} {Phys. Rev.},\ }\textbf {\bibinfo {volume} {D72}},\
  \bibinfo {pages} {082001} (\bibinfo {year} {2005})},\ \Eprint
  {http://arxiv.org/abs/gr-qc/0505041} {arXiv:gr-qc/0505041} \BibitemShut
  {NoStop}%
\bibitem [{\citenamefont {Blanchet}(2002)}]{Bliving}%
  \BibitemOpen
  \bibfield  {author} {\bibinfo {author} {\bibfnamefont {L.}~\bibnamefont
  {Blanchet}},\ }\href@noop {} {\bibfield  {journal} {\bibinfo  {journal}
  {Living Rev. Rel.},\ }\textbf {\bibinfo {volume} {5}},\ \bibinfo {pages} {3}
  (\bibinfo {year} {2002})},\ \Eprint {http://arxiv.org/abs/gr-qc/0202016}
  {arXiv:gr-qc/0202016} \BibitemShut {NoStop}%
\bibitem [{\citenamefont {Hannam}\ \emph {et~al.}(2009)\citenamefont {Hannam}
  \emph {et~al.}}]{Hannam:2009hh}%
  \BibitemOpen
  \bibfield  {author} {\bibinfo {author} {\bibfnamefont {M.}~\bibnamefont
  {Hannam}} \emph {et~al.},\ }\Doi {10.1103/PhysRevD.79.084025} {\bibfield
  {journal} {\bibinfo  {journal} {Phys. Rev.},\ }\textbf {\bibinfo {volume}
  {D79}},\ \bibinfo {pages} {084025} (\bibinfo {year} {2009})},\ \Eprint
  {http://arxiv.org/abs/0901.2437} {arXiv:0901.2437 [gr-qc]} \BibitemShut
  {NoStop}%
\bibitem [{\citenamefont {Abadie}\ \emph
  {et~al.}(2010){\natexlab{e}}\citenamefont {Abadie} \emph
  {et~al.}}]{Abadie:2010px}%
  \BibitemOpen
  \bibfield  {author} {\bibinfo {author} {\bibfnamefont {J.}~\bibnamefont
  {Abadie}} \emph {et~al.} (\bibinfo {collaboration} {LIGO Scientific
  Collaboration}),\ }\Doi {10.1016/j.nima.2010.07.089} {\bibfield  {journal}
  {\bibinfo  {journal} {Nucl. Instrum. Meth.},\ }\textbf {\bibinfo {volume}
  {A624}},\ \bibinfo {pages} {223} (\bibinfo {year} {2010}{\natexlab{e}})},\
  \Eprint {http://arxiv.org/abs/1007.3973} {arXiv:1007.3973 [gr-qc]}
  \BibitemShut {NoStop}%
\bibitem [{\citenamefont {Van Den~Broeck}\ \emph {et~al.}(2009)\citenamefont
  {Van Den~Broeck} \emph {et~al.}}]{Vandenbroeck:2009cv}%
  \BibitemOpen
  \bibfield  {author} {\bibinfo {author} {\bibfnamefont {C.}~\bibnamefont {Van
  Den~Broeck}} \emph {et~al.},\ }\Doi {10.1103/PhysRevD.80.024009} {\bibfield
  {journal} {\bibinfo  {journal} {Phys. Rev.},\ }\textbf {\bibinfo {volume}
  {D80}},\ \bibinfo {pages} {024009} (\bibinfo {year} {2009})},\ \Eprint
  {http://arxiv.org/abs/0904.1715} {arXiv:0904.1715 [gr-qc]} \BibitemShut
  {NoStop}%
\bibitem [{\citenamefont {Fazi}(2009)}]{Fazi:2009}%
  \BibitemOpen
  \bibfield  {author} {\bibinfo {author} {\bibfnamefont {D.}~\bibnamefont
  {Fazi}},\ }\emph {\bibinfo {title} {Development of a physical-template search
  for gravitational waves from spinning compact-object binaries with LIGO}},\
  \href@noop {} {Ph.D. thesis},\ \bibinfo  {school} {Universit\`{a} di Bologna}
  (\bibinfo {year} {2009})\BibitemShut {NoStop}%
\bibitem [{\citenamefont {Abbott}\ \emph {et~al.}(2007)\citenamefont {Abbott}
  \emph {et~al.}}]{LIGOS3S4Tuning}%
  \BibitemOpen
  \bibfield  {author} {\bibinfo {author} {\bibfnamefont {B.}~\bibnamefont
  {Abbott}} \emph {et~al.} (\bibinfo {collaboration} {{LIGO Scientific
  Collaboration}}),\ }\href {http://www.ligo.caltech.edu/docs/T/T070109-01.pdf}
  {\emph {\bibinfo {title} {Tuning matched filter searches for compact binary
  coalescence}}},\ \bibinfo {type} {Tech. Rep.}\ \bibinfo {number}
  {{LIGO}-T070109-01}\ (\bibinfo {year} {2007})\BibitemShut {NoStop}%
\bibitem [{\citenamefont {Itoh}\ \emph {et~al.}(2004)\citenamefont {Itoh},
  \citenamefont {Papa}, \citenamefont {Krishnan},\ and\ \citenamefont
  {Siemens}}]{Itoh:2004zd}%
  \BibitemOpen
  \bibfield  {author} {\bibinfo {author} {\bibfnamefont {Y.}~\bibnamefont
  {Itoh}}, \bibinfo {author} {\bibfnamefont {M.~A.}\ \bibnamefont {Papa}},
  \bibinfo {author} {\bibfnamefont {B.}~\bibnamefont {Krishnan}}, \ and\
  \bibinfo {author} {\bibfnamefont {X.}~\bibnamefont {Siemens}},\ }\Doi
  {10.1088/0264-9381/21/20/009} {\bibfield  {journal} {\bibinfo  {journal}
  {Class. Quant. Grav.},\ }\textbf {\bibinfo {volume} {21}},\ \bibinfo {pages}
  {S1667} (\bibinfo {year} {2004})},\ \Eprint
  {http://arxiv.org/abs/gr-qc/0408092} {arXiv:gr-qc/0408092} \BibitemShut
  {NoStop}%
\bibitem [{lal()}]{lalsuite}%
  \BibitemOpen
  \href@noop {} {\enquote {\bibinfo {title} {L{SC} {A}lgorithm {L}ibrary
  {S}uite},}\ }\bibinfo {howpublished}
  {\url{https://www.lsc-group.phys.uwm.edu/daswg/projects/lalsuite.html}}\BibitemShut
  {NoStop}%
\bibitem [{\citenamefont {Davies}\ \emph {et~al.}(2005)\citenamefont {Davies},
  \citenamefont {Levan},\ and\ \citenamefont {King}}]{DavLevKing04}%
  \BibitemOpen
  \bibfield  {author} {\bibinfo {author} {\bibfnamefont {M.~B.}\ \bibnamefont
  {Davies}}, \bibinfo {author} {\bibfnamefont {A.~J.}\ \bibnamefont {Levan}}, \
  and\ \bibinfo {author} {\bibfnamefont {A.~R.}\ \bibnamefont {King}},\ }\Doi
  {10.1111/j.1365-2966.2004.08423.x/abs/} {\bibfield  {journal} {\bibinfo
  {journal} {Mon. Not. Roy. Astron. Soc.},\ }\textbf {\bibinfo {volume}
  {356}},\ \bibinfo {pages} {54} (\bibinfo {year} {2005})},\ \Eprint
  {http://arxiv.org/abs/astro-ph/0409681} {arXiv:astro-ph/0409681} \BibitemShut
  {NoStop}%
\bibitem [{\citenamefont {Owen}(1996)}]{Owen96}%
  \BibitemOpen
  \bibfield  {author} {\bibinfo {author} {\bibfnamefont {B.~J.}\ \bibnamefont
  {Owen}},\ }\Doi {10.1103/PhysRevD.53.6749} {\bibfield  {journal} {\bibinfo
  {journal} {Phys. Rev.},\ }\textbf {\bibinfo {volume} {D53}},\ \bibinfo
  {pages} {6749} (\bibinfo {year} {1996})},\ \Eprint
  {http://arxiv.org/abs/gr-qc/9511032} {arXiv:gr-qc/9511032} \BibitemShut
  {NoStop}%
\bibitem [{\citenamefont {Babak}\ \emph {et~al.}(2006)\citenamefont {Babak},
  \citenamefont {Balasubramanian}, \citenamefont {Churches}, \citenamefont
  {Cokelaer},\ and\ \citenamefont {Sathyaprakash}}]{Bank06}%
  \BibitemOpen
  \bibfield  {author} {\bibinfo {author} {\bibfnamefont {S.}~\bibnamefont
  {Babak}}, \bibinfo {author} {\bibfnamefont {R.}~\bibnamefont
  {Balasubramanian}}, \bibinfo {author} {\bibfnamefont {D.}~\bibnamefont
  {Churches}}, \bibinfo {author} {\bibfnamefont {T.}~\bibnamefont {Cokelaer}},
  \ and\ \bibinfo {author} {\bibfnamefont {B.~S.}\ \bibnamefont
  {Sathyaprakash}},\ }\Doi {10.1088/0264-9381/23/18/002} {\bibfield  {journal}
  {\bibinfo  {journal} {Class. Quant. Grav.},\ }\textbf {\bibinfo {volume}
  {23}},\ \bibinfo {pages} {5477} (\bibinfo {year} {2006})},\ \Eprint
  {http://arxiv.org/abs/gr-qc/0604037} {arXiv:gr-qc/0604037} \BibitemShut
  {NoStop}%
\bibitem [{\citenamefont {Cokelaer}(2007)}]{Cokelaer:2007kx}%
  \BibitemOpen
  \bibfield  {author} {\bibinfo {author} {\bibfnamefont {T.}~\bibnamefont
  {Cokelaer}},\ }\Doi {10.1103/PhysRevD.76.102004} {\bibfield  {journal}
  {\bibinfo  {journal} {Phys. Rev.},\ }\textbf {\bibinfo {volume} {D76}},\
  \bibinfo {pages} {102004} (\bibinfo {year} {2007})},\ \Eprint
  {http://arxiv.org/abs/0706.4437} {arXiv:0706.4437 [gr-qc]} \BibitemShut
  {NoStop}%
\bibitem [{\citenamefont {Abbott}\ \emph
  {et~al.}(2009){\natexlab{f}}\citenamefont {Abbott} \emph
  {et~al.}}]{Abbott:2009km}%
  \BibitemOpen
  \bibfield  {author} {\bibinfo {author} {\bibfnamefont {B.~P.}\ \bibnamefont
  {Abbott}} \emph {et~al.} (\bibinfo {collaboration} {LIGO Scientific
  Collaboration}),\ }\Doi {10.1103/PhysRevD.80.062001} {\bibfield  {journal}
  {\bibinfo  {journal} {Phys. Rev.},\ }\textbf {\bibinfo {volume} {D80}},\
  \bibinfo {pages} {062001} (\bibinfo {year} {2009}{\natexlab{f}})},\ \Eprint
  {http://arxiv.org/abs/0905.1654} {arXiv:0905.1654 [gr-qc]} \BibitemShut
  {NoStop}%
\bibitem [{\citenamefont {Goggin}(2008)}]{Goggin:2008dz}%
  \BibitemOpen
  \bibfield  {author} {\bibinfo {author} {\bibfnamefont {L.~M.}\ \bibnamefont
  {Goggin}},\ }\href@noop {} { (\bibinfo {year} {2008})},\ \Eprint
  {http://arxiv.org/abs/0908.2085} {arXiv:0908.2085 [gr-qc]} \BibitemShut
  {NoStop}%
\bibitem [{\citenamefont {Allen}\ \emph {et~al.}(2005)\citenamefont {Allen},
  \citenamefont {Anderson}, \citenamefont {Brady}, \citenamefont {Brown},\ and\
  \citenamefont {Creighton}}]{Allen:2005fk}%
  \BibitemOpen
  \bibfield  {author} {\bibinfo {author} {\bibfnamefont {B.}~\bibnamefont
  {Allen}}, \bibinfo {author} {\bibfnamefont {W.~G.}\ \bibnamefont {Anderson}},
  \bibinfo {author} {\bibfnamefont {P.~R.}\ \bibnamefont {Brady}}, \bibinfo
  {author} {\bibfnamefont {D.~A.}\ \bibnamefont {Brown}}, \ and\ \bibinfo
  {author} {\bibfnamefont {J.~D.~E.}\ \bibnamefont {Creighton}},\ }\href@noop
  {} { (\bibinfo {year} {2005})},\ \Eprint {http://arxiv.org/abs/gr-qc/0509116}
  {arXiv:gr-qc/0509116} \BibitemShut {NoStop}%
\bibitem [{\citenamefont {Pan}\ \emph {et~al.}(2004)\citenamefont {Pan},
  \citenamefont {Buonanno}, \citenamefont {Chen},\ and\ \citenamefont
  {Vallisneri}}]{PBCV04}%
  \BibitemOpen
  \bibfield  {author} {\bibinfo {author} {\bibfnamefont {Y.}~\bibnamefont
  {Pan}}, \bibinfo {author} {\bibfnamefont {A.}~\bibnamefont {Buonanno}},
  \bibinfo {author} {\bibfnamefont {Y.}~\bibnamefont {Chen}}, \ and\ \bibinfo
  {author} {\bibfnamefont {M.}~\bibnamefont {Vallisneri}},\ }\Doi
  {10.1103/PhysRevD.69.104017} {\bibfield  {journal} {\bibinfo  {journal}
  {Phys. Rev.},\ }\textbf {\bibinfo {volume} {D69}},\ \bibinfo {pages} {104017}
  (\bibinfo {year} {2004})},\ \bibinfo {note} {erratum-ibid. {\bf D74},
  029905(E) (2006)},\ \Eprint {http://arxiv.org/abs/gr-qc/0310034}
  {arXiv:gr-qc/0310034} \BibitemShut {NoStop}%
\bibitem [{\citenamefont {Abadie}\ \emph
  {et~al.}(2011){\natexlab{a}}\citenamefont {Abadie} \emph
  {et~al.}}]{s6allsky}%
  \BibitemOpen
  \bibfield  {author} {\bibinfo {author} {\bibfnamefont {J.}~\bibnamefont
  {Abadie}} \emph {et~al.},\ }\href@noop {} {\bibfield  {journal} {\bibinfo
  {journal} {In preparation}} (\bibinfo {year}
  {2011}{\natexlab{a}})}\BibitemShut {NoStop}%
\bibitem [{\citenamefont {Abadie}\ \emph
  {et~al.}(2011){\natexlab{b}}\citenamefont {Abadie} \emph {et~al.}}]{s6grb}%
  \BibitemOpen
  \bibfield  {author} {\bibinfo {author} {\bibfnamefont {J.}~\bibnamefont
  {Abadie}} \emph {et~al.},\ }\href@noop {} {\bibfield  {journal} {\bibinfo
  {journal} {In preparation}} (\bibinfo {year}
  {2011}{\natexlab{b}})}\BibitemShut {NoStop}%
\bibitem [{\citenamefont {Abbott}\ \emph
  {et~al.}(2009){\natexlab{g}}\citenamefont {Abbott} \emph
  {et~al.}}]{abbott:122001}%
  \BibitemOpen
  \bibfield  {author} {\bibinfo {author} {\bibfnamefont {B.}~\bibnamefont
  {Abbott}} \emph {et~al.} (\bibinfo {collaboration} {LIGO Scientific
  Collaboration}),\ }\Doi {10.1103/PhysRevD.79.122001} {\bibfield  {journal}
  {\bibinfo  {journal} {Phys. Rev.},\ }\textbf {\bibinfo {volume} {D79}},\
  \bibinfo {eid} {122001} (\bibinfo {year} {2009}{\natexlab{g}})}\BibitemShut
  {NoStop}%
\bibitem [{fer()}]{fermiwebsite}%
  \BibitemOpen
  \href@noop {} {\enquote {\bibinfo {title} {{Fermi Gamma-ray space
  telescope}},}\ }\bibinfo {howpublished}
  {\url{http://www.nasa.gov/mission\_pages/GLAST/science/index.html}}\BibitemShut
  {NoStop}%
\bibitem [{ipn()}]{ipnwebsite}%
  \BibitemOpen
  \href@noop {} {\enquote {\bibinfo {title} {{The Interplanetary Network
  Progress Report}},}\ }\bibinfo {howpublished}
  {\url{http://ipnpr.jpl.nasa.gov/index.cfm}}\BibitemShut {NoStop}%
\bibitem [{\citenamefont {Klimenko}\ \emph {et~al.}(2008)\citenamefont
  {Klimenko}, \citenamefont {Yakushin}, \citenamefont {Mercer},\ and\
  \citenamefont {Mitselmakher}}]{Klimenko:2008fu}%
  \BibitemOpen
  \bibfield  {author} {\bibinfo {author} {\bibfnamefont {S.}~\bibnamefont
  {Klimenko}}, \bibinfo {author} {\bibfnamefont {I.}~\bibnamefont {Yakushin}},
  \bibinfo {author} {\bibfnamefont {A.}~\bibnamefont {Mercer}}, \ and\ \bibinfo
  {author} {\bibfnamefont {G.}~\bibnamefont {Mitselmakher}},\ }\Doi
  {10.1088/0264-9381/25/11/114029} {\bibfield  {journal} {\bibinfo  {journal}
  {Class. Quant. Grav.},\ }\textbf {\bibinfo {volume} {25}},\ \bibinfo {pages}
  {114029} (\bibinfo {year} {2008})},\ \Eprint {http://arxiv.org/abs/0802.3232}
  {arXiv:0802.3232 [gr-qc]} \BibitemShut {NoStop}%
\bibitem [{\citenamefont {Harry}\ and\ \citenamefont
  {Fairhurst}(2010)}]{HFspin}%
  \BibitemOpen
  \bibfield  {author} {\bibinfo {author} {\bibfnamefont {I.}~\bibnamefont
  {Harry}}\ and\ \bibinfo {author} {\bibfnamefont {S.}~\bibnamefont
  {Fairhurst}},\ }\href@noop {} { (\bibinfo {year} {2010})},\ \bibinfo {note}
  {in Preparation}\BibitemShut {NoStop}%
\end{thebibliography}%
\end{document}